\documentclass[fleqn,usenatbib]{mnras}

\usepackage{newtxtext,newtxmath}
\usepackage[T1]{fontenc}
\usepackage[dvipsnames]{xcolor}
\usepackage{hyperref}
\usepackage[normalem]{ulem}
\usepackage{environ,censor,calc}
\DeclareRobustCommand{\VAN}[3]{#2}
\let\VANthebibliography\thebibliography
\def\thebibliography{\DeclareRobustCommand{\VAN}[3]{##3}\VANthebibliography}

\usepackage{graphicx}	
\usepackage{amsmath}	
\usepackage{array} 
\usepackage{rotating}
\usepackage{hyperref}

\title[FLAMINGO: X-ray--cosmic-shear]{The FLAMINGO Project: Exploring the X-ray--cosmic-shear cross-correlation as a probe of large-scale structure}

\author[W. McDonald et al.]{
William McDonald$^{1}$\thanks{E-mail: mcdonald@strw.leidenuniv.nl},
Joop Schaye$^{1}$,
Konrad Kuijken$^{1}$,
John Helly$^{2}$,
\newauthor
Joey Braspenning$^{1,3}$
and Matthieu Schaller$^{1,4}$,
\\
$^{1}$Leiden Observatory, Leiden University, PO Box 9513, 2300 RA Leiden, the Netherlands\\
$^{2}$Institute for Computational Cosmology, Department of Physics, University of Durham, South Road, Durham, DH1 3LE, UK\\
$^{3}$Max-Planck-Institut für Astronomie, Königstuhl 17, D-69117 Heidelberg, Germany\\
$^{4}$Lorentz Institute for Theoretical Physics, Leiden University, PO box 9506, 2300 RA Leiden, the Netherlands
}

\date{Accepted XXX. Received YYY; in original form ZZZ}

\pubyear{\the\year{}}

\begin{document}
\label{firstpage}
\pagerange{\pageref{firstpage}--\pageref{lastpage}}
\maketitle

\begin{abstract}
Baryonic feedback processes associated with galaxy formation directly influence the large-scale structure by redistributing gas. Recent measurements of the kinetic Sunyaev-Zel’dovich (kSZ) effect and eROSITA stacks of X-ray emission from optically-selected galaxy clusters suggest that feedback from Active Galactic Nuclei (AGN) is more efficient at expelling gas from low-mass clusters than previously thought. The measurement of the cross-correlation between cosmic shear and diffuse X-ray emission provides a new probe of the distribution of gas in groups and clusters. We use the FLAMINGO cosmological, hydrodynamical simulations to examine the X-ray--cosmic-shear cross-correlation. The cross-correlation is most sensitive to the distribution of gas in haloes with masses $10^{14}\leq M_{200\mathrm{c}}/\mathrm{M}_{\odot}\leq10^{15}$. It is sensitive to the strength of feedback, but the effects of variations in cosmology and baryonic physics are largely degenerate. We compare the FLAMINGO predictions with the cross-correlation between cosmic shear from the Dark Energy Survey and the ROSAT all-sky X-ray map. We account for AGN that would remain unresolved by ROSAT using either the direct FLAMINGO predictions or by abundance matching to the observed (extrapolated) AGN luminosity function. We find that the fiducial FLAMINGO model is in excellent agreement with the ROSAT--DES-Y3 cross-correlation, while models assuming weaker feedback and models with the significantly stronger feedback suggested by kSZ and eROSITA stacks are both incompatible with the observed X-ray--cosmic-shear cross-correlation. 
\end{abstract}



\section{Introduction}

The current standard cosmological model; $\Lambda$CDM, is a highly successful yet relatively simple model that depicts a spatially flat universe whose energy-density is dominated by dark energy; described as a cosmological constant ($\Lambda$), and a form of cold dark matter (CDM) that only interacts gravitationally. A minority of the energy density budget within the $\Lambda$CDM paradigm is comprised of baryons, photons, and neutrinos. 
The standard cosmological model reproduces a wealth of observational data to an impressive level of accuracy \citep[for a relevant overview see][]{Lahav_2022}. 
One key avenue in developing the $\Lambda$CDM paradigm is to exploit tensions between the best-fitting parameters of different probes, where deviations from the $\Lambda$CDM model may indicate shortcomings of the model \citep[for overviews of such tensions see e.g.][]{Abdalla_2022, Peebles_2022, Perivolaropoulos_2022, Peebles_2025, Efstathiou_2025}.

Modern and next-generation cosmology probes, so-called `Stage IV' probes, are testing the standard model on progressively smaller length scales, well into the non-linear regime ($\lambda \lesssim 10$ Mpc). Part and parcel of testing $\Lambda$CDM is the accurate modelling of the formation and clustering of galaxies within dark matter haloes, which includes a multitude of complex physical processes such as those associated with the radiative heating and cooling of gas, the accretion of matter onto black holes (BH), Active Galactic Nuclei (AGN) feedback, the formation of stars and stellar feedback. 
As these baryonic `feedback processes' are inherently complex in nature and can impact large-scale structure (LSS) over a large dynamical range of scales, they are a leading source of uncertainty in modelling LSS formation \citep[see overviews by e.g.][and references therein]{Chisari_2019, Schneider_2019}.
Cosmological hydrodynamical simulations (and models calibrated on them) have shown that AGN feedback, which redistributes (and ejects) baryonic matter within (and out of) haloes, is a dominant source of baryonic suppression of the matter power spectrum on non-linear scales and is strongly linked to the baryonic mass fraction in clusters \citep[e.g.][]{van_Daalen_2011, Semboloni_2011, Semboloni_2013, Van_Daalen_2020, Schneider_2019,Debackere_2020, Arico_2021, Salcido_2023, Schaller_2024}. Due to the large astrophysical uncertainties, it is clear that in order to further understand baryonic feedback, a data-driven approach is of the utmost importance.

Large-volume cosmological hydrodynamical simulations self-consistently model the coupling between LSS formation and baryonic physics.
However, hydrodynamic simulations have a finite resolution and therefore employ empirical subgrid models (prescriptions or recipes) to `mimic' processes which originate on scales below the resolution limits and re-create their effective impact on the scales resolved by the simulation \citep[see overviews given by e.g,][]{Vogelsberger_2020, Crain_2023}. Subgrid prescriptions typically involve free parameters that can be constrained via calibration of the hydrodynamical simulation to relevant observables. 
X-ray measurements provide some of the highest quality insights into the distribution and thermodynamic state of hot gas within clusters and groups, where the inferred gas mass fractions \citep[e.g.][]{Eckert_2016, Akino_2022, Popesso_2024_gas_frac} have served as a benchmark for the calibration of subgrid prescriptions relevant to the efficiency of AGN feedback and, ultimately, the cluster gas mass fractions in cosmological hydrodynamical simulations (e.g., the BAHAMAS \citep[][]{McCarthy_2017}, FABLE \citep[][]{Henden_2018} and FLAMINGO \citep[][]{FLAMINGO_2023, Kugel_2024} simulation suites). 

Recent developments, such as those emanating from the joint analysis of the kinetic Sunyaev-Zel’dovich (kSZ) effect with galaxy-galaxy lensing measurements \citep[e.g.][]{Hadzhiyska_2025, Bigwood_2024, Bigwood_2025b, Siegel_2025} and stacks of optically-selected eROSITA (X-ray observations of) galaxy groups/clusters by \citet[][]{Popesso_2024_gas_frac}, suggest that AGN feedback is more efficient at expelling gas from low-mass clusters than previously thought. 
However, it has been shown by \citet[][]{Eckert_2025} and \citet[][]{Seppi_2026} that the X-ray luminosity and temperature scaling relations of low-redshift galaxy groups from the \textit{XMM-Newton} Group AGN Project \citep[X-GAP][]{Eckert_2024} do not advocate for such an extreme increase in the efficiency of AGN feedback. Instead, \citet[][]{Eckert_2025} finds the X-GAP sample to be more in line with the predictions of hydrodynamical simulations calibrated on cluster gas fractions inferred from pre-eROSITA X-ray-measurements. Whilst, for the same sample \citet[][]{Seppi_2026} find a preference for a mild increase in AGN feedback efficiency that results in a much smaller deviation from the  pre-eROSITA X-ray-measurements than indicated by kSZ and eROSITA measurements. 
This discrepancy could indicate shortcomings in the simulations, but it may also be a symptom of different observables (and measurements) having different systematic errors and being sensitive to different halo masses, redshifts and radii of groups/clusters \citep[e.g., see relevant discussions in][]{Eckert_2025, Smith_2025, Siegel_2025b, Seppi_2026}.
This emphasises that employing multiple, different observational probes is crucial to effectively examine how baryonic feedback influences LSS.

The angular auto- and cross-power spectra of any given observables that trace properties of the matter density field can, depending on the observables, encode a wealth of information about the role of cosmology and baryonic physics in the formation of LSS.
X-ray measurements are sensitive to the thermodynamic properties and spatial distribution of gas in galaxy groups/clusters, in a manner that is complementary to other low-redshift probes such as the thermal Sunyaev-Zel’dovich (tSZ) and kSZ effect measurements.
Angular auto-spectra of and cross-spectra with the observed background of soft X-ray emission from groups/clusters can provide additional constraints on cosmology and baryonic physics that are complementary to, widely adopted, adjacent low-redshift probes (e.g, with cosmic shear, tSZ and kSZ effect measurements). 

In the recent literature there are only a few instances of auto- and cross-correlations utilising soft X-ray emission as probes of LSS. This is in large part due to the limited sensitivity and redshift depth of the ROSAT all-sky survey which was the only X-ray survey with sufficient on-sky coverage until the more recent eROSITA all-sky survey. 
Using a halo model formalism \citep[for a recent review see e.g.][]{Asgari_2023}, measured X-ray auto-correlations have been utilised to initially place upper bounds \citep[due to limitations of ROSAT; see][]{Diego_2003} and then set constraints (with the greater sensitivity of eROSITA) on cosmology \citep[in particular the $\sigma_8$ and $S_8$ parameters \footnote{$S_8=\sigma_8\sqrt{\Omega_{\mathrm{m}}/0.3}$ where; $\Omega_{\mathrm{m}}$ is the present day matter density and $\sigma_8$ is the variance of the linear matter density field averaged over spheres of radius $8 h^{-1}$ Mpc extrapolated to the present day (redshift $z=0$).};][]{Lau_2024}. 
Likewise, with a halo model, the measured angular cross-spectra of X-rays, the tSZ effect and weak-lensing of the cosmic microwave background (CMB) have been shown to be sensitive to changes in cosmology (in particular the $\sigma_8$ parameter), 
and to provide insights into the thermodynamic properties of gas in galaxy clusters and cluster scaling relations \citep[][]{Hurier_2015, Hurier_2017}. 
As noted by \citet[][]{Hurier_2015}, the uncertainty in the cosmological constraints set by X-ray--tSZ cross-spectra are large relative to other probes due to the limitations of the instruments. 
Furthermore, halo models constrained on the cross-spectra of X-rays and cosmic shear detected by \citet[][]{Ferreira_2024} (F24 from here-on), and the joint analysis of X-ray, cosmic shear and tSZ cross-correlations presented by \citet[][]{Posta_2024} have highlighted the sensitivity of these cross-spectra to the thermodynamic properties and spatial distribution of gas in clusters. Both F24 and \citet[][]{Posta_2024} find a greater level of baryonic suppression than given by cosmological hydrodynamical simulations calibrated on pre-eROSITA cluster gas fractions. This discrepancy can be reduced by increasing the strength of AGN feedback within the simulations, although not to the extent suggested by recent kSZ measurements. 
Additionally, X-ray auto- and cross-spectra have been forecast with semi-analytic models, halo occupation distribution (HOD) and halo model formalisms \citep[e.g,][]{Hurier_2014, Singh_2017, Shirasaki_2020, Lakey_2023}. They have also been measured from small patches of the sky, however this leads to larger uncertainties on any extracted constraints due to cosmic variance \citep[e.g,][]{Hajian_2013, Kolodzig_2016, Kolodzig_2017, Zhou_2022}. 

In this work, we specifically focus on the cross-correlation of diffuse soft X-ray emission and cosmic shear. F24 presented the first -- and, at present, only -- detection of this cross-correlation, using maps from the ROSAT all-sky survey \citep[RASS;][]{Voges_1993, Voges_1999} and cosmic shear data from the Dark Energy Survey year-3 data release \citep[DES-Y3;][]{Gatti_2021, Abbott_2022}.
This probe is of particular interest as it may break degeneracies between the effects of cosmology and baryonic physics given it is a measure of how hot gas (its spatial distribution and energy density) traces the underlying total matter field.
Moreover, this probe is well positioned to further contribute to the discussion of the strength of AGN feedback and cluster gas fractions indicated by recent kSZ measurements and stacks of optically selected eROSITA clusters. 

A potential major source of uncertainty in the interpretation of the observed X-ray--cosmic-shear cross-correlation (and for other X-ray auto- and cross-spectra) is contamination by unresolved point sources, particularly the X-ray emission from unresolved AGN. 
Large area X-ray surveys and X-ray observations of clusters are highly susceptible to contamination from unresolved AGN, and it is commonly accepted that AGN dominate the diffuse X-ray background \citep[e.g.][]{Brandt_2005, 2RXS_2016, Comparat_2019, Helgason_2014, Seppi_2022, Brandt_2022, Comparat_2023}. This poses the question of how much of the detected signal in the X-ray--cosmic-shear cross-correlation can be attributed to X-ray emission from unresolved AGN? This issue is not discussed by F24. However, \citet[][]{Posta_2024} predict that unresolved AGN account for $\approx 20 \%$ of the X-ray--cosmic-shear cross-correlation. They find that X-ray emission from AGN is necessary to prevent internal tensions between their halo models constrained by X-ray, tSZ and cosmic shear data. Similarly, \citet[][]{Comparat_2025}, used a halo model formalism to demonstrate that unresolved AGN can have a noticeable impact on X-ray--galaxy-clustering cross-correlations, specifically on small scales ($<80$ kpc). However, \citet[][]{Lau_2024} found, with a HOD model \citep[following][]{Comparat_2023}, that contamination from unresolved AGN has only a $< 10\%$ effect on their eROSITA X-ray auto-correlation.

In addition to the uncertainty surrounding the contamination from unresolved AGN, there is cause to examine the extent to which X-ray emission from gas outside of haloes contributes to the measured X-ray--cosmic-shear cross-correlation. The contribution from diffuse, extra-halo gas is not explicitly captured within the flexible framework of the halo model formalism that has been exclusivity employed up to this point. It is expected that the cross-correlation is generally insensitive to X-ray emission from extra-halo sources \citep[e.g.][]{Posta_2024}, however, self-consistent modelling using hydrodynamical simulations is required to robustly address this question.

In this work we use full-sky lightcones from the FLAMINGO suite of cosmological hydrodynamical simulations \citep[][]{FLAMINGO_2023, FLAMINGO_Calibration_2023} to examine the efficacy of the X-ray–cosmic-shear cross-correlation as a probe of LSS and compare, for the first time, the predictions from self-consistent hydrodynamical simulations to the X-ray--cosmic-shear cross-correlation observed by F24. 
The FLAMINGO suite includes hydrodynamical simulations with various box sizes, resolutions, different cosmologies, and systematic variations in the strength of baryonic feedback (both stellar and AGN feedback).
Using the X-ray--cosmic-shear cross-correlations computed from the full-sky lightcones of FLAMINGO, we explore the sensitivity of this probe to changes in cosmology and baryonic feedback. We demonstrate that the X-ray–cosmic-shear cross-correlation is especially sensitive to changes in AGN feedback strength, but that this effect is largely degenerate with respect to changes in cosmology. 
Additionally, we demonstrate that the contribution of X-ray emission from unresolved AGN should not be neglected. 
We find that the fiducial FLAMINGO model, which was calibrated to reproduce cluster gas fractions inferred from pre-eROSITA X-ray measurements, is consistent with the measurements of F24. However, depending on the level of contamination from unresolved AGN, somewhat stronger baryonic feedback, as favoured by the X-ray scaling relations of \citet[][]{Seppi_2026}, can be reconciled with the measurements of F24. On the other hand, the X-ray–cosmic-shear cross-correlation rules out both weaker feedback, which results in cluster gas fractions greater than in the fiducial model, and the more extreme feedback preferred by observations of the kSZ effect and optically-selected eROSITA clusters.

This paper is organized as follows; In \S \ref{sec:X-ray Lensing cross-spectra using FLAMINGO} we introduce the FLAMINGO suite of simulations and outline the lightcone data products that were used in this work.
In \S \ref{sec:methods} we describe our 2D map based approach for computing X-ray--cosmic-shear cross-correlations from simulated weak-lensing and X-ray all-sky maps. 
In \S \ref{sec:results} we present our main results, which include: the predicted cross-correlations, the dependence of said cross-correlations on cosmology and baryonic feedback, the impact of the X-ray emission from unresolved AGN on the cross-correlation, and comparisons to the observational data.  
In \S \ref{sec:discussion} we discuss our results in the context of other low-redshift probes, before concluding in \S \ref{sec:summary_and_conclusions}.

\section{Simulation datasets} \label{sec:X-ray Lensing cross-spectra using FLAMINGO}

In this section, we introduce the FLAMINGO (Full-hydro Large-scale structure simulations with All-sky Mapping for the Interpretation of Next Generation Observations) suite of simulations used in this paper (\S \ref{sec:flamingo_overview}) and provide an overview of the different data structures of FLAMINGO lightcones utilised in this study (\S \ref{sec:lightcone_overview}).

\subsection{FLAMINGO simulations} \label{sec:flamingo_overview}
Here we provide a short overview of the FLAMINGO simulations, for a complete description of the FLAMINGO simulations, the calibration methods used, and the resultant datasets\footnote{Publicly available at \url{https://flamingo.strw.leidenuniv.nl/data.html}} we refer the reader to \citet[][]{FLAMINGO_2023}, \citet[][]{FLAMINGO_Calibration_2023} and \citet[][]{Helly_2026}, respectively.

The FLAMINGO suite contains both hydrodynamical (N-body+Smooth Particle Hydrodynamic (SPH)) and gravity-only (N-body) cosmological simulations \citep[][]{FLAMINGO_2023} run with the open-source simulation code \textsc{SWIFT}\footnote{Publicly available at \url{https://swift.strw.leidenuniv.nl/}} \citep[][]{SWIFT_2024}{}{} and the \textsc{sphenix} \citep[][]{Borrow_2022}{}{} SPH scheme. Massive neutrinos are included as particles in \textsc{SWIFT} via the $\delta f$ method \citep[][]{Elbers_2021}{}{}. Initial conditions are generated by \textsc{monofonIC} \citep{Hahn_2021}.  
The FLAMINGO suite includes simulations with different cosmological volumes and mass resolutions, in this work we focus on the $(1 ~\mathrm{Gpc})^3$ volume (denoted as `L1') intermediate resolution (labelled as `m9', the number indicates $\log_{10}$ of the mean baryonic particle mass) hydrodynamic simulations, hereafter referred to as L1$\_$m9. For the remainder of this paper, unless specified otherwise, L1$\_$m9 implies that the fiducial cosmological and hydrodynamical models have been assumed. 
The hydrodynamical simulations within FLAMINGO are calibrated on pre-eROSITA\footnote{The cluster gas mass fractions inferred from the stacking of optically selected eROSITA clusters are significantly lower than those FLAMINGO has been calibrated on for haloes with masses in the range of $10^{13.5}\lesssim M_{500\mathrm{c}} / \mathrm{M}_\odot \lesssim10^{14.5}$ \citep[][]{Popesso_2024_gas_frac}}, low-redshift cluster gas fractions ($f_{\mathrm{gas}}$) at $R_{500\mathrm{c}}$ and galaxy stellar mass functions ($\mathrm{M}^{*}$) at $z=0$ via tuning the AGN and stellar feedback through a machine learning assisted approach. 
The mass range that the cluster gas fraction is calibrated over has a lower limit of $M_{500\mathrm{c}}=10^{13.5}~\mathrm{M}_\odot$ and a resolution dependent upper limit. The upper limit for the low (m10), intermediate (m9) and high (m8) resolutions are $10^{13.73}, 10^{14.46}$ and $10^{14.53} ~\mathrm{M}_\odot$ respectively. Conversely, the stellar mass range used for the calibration of the galaxy stellar mass function has a constant upper limit of, $M_{*}=10^{11.5}~\mathrm{M}_\odot$ and a resolution dependent lower limit. For the low (m10), intermediate (m9) and high (m8) resolutions this is $M_{*}>10^{11.17}, 10^{9.92}$ and $10^{8.67} ~\mathrm{M}_\odot$.
We refer the reader to \citet{FLAMINGO_Calibration_2023} for a complete description of the calibration methods. 

The FLAMINGO hydrodynamical models include sub-grid prescriptions of radiative cooling \citep[][]{Ploeckinger_2020}{}{}, star formation \citep[][]{Schaye_Dalla_2008}{}{}, stellar mass loss \citep[][]{Wiersma_2009, Schaye_2015}{}{}, supernova feedback \citep[][]{Schaye_Dalla_2008, Chaikin_2022}{}{}, seeding and black hole growth \citep[][]{Springel_2005, Bahe_2022}{}{}.
Additionally, the FLAMINGO suite contains two different modes of AGN feedback; the fiducial thermally-driven AGN feedback \citep[][]{Booth_Schaye_2009}{}{} and the alternative kinetic AGN jet feedback \citep[][]{Husko_2022}{}{}. 
Note that, importantly for this study, the galaxy clusters from the fiducial FLAMINGO simulation have temperature, density, pressure, and entropy profiles that reproduce those derived from X-ray measurements  \citep[see][]{Joey_2023}.

\subsubsection{Cosmology Variations} \label{sec:flamingo_cosmo_overview}
The assumed fiducial $\Lambda \mathrm{CDM}$ cosmology includes a single massive neutrino species, with a minimal neutrino mass $\Sigma m_\nu = 0.06 ~\mathrm{eV}$ and two massless species. The cosmological parameters are derived from the maximum posterior likelihood values of the Dark Energy Survey Year 3 (DES-Y3) $3\times2\mathrm{pt} + \mathrm{All~External Constraints}$ \citep[see][]{Abbott_2022}. The fiducial cosmological model is labelled as D3A. 
Table \ref{tab:cosmo_table} \citep[reproduced from Table 4 of][]{FLAMINGO_2023} presents the parameters associated with the fiducial, D3A, cosmological model along with those of the six other alternate cosmological models used in this work. 

Of the five cosmologies considered, three are variations of the best-fitting model of \citet[][]{Planck_2020}. The \citet[][]{Planck_2020} model that adopts the minimum neutrino mass, $\Sigma m_\nu = 0.06 ~\mathrm{eV}$, is labelled `Planck'. We include two additional models based on the FLAMINGO Planck cosmology, where the massive neutrino species has a mass of $\Sigma m_\nu = 0.24 ~\mathrm{eV}$ and $0.48 ~\mathrm{eV}$, whilst all other parameters are fixed to the values of the Planck model, except $\Omega_{\mathrm{CDM}}$, which is reduced to accommodate the increased neutrino mass whilst maintaining $\Omega_{\mathrm{m}}$. We denote these models as `PlanckNu0p24Fix' and `PlanckNu0p48Fix' \citep[as described in][]{Elbers_2024}, the neutrino mass (in eV) is indicated by the number value of each label. 

Lastly, we include a `lensing cosmology' from \citet[][]{Amon_2023}{}{} that has a reduced amplitude of the matter power spectrum, giving a lower $S_8$ value, $S_8=0.766$, compared to 0.815 of the fiducial D3A cosmology and 0.833 from Planck. We denote this low $S_8$ simulation as `LS8'. The LS8 simulation has been shown, at the power spectrum level, by \citet[][]{Amon_2023}{}{} to be consistent with observations of galaxy clustering from BOSS DR12 \citep[][]{Reid_2016} and galaxy-galaxy lensing from DES-Y3 \citep[][]{Abbott_2022}{}{}, the Hyper Suprime-cam Survey \citep[HSC;][]{Aihara_2018}{}{} and the Kilo Degree Survey 1000 square degree sample \citep[KiDS-1000][]{Kuijken_2019}{}{}.

\begin{table*}
\centering
        \caption{The values of the cosmological parameters used in different simulations. The columns list the prefix used to indicate the cosmology in the simulation name (note that, for brevity the prefix `D3A' that indicates the fiducial cosmology is omitted from the simulation identifiers); the dimensionless Hubble constant, $h$; the total matter density parameter, $\Omega_\text{m}$; the dark energy density parameter, $\Omega_\Lambda$; the baryonic matter density parameter, $\Omega_\text{b}$; the sum of the particle masses of the neutrino species, $\sum m_\nu c^2$; the amplitude of the primordial matter power spectrum, $A_\text{s}$; the power-law index of the primordial matter power spectrum, $n_\text{s}$; the amplitude of the initial power spectrum parametrized as the r.m.s. mass density fluctuation in spheres of radius $8~h^{-1},\mathrm{Mpc}$ extrapolated to $z=0$ using linear theory, $\sigma_8$; the amplitude of the initial power spectrum parametrized as $S_8\equiv \sigma_8\sqrt{\Omega_\text{m}/0.3}$; the neutrino matter density parameter, $\Omega_\nu \cong \sum m_\nu c^2/(93.14~h^2\,\mathrm{eV})$. Note that the values of the Hubble and density parameters are given at $z=0$.}
	\label{tab:cosmo_table}
	\begin{tabular}{lccccccccccc} 
		\hline
		Identifier & $h$ & $\Omega_\text{m}$ & $\Omega_\Lambda$ & $\Omega_\text{b}$ & $\sum m_\nu c^2$ & $A_\text{s}$ & $n_\text{s}$ & $\sigma_8$ & $S_8$ & $\Omega_\nu$ \\
		\hline
	D3A                & 0.681 & 0.306 & 0.694 & 0.0486 & 0.06~eV & $2.099\times 10^{-9}$ & 0.967 & 0.807 & 0.815 & $1.39\times 10^{-3}$ \\
	Planck          & 0.673 & 0.316 & 0.684 & 0.0494 & 0.06~eV & $2.101\times 10^{-9}$ & 0.966 & 0.812 & 0.833 &   $1.42\times 10^{-3}$ \\
        PlanckNu0p24Fix & 0.673 & 0.316 & 0.684 & 0.0494 & 0.24~eV & $2.101\times 10^{-9}$ & 0.966 & 0.769 & 0.789 & $5.69\times 10^{-3}$   \\
        PlanckNu0p48Fix & 0.673 & 0.316 & 0.684 & 0.0494 & 0.48~eV & $2.101\times 10^{-9}$ & 0.966 & 0.709 & 0.728 & $11.4\times 10^{-3}$   \\
	LS8             & 0.682 & 0.305 & 0.695 & 0.0473 & 0.06~eV & $1.836\times 10^{-9}$ & 0.965 & 0.760 & 0.766 & $1.39\times 10^{-3}$   \\
		\hline
	\end{tabular}
\end{table*}

\subsubsection{Baryonic feedback Variations} \label{sec:flamingo_astro_overview}

Variations of the fiducial model for baryonic feedback are created via shifting $f_{\mathrm{gas}}$ and $\mathrm{M}^{*}$ by a multiple of the uncertainty ($\sigma$) ascribed to the value ($f_{\mathrm{gas}}$ and $\mathrm{M}^{*}$) and then re-calibrating the model to fit these new points. We denote these variations from the fiducial values with fgas$\pm N\sigma$ and M$^{*}\pm N\sigma$.

We consider all the hydrodynamic variations for L1$\_$m9 specified in Tables 1 and 2 of \citet[][]{FLAMINGO_2023}. We utilise simulations with $N_{\sigma}$ variations from the fiducial model in the cluster gas fraction; `fgas$\pm2\sigma$', `fgas$-4\sigma$', `fgas$-8\sigma$', stellar mass function:`$\mathrm{M}^{*}-\sigma$', and a combination of the two; `$\mathrm{M}^{*} - \sigma\_f_{\mathrm{gas}}-4\sigma$'.
Furthermore, we incorporate simulations with the AGN jet feedback models (as opposed to the fiducial thermal feedback model) calibrated on the fiducial cluster gas fractions: `Jet'.
We also include two simulations introduced by \citet[][]{McCarthy_2024}; the first combines the lensing cosmology with strong baryonic feedback ('LS8$\_$fgas$-8\sigma$'), whilst the other is an Adiabatic (`no-cooling') simulation using the fiducial D3A cosmology. 
The LS8$\_$fgas$-8\sigma$ simulation combines the LS8 cosmological (see \S \ref{sec:flamingo_cosmo_overview}) model with sub-grid parameters calibrated to reproduce a significantly reduced cluster gas fraction, fgas$-8\sigma$. The intention of this simulation is to explore degeneracies between astrophysical parameters and cosmology through comparing it to both the LS8 and fgas$-8\sigma$ variations.
The `no-cooling' simulation, sets the net radiative heating$+$cooling rate to zero for gas where the net rate would be negative. The consequence is that there is no-cooling, star formation, BHs and feedback present in the simulation. This simulation is non-physical and serves as an extreme boundary condition.

\subsection{FLAMINGO Lightcones} \label{sec:lightcone_overview}

A key feature of FLAMINGO is the lightcones, which can be used to construct all-sky maps and are therefore an ideal dataset for creating our X-ray--shear cross-correlations. 
These lightcones are built on-the-fly (OTF) and store information in two formats: either as raw particle data (i.e. akin to a simulation snapshot) or compiled as \textsc{HEALPix} maps \citep[][]{Górski_2005}{}{} for various observable features (e.g. X-ray photon count rates, total mass, gas temperature) along the line of sight.
These lightcones are constructed by recording particles crossing the past lightcone of an observer (excluding a sphere with a radius of 3 Mpc centred on the observer). The crossing particles are stored in the particle lightcone (PLC) and \textsc{HEALPix} maps are constructed by projecting said particles' properties onto the sky in concentric spherical shells that are equally spaced in redshift ($\Delta z_{\mathrm{shell}}=0.05$). These \textsc{HEALPix} maps have $\mathrm{N}_{\mathrm{side}}=16384$, corresponding to $12\times (\mathrm{N}_{\mathrm{side}})^2$ pixels with a maximum radius of 13.46 arcseconds.
We point the reader to Appendix A of \citet[][]{FLAMINGO_2023}{}{} for further information on lightcone generation and the construction of \textsc{HEALPix} maps within FLAMINGO. 
In this work, we utilise both forms of the lightcones: the shells of \textsc{HEALPix} maps and the particle data. 

The box sizes of the L1$\_$m9 simulations are not large enough to contain the cosmological volume of a singular lightcone over the redshift ranges required for this work. To counter this, when the comoving radius of the lightcone surpasses the box side length, the box is replicated, which leads to the same structures being observed multiple times, albeit at different cosmological times and thus at different stages in their evolution. This can increase the correlation within full-sky maps on large scales. To account for this box replication we rotate the lightcone shell (either the \textsc{HEALPix} maps or the particle's coordinates on the sky) via a randomised latitude and longitude each time the lightcone's diameter crosses a single box length \citep[e.g.][]{Bernyk_2016, Amol_2024,Broxterman_2024}.

\subsubsection{X-ray models and X-ray lightcone data} \label{sec:FLAMINGO_base_Xray_Lightcones}

The FLAMINGO hydrodynamical datasets contain both X-ray intrinsic-energy and photon flux estimates for gas particles in three different energy bands; eROSITA low (0.2-2.3 keV), eROSITA high (2.3-8.0 keV) and ROSAT (0.5–2 keV). In this paper we make use of the ROSAT band. 
The lightcones contain all-sky maps (in shells of width $\Delta z_{\mathrm{shell}}=0.05$) of each energy band for redshift, z$\leq 3$. The individual photon flux maps are generated --the same way as all smoothed \textsc{HEALPix} maps \citep[see appendix A of][]{FLAMINGO_2023}-- from the X-ray luminosities of `hot' gas particles, as detailed by \citet[][]{Joey_2023}, using emissivity tables\footnote{The UV background assumed in the emissivity tables used to compute the OTF X-ray all-sky maps did not evolve with redshift and was constantly set at redshift $z=0$. %
This has been corrected for the emissivity tables used in this paper. We find the constant UV background has a negligible impact on the  X-ray all-sky maps and X-ray--cosmic-shear cross-correlations} from \textsc{cloudy} \citep[version 17.02][]{cloudy_2017} and are consistent with the radiative cooling rates of \citet[][]{Ploeckinger_2020} that are used in the FLAMINGO simulations.
X-ray luminosities are only determined for `hot' gas particles that are not star-forming (i.e. the star formation rate is zero) and have not recently been heated by AGN feedback. 
Gas particles with temperatures of $10^{5} \mathrm{K} < T/\mathrm{K} < 10^{9.5}$ are `hot' or of a relevant temperature for X-ray groups and clusters. 
As AGN feedback is unrealistic on small scales, gas particles that have been directly heated by AGN feedback can temporarily have extreme temperatures and densities, leading to them being unrealistically X-ray bright. A gas particle is considered recently heated if it has been directly heated by the energy from AGN feedback within the last $15 ~\mathrm{Myr}$ and with has a temperature of $10^{-1} \Delta T_{\mathrm{AGN}} < T/\mathrm{K} < 10^{0.3} \Delta T_{\mathrm{AGN}}$, where the AGN feedback temperature ($\Delta T_{\mathrm{AGN}}$) is a calibrated sub-grid parameter for the thermal model of AGN feedback ($\Delta T_{\mathrm{AGN}}=10^{7.78}~\mathrm{K}$ in the fiducial hydrodynamical model).

\subsubsection{Halo-lightcones and Halo catalogues} \label{sec:Halo_lightcone_overview}
In addition to the PLC data and shells of all-sky maps, we make use of the FLAMINGO halo-lightcones. The FLAMINGO suite employs the modified version of the Hierarchical Bound-Tracing \citep[HBT+, ][]{Han_2012,Han_2018} structure finding scheme HBT-HERONS \citet[][]{Moreno_2025}. Furthermore, the FLAMINGO data products contain a larger catalogue of halo properties computed using the Spherical Overdensity Aperture Processor \citep[SOAP; ][]{McGibbon2025}. The SOAP catalogues contain properties measured from either particles bound to subhaloes or all particles within apertures of a specified radius about a given subhalo. The halo-lightcones link both HBT-HERONS and SOAP catalogues to on-sky coordinates, which allows for a comparison between a subhalo's three-dimensional quantities (from the simulation snapshots) and the two-dimensional, on-sky, counterparts (e.g. as seen in the lightcones). 
The halo-lightcones are constructed (using the \textsc{LightconeIO}\footnote{\url{https://lightconeio.readthedocs.io}} package) by linking each HBT-HERONS output from a given snapshot to the PLC via a tracer particle. In this case the tracer particle is the most bound black hole (BH) particle in a subhalo that exists in both the preceding and proceeding snapshots\footnote{This choice is made to handle the fact that most subhaloes contain multiple BHs. If the most bound BH is not present in the proceeding snapshot, then the second most bound BH is identified as the tracer particle. This iterative searching for the tracer particle continues until a suitable BH is identified.}. 
For each snapshot (and its associated HBT-HERONS catalogue) we read in a spherical shell of the BH PLC that spans half the redshift range to the adjacent snapshots, then for each subhalo in the HBT-HERONS catalogue, each time its tracer particle appears in the PLC the subhalo is placed onto the lightcone at the position of the particle. 
For redshifts $z\leq 3$ the snapshots and therefore the halo-lightcones have the same output cadence as the lightcone shells; $\Delta z =0.05$.
The same rotation scheme as mentioned above (\S \ref{sec:lightcone_overview}) is applied to the halo-lightcones. 

\section{Methods: computation of FLAMINGO observables and cross-correlations} \label{sec:methods}

In this section we describe the various simulated all-sky maps used in this work and provide an overview of the 2D map based approach we use to compute the X-ray--cosmic--shear cross-correlation, $C^{\mathrm{x}\kappa}_\ell$.
In this paper we assume the observed diffuse X-ray background ($\mathbf{X}(\theta)$) is comprised of X-ray emission from: only hot gas, observationally unresolved AGN or a combination of both hot gas and unresolved AGN. We assume that XRBs (X-ray binaries) have a negligible impact on the measured X-ray--cosmic-shear cross-correlation (see Appendix \ref{appendix_diffuse_gas_lightcones}) and given that the total X-ray luminosity from point sources is dominated by AGN \citep[e.g., see relevant discussions in][]{Comparat_2025}. Additionally, we do not consider any effects related to the lensing of X-ray emission from hot gas or point sources \citep[e.g., see relevant discussions in][]{Yang_2025}.  
In \S \ref{sec:Xray_all_sky_maps_methods} we describe how each of the simulated X-ray observables are constructed and introduce a simple model to estimate the intrinsic photon flux expected from AGN from BH particles within a lightcone. In \S \ref{sec:BH_luminosities_and_selection} we describe the different populations of BHs used in this paper and in \S \ref{sec:cross-corr_method} we provide the framework required to estimate the various X-ray--cosmic-shear cross-correlations utilised in this study.

\subsection{X-ray all-sky maps} \label{sec:Xray_all_sky_maps_methods}

\subsubsection{Hot gas all-sky maps} \label{sec:Xray_diffuse_gas_lightcone_methods}

Two of the core goals of this paper are predicated on making comparisons between the X-ray--cosmic-shear cross-correlation predicted by the FLAMINGO simulation suite and the measurements presented by F24. In order to make these comparisons the simulated X-ray all-sky maps need to reflect the RASS data \citep[ROSAT all-sky survey][]{Voges_1993, Voges_1999} utilised by F24. As such, the diffuse gas component of the FLAMINGO X-ray all-sky maps must, at a bare minimum, meet two basic requirements: first, it needs to be in the same energy band (0.5–2 keV) and second, it must account for the effective detector area along with the beam of ROSAT. FLAMINGO's on-the-fly lightcones, as outlined in \S \ref{sec:FLAMINGO_base_Xray_Lightcones}, meet the first requirement, but not the second.

To account for the effective detector area of ROSAT, we construct a new `ROSAT-like' set of all-sky photon flux \textsc{HEALPix} maps in the ROSAT soft X-ray band (0.5–2 keV) from the gas particle data. For each redshift shell ($\Delta z_{\mathrm{shell}} = 0.05$) we recompute the intrinsic photon flux of each gas particle using a slightly modified approach to that used in the FLAMINGO model \citep[][]{Joey_2023}, by convolving the emitted photon energy with the effective area information from the publicly available ROSAT\footnote{\url{https://heasarc.gsfc.nasa.gov/docs/rosat/ruh/handbook/node122.html\#tabeffarea}} on-axis response function and the expected Galactic foreground absorption from the Galaxy. For the latter we make use the of the \textit{tbabs} \citep[][]{Wilms_2000} model from  \textsc{Xspec} \citep[][]{Arnaud_1996} and assume a typical column density of $2\times10^{20}\,\mathrm{cm}^{-2}$ \citep[e.g., as adopted by][]{Biffi_2018, Seppi_2026}, which is constant across the sky. For the fiducial L1$\_$m9 simulation we find this implementation of Galactic foreground absorption decreases the contribution of X-ray emission from hot gas to the X-ray--cosmic-shear cross-correlation in each tomographic bin by $4\%$ on all angular scales we consider. 
We then construct a \textsc{HEALPix} all-sky map for each recomputed shell with the same approach as used for all other FLAMINGO smoothed maps \citep[see appendix A of][]{FLAMINGO_2023}, aside from two minor changes, described in Appendix \ref{appendix_diffuse_gas_lightcones}, which have a negligible impact on our results. 
We refer to these new maps (lightcones and resultant cross-correlations) as `ROSAT-like' given that they account for the effective area of the detector. The ROSAT-like all-sky maps give the photon flux in terms of $\mathrm{photon}~\mathrm{s}^{-1}~(\mathrm{pixel~area})^{-1}$, whereas the base, non-instrument specific photon flux all-sky maps have units of $\mathrm{photon}~\mathrm{s}^{-1}~\mathrm{cm^{-2}}~(\mathrm{pixel~area})^{-1}$. Throughout this paper, unless stated otherwise, the reader can assume that all comparisons to observations are made with the ROSAT-like lightcones, all-sky maps and cross-correlations (convolved with the ROSAT response matrix).

We construct a synthetic all-sky map of the soft X-ray emission from gas ($\mathbf{X}_{\mathrm{gas}}(\theta)$) by integrating along the line of sight,  
\begin{equation}
   \mathbf{X}_{\mathrm{gas}}(\theta)=\int_{0} ^{\chi (z_{\mathrm{max}})} \mathbf{x}(\chi, \theta)\, \mathrm{d}\chi, 
	\label{eq:sum_xray}
\end{equation}
where $\mathbf{x}(\chi, \theta)$ is the photon count rate on the sky at a given comoving distance from the observer, $\chi$. 
When computing $\mathbf{X}_{\mathrm{gas}}(\theta)$, we downgrade (or create, in the case of the ROSAT-like maps) each of the constituent $\mathbf{x}(\chi, \theta)$ maps from $\mathrm{N}_{\mathrm{side}}=16384$ to $\mathrm{N}_{\mathrm{side}}=4096$ (corresponding to an angular resolution of $0.85$ arcmin) using the `ud$\_$grade' function from the \textsc{HEALPix} package. This downgrading is done for the sake of computational efficiency, whilst maintaining sufficient angular resolution to compare to the observational data.

In the process of integrating along the line of sight, we rotate each shell, $\mathbf{x}(\chi, \theta)$, on the sky (as given in \S \ref{sec:lightcone_overview}) via the \textsc{rotator} functions within the \textsc{HEALPix} package and set a maximum multipole (or minimum angular scale) of $\ell_{\mathrm{max}} = 2 \mathrm{N}_{\mathrm{side}}$. This scale cut is made to mitigate the additional noise over smaller angular scales (larger $\ell$ values) that is introduced by rotating the $\mathbf{x}(\chi, \theta)$ maps. The remaining angular scales, $\ell \leq 2 \mathrm{N}_{\mathrm{side}}$, where $\mathrm{N}_{\mathrm{side}}=4096$, are more than sufficient for this study.
Lastly, we note two limitations of gas particle data contained within the PLC of the (1Gpc)$^3$ FLAMINGO simulations. 
Firstly, the lightcones of the Planck, LS8 and massive neutrino simulations do not contain any gas particle data. 
Secondly, for the remaining simulations, the redshift range of the gas particle data stored in the PLC varies with the simulations resolution and box size. The L1$\_$m9 PLC contains all gas particles for redshifts $0 \lesssim z \leq 0.25$ , whilst only storing the X-ray bright gas particles between redshifts $0.25 \leq z \leq 0.5$ \citep[Appendix A of][]{FLAMINGO_2023}. Given the shallow redshift depth of the RASS data, the redshift range of $0 \lesssim z \leq 0.5$ is more than sufficient for this study, as verified  in Fig. \ref{figure:halo_mass_decomposition_example}.

\subsubsection{AGN all-sky maps} \label{sec:AGN_lightcone_methods}

To account for the X-ray emission from AGN, we estimate the photon flux (or photon count rate) from the BH (AGN) particles within the PLC and construct an AGN `ROSAT-like' all-sky map, $\mathbf{X}_{\mathrm{AGN}}(\theta)$, as done for the hot gas in \S \ref{sec:Xray_diffuse_gas_lightcone_methods}. 

The photon count rate specific to ROSAT ($\dot{n}_{\mathrm{ROSAT}}$) is determined by integrating the expected AGN photon emission spectra, $f_{\mathrm{E}}$, over the soft band (0.5 - 2 keV) convolved with the ROSAT response matrix,
\begin{equation}
        \dot{n}_{\mathrm{ROSAT}} = \int_{0.5\,\mathrm{keV}} ^{2.0\,\mathrm{keV}} A_{\mathrm{eff}}(E_{o})\,  f_{E}(N_{H}, E_{o}, \alpha, z, A_{\rm{pl}})\,\mathrm{d}E_{o}.
    \label{eq:count_rate_basic}
\end{equation}
Here, $z$ is the redshift at which the BH entered the lightcone, $\mathrm{E}_{o}$ is the energy of an emitted photon in the observer-frame, $E_\gamma$ is the energy of the emitted photon in the rest-frame of the emitting source ($E_\gamma = E_{o}(1+z)$), $A_{\mathrm{eff}}$ is the effective area of the ROSAT response matrix (hence it is a function of $\mathrm{E}_{o}$) and $\alpha$ is the photon spectral index. 
We construct a simple template for the AGN spectrum by modelling an absorbed power-law, which includes a reflection component, with \textsc{Xspec} \citep[][]{Arnaud_1996}. 
The intrinsic AGN emission spectrum is assumed to be a power-law ($f_{\rm{AGN}} = A_{\mathrm{pl}}\,(E_\gamma /\mathrm{keV}) ^ {-\alpha}~\mathrm{photon\,cm}^{-2} \,\mathrm{s}^{-1}\,\mathrm{keV}^{-1}$) with a typical photon spectral index of $\alpha=1.85$\footnote{We find that the difference between assuming $\alpha=1.8$, 1.85 and 1.9 is negligible and secondary to other assumptions made when estimating BH luminosities from FLAMINGO, as will be discussed in \S \ref{sec:BH_luminosities_and_selection}. A similar conclusion is reached by \citet[][]{Lau_2024} for the range $\alpha=1.5- 1.9$ when examining X-ray auto-correlations} \citep[e.g.,][]{Nandra_2007, 2RXS_2016}  and a normalisation constant, $A_{\mathrm{pl}}$, such that the total intrinsic (de-absorbed) flux $F_{\mathrm{x}}$ ($\mathrm{erg\,s}^{-1}\,\mathrm{cm}^{-2}$) of an AGN is given by the integral 
\begin{equation}
    F_{\mathrm{x}} = A_{\rm{pl}}\, \int_{0.5\,\rm{keV}} ^{2.0\,\rm{keV}} \left[E_{o}(1+z)\right]^{1-\alpha} \, (1+z)^{-1} \,\mathrm{d}E_{o}.
    \label{eq:Fx_intergral}
\end{equation}

The prime emission spectrum is attenuated by the intervening neutral hydrogen column density intrinsic to the host galaxy, $N_\text{H}$. We account for the effects of this intrinsic photoelectric and dust grain absorption with the \textit{ztbabs} model from \textsc{Xspec}. We include an additional reflection component, which is the reflection of the AGN power-law spectrum on cold, optically thick matter, using the \textit{pexmon} model \citep[][]{Nandra_2007}, which gives the spectrum its characteristic Compton hump. We assume a constant relative reflection strength of $R\equiv \Omega/2\pi = 1$, an inclination of $60^\circ$ \citep[e.g.][]{Nandra_2007, Buchner_2014, Buchner_2015, Shen_2020} and solar abundances.
The reflection component is attenuated by the same column density as the intrinsic AGN emission spectrum. 
The absorbed spectra are subject to Galactic foreground absorption from the Milky Way, which we model with \textit{tbabs} and assume a typical constant column density of $2\times10^{20}\,\mathrm{cm}^{-2}$ \citep[e.g., as adopted by][]{Biffi_2018, Seppi_2026}. The absorption in the Milky Way is modelled in the observer frame. When constructing the spectrum templates we assume the solar abundance ratios given by \citet[][]{Asplund_2009}, as done for the X-ray spectra computed for gas particles in FLAMINGO \citep[see][]{Joey_2023}.

It is impractical to recompute the photon count rate for each BH in the lightcone as per equations \ref{eq:count_rate_basic} and \ref{eq:Fx_intergral}. Instead, we construct a table of precomputed photon count-rates given by equation \ref{eq:count_rate_basic}, where $A_{\rm{pl}}$ is defined such that equation \ref{eq:Fx_intergral} results in $F_{\mathrm{x}}=1\, \mathrm{erg\,s}^{-1}\,\mathrm{cm}^{-2}$, for intrinsic column densities of $20 \leq \log_{10}(N_\text{H} /\rm{cm}^{-2}) \leq 26$ in intervals of 0.1 dex and redshifts $z=0-6$ in intervals of $\Delta z=0.02$. 
For each BH in the lightcone the observed photon count rate is then given by multiplying the table value at a given intrinsic column density and redshift by the BH's observed, intrinsic (de-absorbed) flux, $F_{\mathrm{x}}$. 

The total intrinsic (de-absorbed) X-ray flux, $F_{\mathrm{x}}$, in the observer frame is estimated from the intrinsic X-ray luminosity ($L_{\mathrm{x}}$) in the ROSAT band (0.5 - 2.0 keV) as $F_{\mathrm{x}} \, = L_{\mathrm{x}} / (4 \pi \chi^2 (1+z)^2)$ $\mathrm{erg\,cm^{-2}\,s^{-1}}$, where $\chi$ is the comoving distance from the BH to the observer. Since $L_{\mathrm{x}}$ is not directly measurable from the BH particles, we instead compute the total bolometric luminosity ($L_{\mathrm{bol}}$) of each BH particle and then estimate $L_{\mathrm{x}}$ as $L_{\mathrm{x}}= L_{\mathrm{bol}} / K(L_{\mathrm{bol}})$, where $K(L_{\mathrm{bol}})$ is a bolometric correction factor.
We adopt the correction function (equation 5) and the best-fitting parameters (Table 1) of \citet[][]{Shen_2020}. Additionally, to allow for a wider range of bolometric luminosities than fit by \citet[][]{Shen_2020}, we extend the bolometric correction function by ensuring the function asymptotes towards $K(L_{\mathrm{bol}}=10^{42}\,\mathrm{erg}\,\mathrm{s}^{-1})$ and $K(L_{\mathrm{bol}}=10^{49}\,\mathrm{erg}\,\mathrm{s}^{-1})$ for $L_{\mathrm{bol}} < 10^{42}\, \mathrm{erg}\,\mathrm{s}^{-1}$ and $L_{\mathrm{bol}} > 10^{48}\,\mathrm{erg}\,\mathrm{s}^{-1}$, respectively, following the correction functions given by \citet[][]{Lopez_2024} and \citet[][]{Zou_2024}.

The intrinsic column density $N_\text{H}$, for each BH particle, is sampled from the distribution model given by equations 2-6 of \citet[][]{Ueda_2014}, where $f(L_{\mathrm{x}, z; N_{\mathrm{H}}})$ (see appendix \ref{appendix_diffuse_gas_lightcones}) defines the probability distribution of $N_\text{H}$ as a function of redshift and the intrinsic X-ray luminosity in the 2-10 keV band. We apply the bolometric correction function for the hard band (2--10 keV) from \citet[][]{Shen_2020} to correct from hard X-ray luminosities to bolometric luminosities when sampling from this distribution.

The AGN spectrum template used in this work is intentionally simplistic and relies on some basic assumptions (e.g., assuming a constant photon index and angle of inclination) that limit the variability between different BHs. 
Additionally, our adopted template does not account for the soft excess \citep[e.g.][]{Boissay_2016, Waddell_2024} or a soft scattering component (the prime spectra that is scattered back into the line of sight by ionized gas surrounding the AGN), which is typically assumed to be $\approx 2\%$ of the prime spectra \citep[e.g.][]{Murphy_2009, Buchner_2014, Aird_2015, Seppi_2025}. We find that these additional components lead to, at most, a $\approx1.63 \, \%$ (or $\approx0.7\%$ on average) increase in amplitude of the final cross-correlation for the fiducial L1$\_$m9 simulation, which does not affect our conclusions.
Moreover, the assumptions required for these additional components are secondary to the those made in creating our spectrum template, which are in turn much weaker than the assumptions associated with computing the BH X-ray luminosities, which we will introduce in \S \ref{sec:BH_luminosities_and_selection}. Hence, we neglect these additional components as the impact on our results does not outweigh the increased uncertainties and computational cost. 
In upcoming work we will further explore the variability of BH properties (e.g., the assumed photon index and angle of inclination), the intricacies of the models related to the spectrum template (e.g. the geometry of the medium related to Compton scattering of the AGN spectrum) and incorporating more sophisticated templates, such as \textsc{MyTorus} \citep[][]{Murphy_2009} and \textsc{BNTorus} \citet[][]{Balokovic_2018} when estimating the X-ray emission of AGN from BH particles in a FLAMINGO.

\subsubsection{AGN contaminated X-ray all-sky maps} \label{sec:AGN_contaminated_all_sky_map_method}
It is well understood that AGN dominate the soft X-ray background (CXB) and that unresolved AGN are the dominant contaminant in X-ray all-sky surveys \citep[for overviews see][and the references therein]{Brandt_2005, Brandt_2022, Brandt_Yang_2023}. Hence, to compare the predictions from FLAMINGO with measurements and examine the impact of contamination from unresolved AGN on the measured cross-correlation, we must incorporate some level of contamination from unresolved AGN in the X-ray all-sky maps.

We create an AGN contaminated X-ray all-sky map by combining both the AGN ($\mathbf{X}_{\mathrm{AGN}}(\theta)$) and hot gas ($\mathbf{X}_{\mathrm{gas}}(\theta)$) maps with the binary point source masking scheme employed by F24 \footnote{Whilst not out-right specified in F24, we have confirmed via private communication that the same masking procedure and data is replicated in \citet[][]{Posta_2024}.}. 

We initially downgrade both $\mathbf{X}_{\mathrm{AGN}}(\theta)$ and $\mathbf{X}_{\mathrm{gas}}(\theta)$ from $\mathrm{N}_{\mathrm{side}}~4096$ to $2048$ before summing them and masking all pixels that contain a point source above the threshold of $0.02 ~\mathrm{photon}~\mathrm{s^{-1}}$, matching the limits imposed by both F24 and the RASS diffuse X-ray background maps of \citet[][]{Snowden_1997}. The final masked map contains both the X-ray emission from hot gas and unresolved AGN. Pixels that contain point sources (e.g. resolved AGN) are identified using the pixel values of $\mathbf{X}_{\mathrm{AGN}}(\theta)$. 

\begin{figure*}
    \centering
    \includegraphics[width=\textwidth]{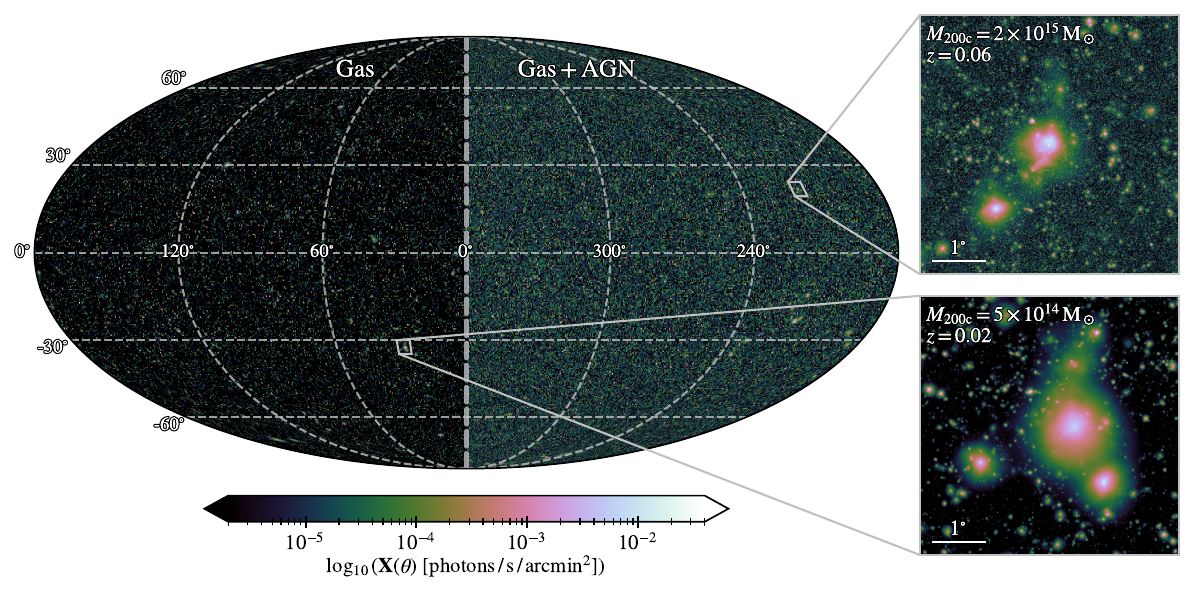}
    \vspace*{-0.5cm}
    \caption{Integrated ROSAT-like X-ray all-sky map ($\mathbf{X}(\theta)$) and $5 \times 5\,\mathrm{deg}^2$ zoom-in regions for the fiducial L1$\_$m9 simulation. The western hemisphere contains only the X-ray emission from hot gas, whilst the eastern hemisphere includes X-ray emission from hot gas and from the base AGN as described in the text (\S \ref{sec:Xray_all_sky_maps_methods} and \S \ref{sec:BH_luminosities_and_selection}). 
    The mass ($M_{200\mathrm{c}}$) and redshift of the clusters shown in the zoom-in regions are printed in the upper left corner of the corresponding panels. }
    \label{figure:all_sky_map_example}
\end{figure*}

An example ROSAT-like X-ray all-sky map ($\mathbf{X}(\theta)$) of the fiducial L1$\_$m9 simulation is shown in Fig. \ref{figure:all_sky_map_example}. The western hemisphere of the map contains only the X-ray emission from hot gas, whilst the eastern hemisphere includes X-ray emission from both hot gas and BH particles (i.e.\ the base AGN, which will be introduced in \S \ref{sec:abundance_matching_motivation_and_method}). Note that the map is shown at $\mathrm{N}_{\mathrm{side}}=4096$, which corresponds to pixels with an area of $\approx0.73~\mathrm{arcmin}^2$. The panels to the right of the all-sky map show $5 \times 5\,\mathrm{deg}^2$ regions of the map centred on clusters that were randomly selected from all-haloes with mass $M_{200\mathrm{c}}\geq5\times10^{14}\,\mathrm{M}_\odot$ and redshift $z < 0.1$.

\subsection{Black Hole Luminosities and Abundances} \label{sec:BH_luminosities_and_selection}

We estimate the X-ray emission from AGN using three different methods: \textit{i)} a base model, which includes all BH particles as they `naturally' exist within the FLAMINGO PLCs, \textit{ii)} a simple abundance matching (AM) model, in which we reassign BH bolometric luminosities to match observations; and \textit{iii)} an extension of this abundance matching model, $\Delta \mathrm{AM}$, where we intentionally perturb the scatter in the BH Luminosity -- maximum circular velocity relation.
We adopt these three different approaches to: \textit{i)} attain a `natural' FLAMINGO prediction of the X-ray emission from AGN; \textit{ii)} correct the abundances of AGN luminosities in FLAMINGO to agree with observed luminosity functions; and \textit{iii)} explore the extent to which X-ray contamination from unresolved AGN can alter the predicted X-ray–lensing cross-correlations. 

\subsubsection{The Luminosities and Abundances of Black Holes in FLAMINGO} \label{sec:base_black_holes_motivation_and_method}

The bolometric luminosities of the `base' BH particles in FLAMINGO are computed from the BH mass accretion rates ($\dot{M}_{\mathrm{accr}}$) as $L_{\mathrm{bol}} = \dot{M}_{\mathrm{accr}} \epsilon_{\mathrm{r}} \mathrm{c}^2$, where we assume a constant radiation efficiency of $\epsilon_{\mathrm{r}}=0.1$ (the relations between $L_{\mathrm{bol}}$, $L_{\mathrm{x}}$ and $\dot{M}_{\mathrm{accr}}$ are depicted in Appendix \ref{appendix_diffuse_gas_lightcones}). The mass accretion rates are given by the FLAMINGO BH subgrid prescriptions \citep[see][]{FLAMINGO_2023}. In this paper, we refer to these BHs as the base BHs and, by extension, the base AGN. 

As shown by \citet[][]{Ding_2025}, these base BHs have an underabundance of both bright and faint quasars relative to observations for redshifts $z\gtrsim1$ (we confirm the same underabundance is replicated by the PLCs in Appendix \ref{appendix_QLF_and_abundance_match}).
The lack of bright quasars is predominantly driven by the spatial and temporal resolution of FLAMINGO being insufficient to resolve the small-scale variability of the processes surrounding BH mass accretion, which drive fluctuations in quasar luminosity. Hence, FLAMINGO is unable to capture the peaks of these fluctuations and instead underestimates the variability in quasar bolometric luminosity, which contributes to the observed underabundance of bright quasars. 
Given these `missing' bright quasars occupy sufficiently high redshifts, the observed photon flux from these quasars may then fall below the minimum flux at which point sources can be resolved. As such, the base AGN may underestimate the total X-ray emission from unresolved AGN.
Accounting for these `missing' bright quasars may increase the impact of AGN contamination on the measured cross-correlation.

\subsubsection{Matching Black Hole Abundances with Observations} \label{sec:abundance_matching_motivation_and_method}

Here, we introduce a simple abundance matching approach to repopulate the lightcones with quasars such that they reproduce, at individual redshifts, the quasar luminosity functions (QLFs) given by \citet[][equation 11 and Table 3, local `polished' fits]{Shen_2020}. 
These QLFs are provided at multiple redshifts from $z=0.2$ to $6.0$ in intervals of $\Delta z=0.2$ and are fit independently at each redshift \citep[see the description of the local fits in Table 2 of][]{Shen_2020}. 

For a given model QLF, at redshift $z_{\mathrm{qlf}}$, we sample bolometric luminosities and assign them to BHs based on their host halo properties as follows:
\begin{enumerate}
    \item For each halo in the lightcone with at least one BH particle\footnote{In the L1$\_$m9 FLAMINGO simulations BHs are seeded in haloes more massive than $\approx 2.76 \times 10^{11}\mathrm{M}_{\odot}$ \citep[][]{FLAMINGO_2023}.} at redshift $z = z_{\mathrm{QLF}}\pm0.1$,\footnote{For the lowest redshift QLF, $z_{\mathrm{QLF}} = 0.2$, we use all haloes within the lightcone at redshifts $z<0.3$.} we place a quasar in the lightcone at the location of the halo's most massive BH (MMBH). 
    
    \item We then use the observed QLF to iteratively sample new bolometric luminosities in discrete logarithmic luminosity bins of width 0.1~dex. In each luminosity bin, $N_i$ new bolometric luminosities are sampled from a uniform distribution that spans the width of the bin. $N_i$ is determined by sampling a Poisson distribution with a mean of $\lambda = \langle N_i \rangle$, where $\langle N_i \rangle$ is the expected number of quasars with luminosities that fall within bin $i$ and is determined from the integral of the observed QLF over the range of the luminosity bin. Starting with the maximum luminosity bin, 
    which is centred on $L_{\mathrm{max}} = 1\times 10^{50}~\mathrm{erg}\mathrm{~s}^{-1}$
    we then continuously repeat this sampling, shifting the centre of each bin by 0.1 dex for each iteration until the total number of sampled bolometric luminosities is equal to the number of selected MMBHs.
    Across all redshifts, we find that the maximum bolometric luminosity sampled does not exceed $L_{\mathrm{max}} = 2\times 10^{49}~\mathrm{erg}\mathrm{~s}^{-1} $ (this corresponds to a BH mass of $\sim 10^{11} \mathrm{M}_{\odot}$ given the Eddington rate limited Bondi-Hoyle rate used within FLAMINGO) whilst the minimum luminosity fluctuates within the range $1\times10^{39} \lesssim L_{\mathrm{min}}/\mathrm{erg~s}^{-1} \lesssim 1\times10^{41}$.
    
    \item The sampled bolometric luminosities are ranked in descending order, whilst the selected MMBHs are ranked in descending order of mass accretion rate. The sampled luminosities are assigned to the BHs via rank, i.e. the MMBH with the greatest mass accretion rate is assigned the highest bolometric luminosity. 
\end{enumerate}
In this paper we will refer to this method and these abundance matched BHs (and by extension AGN), where bolometric luminosities are assigned based on mass accretion rates, as the AM model and AM BHs (and AM AGN).

By design, the AM model imparts the BH Luminosity--halo mass relation inherent to each FLAMINGO simulation onto the AM BHs.
The BH Luminosity–halo mass relation is tied to the contribution from unresolved AGN to the contaminated X-ray--lensing cross-correlation. BH Luminosity is explicitly related to BH X-ray emission (see \S \ref{sec:AGN_lightcone_methods}) and as halo mass scales with density, it is therefore linked to the strength of the lensing signal (see equation \ref{eq:convergence}).
In general, decreasing (increasing) the intrinsic scatter in the BH Luminosity–halo mass relation will strengthen (diminish) the correlation between the X-ray emission from unresolved AGN and the lensing signal therefore, elevating (lowering) the impact of unresolved AGN on the resultant X-ray--lensing cross-correlation. 
Predicated on this idea we now expand on the aforementioned abundance matching approach and intentionally perturb the BH Luminosity–halo mass relation in order to explore the extent to which AGN contamination can impact FLAMINGO's X-ray--lensing cross-correlations. 

In this alternate approach we now rank selected BHs by the host haloes' maximum circular velocity\footnote{Maximum circular velocity is calculated as $v_{\mathrm{max}} = \sqrt{GM(\leq r) / r}$ using all particles bound to the subhalo within the radius $r$, that are further than one softening length from the centre of potential and $r=0$ at the centre of potential.} instead of the mass accretion rate. Prior to ranking each halo, we now perturb the intrinsic scatter in the BH Luminosity -- maximum circular velocity relation by inducing a log-normal scatter onto the haloes' maximum circular velocities. The level of log-normal scatter is given by a free parameter, $\sigma_r$. We refer to these abundance matched BHs and the resultant AGN as $\Delta \mathrm{AM}~\mathrm{BHs}$ and $\Delta \mathrm{AM}~\mathrm{AGN}$. 

The maximum circular velocity is a robust proxy for the total dynamical mass within a halo (see Appendix \ref{appendix_QLF_and_abundance_match}).
Of course, a natural choice would be to rank via halo mass; however, this is rather ambiguous to define in a physically meaningful way (e.g. in terms of $M_{200\mathrm{c}}$), given we use all gravitationally bound haloes with at least one BH particle and the SOAP catalogues only contain $M_{200\mathrm{c}}$ values for central haloes. Hence, we adopt maximum circular velocity as a proxy for halo mass. 
Additionally, the choice to induce a log-normal scatter on only the maximum circular velocity does not fully constrain the BH Luminosity - halo maximum circular velocity relation (or, by extension the BH Luminosity - halo mass relation), which can be done with a more sophisticated model \citep[e.g.][]{Ren_2020, Ren_2021, Elia_2023}. However, we do not require this level of specificity, given the intended use of this abundance matching model is to simply examine what is possible when it comes to AGN contamination (as will be shown in \S \ref{sec:comparison_with_direct_measurements}) and not to constrain an exact level of intrinsic scatter in the BH Luminosity–halo mass relation. 

X-ray--cosmic-shear cross-correlations that incorporate the X-ray emission of unresolved AGN will be affected by the clustering of BHs in FLAMINGO. In our abundance matching approach we have utilised the positions of the most massive BH per halo, which has been shown by \citet[][]{Ding_2025} to reproduce the observed spatial clustering of quasars (with $\log_{10} L_{\mathrm{bol}}\,/\mathrm{erg}\,\mathrm{s}^{-1}\,\gtrsim45.5$) for redshifts $z \lesssim 3$. 

We choose to include all three sets of BHs in our analysis, as each has their own pros and cons. The unresolved AGN given by the base BHs are self-consistent with the FLAMINGO simulations and maintain a `natural' distribution of bolometric luminosities that reflects the different cosmologies and baryonic feedback variations in FLAMINGO. However, the base BHs do not reproduce the observed QLFs. Both sets of abundance matched BHs (AM and $\Delta\mathrm{AM}$) are fixed to reproduce the observed quasar abundances and offer an increased flexibility to explore the potential effects of AGN contamination, but are reliant on some strong assumptions, e.g., only allowing one black hole per halo or that we can sample the model QLF models below the minimum luminosity they are fit to.

\subsection{Computing X-ray -- lensing cross-correlations from FLAMINGO} \label{sec:cross-corr_method}
Here we describe the map-based methodology used in this paper to compute X-ray--lensing cross-correlations from the FLAMINGO lightcones. In \S \ref{sec:Xray_all_sky_maps_methods} we discussed the various integrated X-ray all-sky maps ($\mathbf{X}(\theta)$) that will be cross-correlated with our FLAMINGO based weak-lensing observable, an all-sky map of the convergence field, $\kappa (\theta)$\footnote{We make the distinction of using convergence instead of cosmic shear ($\gamma$), as cosmic shear is not directly measurable from the lightcones 2D maps, whilst the convergence ($\kappa (\theta)$) can be easily computed from the matter overdensity ($\delta$). Given the flat-sky limit implicit in our 2D map-based analysis the resulting E-mode auto(cross)-correlations, $C_{\ell}$, with convergence are equivalent to those obtained from cosmic shear \citep[e.g][]{Kilbinger_2015, Kilbinger_2017}.}. 

We construct convergence maps ($\kappa (\theta)$) from the lightcone total mass maps (total mass implying all matter along the line of sight, including neutrinos) using the Born approximation. As outlined by \citet[][]{McCarthy_2018}, $\kappa (\theta)$ can be computed by integrating $\delta(\theta)$ along the line of sight weighted by the lensing kernel (window Function), $\mathbf{\mathrm{W}^{\gamma _i}}(\chi)$, of the i-th tomographic source sample, 
\begin{equation}
    \mathbf{\kappa} (\theta)=\int_{0} ^{\chi (z_{\mathrm{max})}} \mathbf{\mathrm{W}^{\gamma _i}}(\chi) \delta (\chi, \theta) d\chi,  
	\label{eq:convergence}
\end{equation}
where $\delta (\chi, \theta)$ is the 2D projected density contrast map at comoving distance $\chi$, which can be computed from the surface mass density ($\Sigma(\chi, \theta)$) as $\delta(\chi, \theta) \equiv (\Sigma(\chi, \theta) - \overline{\Sigma}(\chi)) / \overline{\Sigma}(\chi)$ where $\overline{\Sigma}(\chi)$ is the mean surface density.
The window function $\mathbf{\mathrm{W}^{\gamma _i}}(\chi)$ is given by 
\begin{equation}
    \mathbf{\mathrm{W}^{\gamma _i}}(\chi) =\frac{3}{2}  \left(\frac{H_0}{c} \right)  ^2 \Omega_{\mathrm{m}} \frac{\chi}{a({\chi)}} \int_{z(\chi)} ^{z_{\mathrm{max}}} n_{\mathrm{s,i}}(z^{\prime}) \left(\frac{\chi(z^{\prime})-\chi(z)}{\chi(z^{\prime})}\right) d z^{\prime}~,  
	\label{eq:window_function}
\end{equation}
where $a({\chi})$ is the scale factor at a comoving distance $\chi$, $z_{\mathrm{max}}=3$ is the maximum redshift of the L1$\_$m9 lightcone and $n_{\mathrm{s,i}}(z)$ is the normalised redshift source distribution of the i-th sample. We adopt the redshift source distribution of DES-Y3 as given by \citet[][]{Doux_2022} \citep[see the inset panel of Fig. \ref{figure:plot_1} for a normalised version of the source distribution given in figure 2][]{Doux_2022}. As in \S \ref{sec:Xray_all_sky_maps_methods}, we compute $\mathbf{\kappa}$ by downgrading each constituent over-density map ($\delta (\chi, \theta)$) to $\mathrm{N}_{\mathrm{side}}=4096$ and applying the same rotation scheme. The $\kappa (\theta)$ and $\mathbf{X}(\theta)$ all-sky maps (including the binary mask for resolved point sources, if applicable) are then downgraded to $\mathrm{N}_{\mathrm{side}}=1024$ in order to match the angular resolution of the maps used by F24. We then compute the X-ray--convergence cross-correlation with the \textsc{compute$\_$full$\_$master} function of the \textsc{NaMaster}\footnote{\url{https://namaster.readthedocs.io/en/latest/}} package \citep[][]{Alonso_2019}, using the same binning scheme as \citet[][]{Garcia_Garcia_2021} \citep[i.e. Table 2 of][]{Nicola_2021} over the angular scales $30 \leq \ell \leq 3 \mathrm{N}_{\mathrm{side}}$, giving us 24 binned points per $C_{\ell}^{\mathrm{x} \kappa}$ computed. We assume a flat-sky when computing the angular cross-power spectrum, which is sufficiently accurate for angular scales given by $\ell \gtrsim 30$.

In order to make a fair comparison and match the analysis of F24, for all ROSAT-like X-ray--lensing cross-correlations we do not deconvolve the pixel window function from the cross-correlation, but instead apply a Gaussian beam (using the \textsc{gauss$\_$beam} function from the \textsc{HEALPix} package) with a full-width-half-max of 1.8' to account for the ROSAT point-spread function \citep[i.e.][]{Snowden_1997}. In contrast, for all normalised X-ray--lensing cross-correlations (which are not compared to the observational datasets) we deconvolve the pixel window function using the \textsc{pixwin} function in the \textsc{HEALPix} package and do not apply a Gaussian beam.
Lastly, F24 correct for the large variations in the exposure time across RASS \citep[see exposure maps of][]{Snowden_1995, Voges_1999} when computing their ROSAT--DES-Y3 cross-correlation. As such, we are able to make fair comparisons to F24 without accounting for these variations in exposure time in our FLAMINGO ROSAT-like maps. Large variations in exposure time result in significant variations in the density of the detected X-ray sources across the sky, which if not accounted for, would be apparent in the resultant cross-correlation. 

\section{Results} \label{sec:results}

In this section we present the primary results of this study in three parts. 
In \S \ref{sec:results_the_fiducial_model} we present the X-ray--lensing cross-correlation for the fiducial L1$\_$m9  simulation and demonstrate the FLAMINGO simulation's ability to reproduce the directly measured cross-correlation of F24.
In \S \ref{sec:diffuse_gas_results} we identify the contribution from hot gas in different halo masses and examine the cross-correlation's sensitivity to both cosmology and baryonic feedback. 
In \S \ref{sec:comparison_with_direct_measurements} we compare the predictions from FLAMINGO to the measurements of F24.
Through these comparisons we examine the impact of contamination from unresolved AGN on the measured cross-correlation and constrain the suppression of structure formation by baryonic feedback.
We then repeat these comparisons, in \S \ref{sec:comparison_with_direct_measurements_cosmo}, but with a specific focus on the cosmology variations in FLAMINGO.

\subsection{Comparison of the fiducial model to observations} \label{sec:results_the_fiducial_model}

\begin{figure*}
    \centering
    \includegraphics[width=\textwidth]{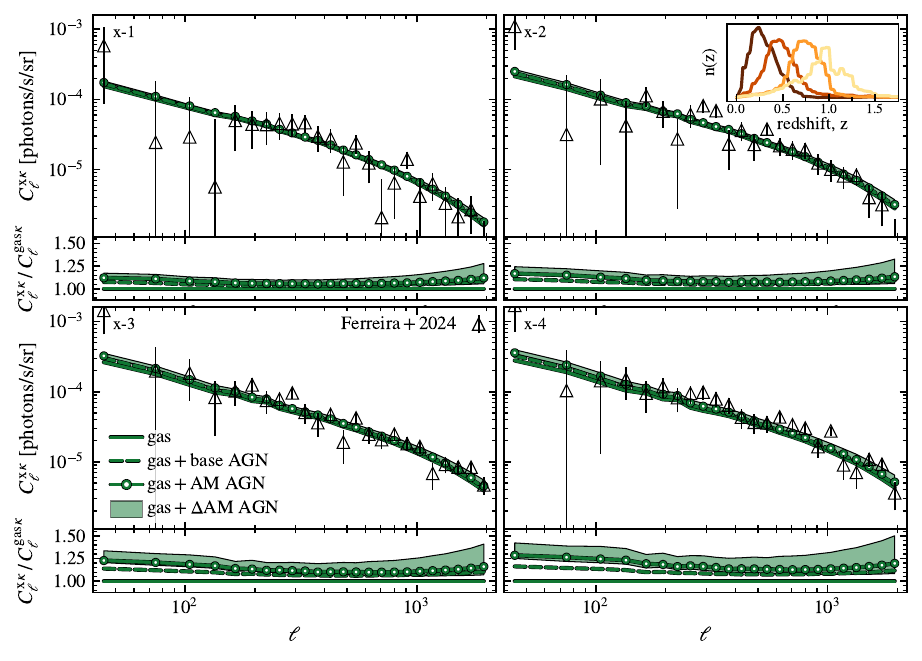}
    \vspace*{-0.5cm}
    \caption{The predicted ROSAT--DES-Y3 X-ray--lensing cross-correlation from the L1$\_$m9 fiducial FLAMINGO simulation (green) compared to the observational data points of F24 (black points and error bars). The numbers in the upper left corner of each panel indicate the DES-Y3 tomographic bin of the shown cross-correlations (e.g. x-1, indicates the cross-correlation of the X-ray signal with the first tomographic bin of the lensing data). Here, the X-ray emission of the L1$\_$m9 simulation is convolved with the ROSAT response matrix (see \S \ref{sec:Xray_diffuse_gas_lightcone_methods}). The solid curves correspond to the predicted cross-correlation if only the X-ray emission from hot gas is included, i.e. it is contamination free. The dashed curves include the X-ray emission from unresolved AGN given by the base FLAMINGO BHs. The circular markers (AM) depict the AGN contaminated cross-correlations when the unresolved AGN are given by the AM BHs. The shaded regions ($\Delta{\mathrm{AM}}$) indicate the full range of possible cross-correlations from minimising and maximising the AGN contamination given by the $\Delta{\mathrm{AM}}$ BHs as described in the text. All sets of BHs (base, AM and $\Delta\mathrm{AM}$) are described in \S \ref{sec:BH_luminosities_and_selection}). The sub panels show each cross-correlation relative to when only the X-ray emission from hot gas is included ($C_{\ell}^{\mathrm{gas}\, \kappa}$). The inset in the top right panel depicts the normalised DES-Y3 source distribution, given by figure 2. of \citet[][]{Doux_2022}, and each tomographic bin (1 through 4, moving left to right) is individually coloured. Without the inclusion of unresolved AGN the FLAMINGO model reproduces the measurements reasonably well. However, AGN contamination increases the power of the cross-correlation over all angular scales considered and can improve the agreement with the data.}
    \label{figure:plot_1}
\end{figure*}

We begin by presenting the predicted ROSAT--DES-Y3 X-ray--lensing cross-correlation ($C_{\ell}^{\mathrm{x} \kappa}$) for the fiducial L1$\_$m9 simulation and making a simple comparison to the measurements of F24. With this comparison we will demonstrate FLAMINGO's capacity to reproduce observations and introduce the different ROSAT-like cross-correlations that we will utilise throughout this paper. 
We compute a ROSAT-like X-ray--lensing cross-correlation, as outlined in \S \ref{sec:cross-corr_method}, for two different scenarios: \textit{i}) where we only include the X-ray emission from hot gas (see \S \ref{sec:Xray_diffuse_gas_lightcone_methods}) and \textit{ii}) when the soft X-ray emission from hot gas is contaminated by unresolved AGN (see \S \ref{sec:AGN_contaminated_all_sky_map_method}). 
In the latter scenario we compute an AGN contaminated cross-correlation using the base AGN and abundance matched AGN given by both the AM BHs and $\Delta{\mathrm{AM~BHs}}$ (see \S \ref{sec:BH_luminosities_and_selection}). 

In Fig. \ref{figure:plot_1} we show the predicted ROSAT--DES-Y3 cross-correlation for the L1$\_$m9 simulation when the soft X-ray background is uncontaminated (solid curve) and when it is contaminated by unresolved AGN given by either the base FLAMINGO BHs (dashed curve) or abundance matched BHs (circle markers and shaded regions) in comparison to the measurements of F24 (triangle markers and error bars). 
The curve indicated by circular markers corresponds to the AGN contamination from the AM BHs (see \S \ref{sec:BH_luminosities_and_selection}). 
The shaded regions correspond to the range of cross-correlations given from the $\Delta\mathrm{AM}$ model (see \S \ref{sec:abundance_matching_motivation_and_method}), where we intentionally perturb the intrinsic scatter in the BH Luminosity -- maximum circular velocity relation whilst constructing a set of abundance matched BHs; $\Delta{\mathrm{AM}}$ BHs. 
The upper (and lower) bound of the shaded regions correspond to when the intrinsic scatter is minimised (and maximised) within the confines of the model. 
The tomographic bin number is shown in the upper left hand corner of each panel (X-ray source distribution -- lensing source distribution) and in the top right panel we display (inset) the normalised form of the DES-Y3 redshift source distribution of each tomographic bin given by Fig. 2 of \citet[][]{Doux_2022}.
Furthermore, in Table \ref{tab:fiducial_model_Nsigma_values_table}, we show the $\chi^2$ values computed in each tomographic bin and, for all tomographic bins combined, the number of standard deviations ($N_{\sigma}$) offset of the predicted L1$\_$m9 cross-correlations with respect to the measurements.
We compute the $\chi^2$ values using the covariance matrices of F24\footnote{The covariance matrices corresponding to the measurements of F24 have been made publicly available by \citet[][]{Posta_2024}} 
with 24 degrees of freedom per tomographic bin and 96 for the whole dataset. The $N_{\sigma}$ values are computed from the sum of the $\chi^2$ values.

Fig. \ref{figure:plot_1} shows that the uncontaminated (hot gas only; solid curve) cross-correlation reproduces the measurements of F24. As summarised in Table~\ref{tab:fiducial_model_Nsigma_values_table}, the predicted cross-correlation is within $1.1 \sigma$ of the observational data points. The cross-correlation increases in power for more distant (i.e. higher redshift) tomographic bins (depicted by the inset in Fig. \ref{figure:plot_1}). The longer path to more distant galaxies allows for a greater lensing effect, which in-turn increases the measured cross-correlation.  Moreover, in higher redshift tomographic bins there is a greater spatial overlap between the the sources of X-ray emission and the intervening mass that produces the lensing signal, allowing for a greater signal of the cross-correlation. For the predicted cross-correlations in bins x-3 and x-4 a majority of the lensing sources are found at redshifts higher than (i.e.\ behind) the highest-redshift X-ray source, which may contribute to the higher signal seen in these tomographic bins. In Table \ref{tab:fiducial_model_Nsigma_values_table} we find that the cross-correlation in the x-1 bin has a significantly lower $\chi^2$ value (i.e.\ a better match between L1$\_$m9 and the measurements) compared to other tomographic bins. However, we may assume that this is primarily driven by the larger uncertainty from the DES-Y3 cosmic shear data for the lowest redshift bin, which is visible in Fig. \ref{figure:plot_1}. 
\begin{table}
\centering
\caption{The $\chi^2$ values ($\chi^2 $), within each tomographic bin, and number of standard deviations ($N_\sigma$) of the predicted X-ray--lensing cross-correlation (as shown in Fig. \ref{figure:plot_1}) with respect to the measurements of F24. We list the $\chi^2$ for each tomographic bin (rows 1 through 4) and $N_\sigma$ values from the sum of the $\chi^2$ values across all tomographic bins (bottom row). There are a total of 24 data points per tomographic bin and 96 in total, the $\chi^2$ values are computed from the diagonal covariance matrices of F24. From left to right we give the summary statistics for the predicted cross-correlations when we only consider the X-ray emission from hot gas (Gas) and when we include the X-ray emission from the base unresolved AGN (Base AGN, dashed curve in Fig. \ref{figure:plot_1}) and abundance matched unresolved AGN (${\mathrm{AM}}\pm~\Delta{\mathrm{AM}}$). We show the summary statistics computed for the contaminated cross-correlations given by the AM BHs (AM, circular markers in Fig. \ref{figure:plot_1}), whilst the super (sub)-scripts given indicate the upper (lower) bounds of the cross-correlations for when the impact of the $\Delta{\mathrm{AM}}$ BHs (the shaded regions in Fig. \ref{figure:plot_1}) is maximised (minimised) within the confines of the model.}
\label{tab:fiducial_model_Nsigma_values_table}
\renewcommand{\arraystretch}{1.4}
\begin{tabular}{l rrr}
\hline
\multicolumn{1}{l}{Bin} & 
\multicolumn{1}{r}{Gas} & 
\multicolumn{1}{r}{Base AGN} & 
\multicolumn{1}{r}{AM$ \pm ~\Delta\mathrm{AM}$} \\
\hline

x-0     &   15.8   &   16.0   & \(   16.1^{+1.2}_{-0.2}\) \\
x-1     &   24.9   &   24.0   & \(   23.6^{+2.2}_{-0.2}\) \\
x-2     &   29.3   &   24.3   & \(   22.3^{+0.6}_{+1.8}\) \\
x-3     &   42.0   &   33.9   & \(   29.5^{-3.3}_{+3.5}\) \\
\hline
$N_{\sigma}$ &    1.1   &   0.2   &\( 0.0^{+0.0}_{+0.1}\) \\
\hline
\end{tabular}
\end{table}

The comparison of the predicted gas-only cross-correlation does not account for AGN that remain unresolved in the RASS dataset. 
As illustrated by the `base AGN' and `AM AGN' curves in Fig. \ref{figure:plot_1}, the inclusion of X-ray emission from unresolved AGN sightly increases the power of the cross-correlation over all angular scales shown, which, as summarised in Table \ref{tab:fiducial_model_Nsigma_values_table} leads to improved agreement with the observed data. 
The `boost' in power from unresolved AGN increases towards both larger and smaller angular scales on either side of $\ell \approx600$, as seen in the sub-panels of Fig. \ref{figure:plot_1}. 
Across all four tomographic bins the unresolved AGN given by the base and AM AGN, on average, account for $\approx7\%$ (ranging between $\approx 3-16 \%$) and $\approx10\%$ (ranging between $\approx 5-22 \%$) of the measured AGN-contaminated cross-correlation's power, respectively.

We can view the limits of the $\Delta\mathrm{AM}$ AGN cross-correlations as an estimate of the possible range of the contamination of the cross-correlation by unresolved AGN (see \S \ref{sec:BH_luminosities_and_selection}), where the lower (and upper) bounds of the shaded regions in Fig.~\ref{figure:plot_1} correspond to the minimum (maximum) levels of AGN contamination. Maximising the level of AGN contamination has a negligible effect on the over-all fit to the observed cross-correlation, whilst minimising the AGN contamination (reducing the signal) leads to a $0.1\sigma$ shift. 
We find that in the first and second tomographic bins the agreement with observations improves slightly (decreases) when AGN contamination is minimised (maximised), as shown by the sub (super)-scripts in Table \ref{tab:fiducial_model_Nsigma_values_table}. The reverse is true for the fourth tomographic bin. However, for the cross-correlations in the third tomographic bin, increasing or decreasing the amplitude of the AM AGN contaminated cross-correlation results in a worse fit to observations.
We leave further discussion of unresolved AGN for \S \ref{sec:comparison_with_direct_measurements}.

\subsection{Origin and robustness of the uncontaminated signal}\label{sec:diffuse_gas_results}

Here we examine the sensitivity of the predicted FLAMINGO X-ray--lensing cross-correlation to baryonic feedback and develop an understanding of where the hot gas (baryonic matter) traced by this probe resides within the LSS.
Contrary to \S \ref{sec:results_the_fiducial_model}, we will use idealised X-ray--lensing cross-correlations (computed as outlined in \S \ref{sec:cross-corr_method}), meaning the X-ray all-sky map, $\mathbf{X}(\theta)$ [$\mathrm{photon}~\mathrm{s}^{-1}~\mathrm{cm^{-2}}~(\mathrm{pixel~area})^{-1}$] is not convolved with the ROSAT response matrix and only includes the X-ray emission from hot gas (i.e. uncontaminated by unresolved AGN) that is not attenuated by the Galactic foreground.

\subsubsection{Decomposition of the soft X-ray flux}\label{sec:decomposition_results}

As shown by \citet[][]{McCarthy_2023}, the cosmic shear auto-correlation has a very weak dependence on baryonic physics for the angular scales we consider. Hence, we can expect the majority of the X-ray--lensing cross-correlation's dependence on baryonic physics to stem from the X-ray component. The range of angular scales, halo masses and redshifts that the angular (and 3D) power spectra from cosmic shear probes are sensitive to is well known \citep[see relevant discussions in e.g.][]{Joachimi_2021, Broxterman_2024, Doux_2025}, however, we have less intuition when it comes to diffuse X-ray emission. 
To understand where the hot gas traced by our FLAMINGO X-ray--lensing cross-correlation comes from, we decompose the X-ray emission by both redshift and halo mass before recomputing the cross-correlation.

\begin{figure}
    \includegraphics[width=\columnwidth]{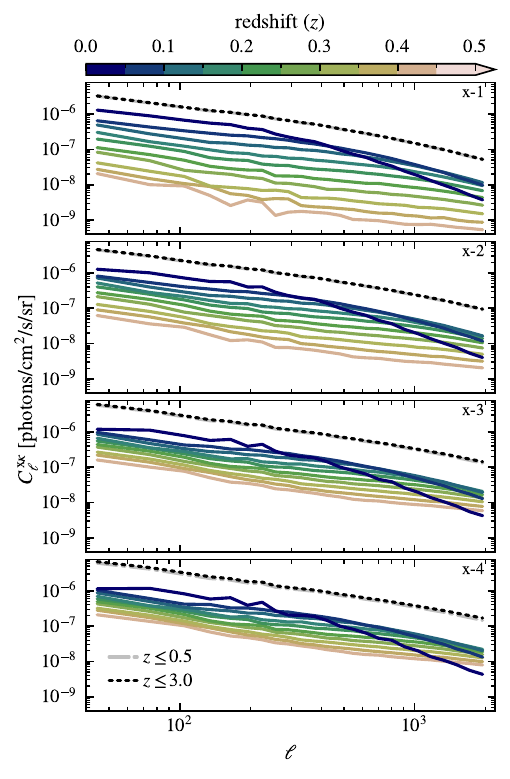}
    \vspace*{-0.5cm}
    \caption{
    The contribution of the diffuse X-ray emission to the X-ray--lensing cross-correlation, predicted by the L1$\_$m9 simulation, decomposed by the redshift of the X-ray emitting hot gas. It is assumed that the X-ray signal is uncontaminated by unresolved AGN. The redshift range of the hot gas is indicated by the colour bar. Additionally, the cross-correlations computed from the total diffuse X-ray signal integrated over the redshift ranges $0\lesssim z \leq 0.5$ (grey dot-dashed) and $0\lesssim z \leq 3.0$ (black dotted) are overlaid for a point of reference. 
    Note that the lensing signal is computed for the maximum redshift depth of the L1$\_$m9 lightcone  ($z\leq 3.0$). The tomographic bin number of each panel is displayed in the upper right corner.
    The contribution to the cross-correlation decreases with the redshift of the X-ray emitting gas and more than $90\%$ of the cross-correlation is due to gas at $z\leq 0.5$. }
    \label{figure:redshift_decomposition}
\end{figure}

We decompose the X-ray all-sky map by the redshift of the gas into 10 shells of equal width ($\Delta z =0.05$; i.e. the same as the concentric shells of the FLAMINGO lightcones) for all gas along the line of sight between $z=0$ and $z=0.5$. With the L1$\_$m9 simulation, we compute the X-ray--lensing cross-correlation (as described in \S \ref{sec:cross-corr_method}) between the photon flux from each shell ($\mathbf{x}(\chi(z), \theta)$) 
and the total convergence map ($\mathbf{\kappa}(\theta)$; equation \ref{eq:convergence}). The total convergence map is computed from the full redshift range of the lightcones, $z \leq 3$, as done in \S \ref{sec:cross-corr_method}.

The resultant cross-correlations are presented in Fig. \ref{figure:redshift_decomposition}. 
The colour of each cross-correlation shown corresponds to the redshift range of the gas as given by the colour bar (each colour indicates a different redshift shell within the lightcone). The tomographic bin number is indicated in the upper right corner of each panel. 
As a point of comparison, we overlay the X-ray--lensing cross-correlations corresponding to the total X-ray emissions from gas for all redshifts $z\leq 0.5$  (grey dot-dashed curve) and $z\leq3.0$ (black dotted curve). These redshift ranges correspond to the maximum redshift of hot gas particles in the PLC and the on-the-fly \textsc{HEALPix} maps of the lightcone.

There are a few notable trends displayed in Fig \ref{figure:redshift_decomposition}. Firstly, across all tomographic bins the power of the cross-correlation decreases as the redshift of the gas increases, which is expected given that photon flux scales as $\propto (\chi(z)^2 (1+z))^{-1}$, where $\chi$ is the comoving distance of the source from the observer. This is made apparent in Fig. \ref{figure:redshift_decomposition}, where $\gtrsim94\%$ of the total soft X-ray contribution to the cross-correlation comes from gas at $z\leq0.5$. 
Secondly, the contribution from gas at $z\leq0.1$ decreases with $\ell$ for $\ell \gtrsim 200$. 
For $z\leq0.1$ there is simply less structure that is able to contribute to the cross-correlation on these scales given the smaller cosmic volume and at such low redshifts, a portion of these angular scales ($\ell > 200$) correspond to physical scales which are unresolved in the lensing signal. 
Note, the lack of structure is due to the cosmological volume, the halo mass function remains consistent for each shell of the lightcone \citep[see FLAMINGO halo mass functions given by][]{FLAMINGO_2023}.
Lastly, with increasing tomographic bin number, the contribution to the measured cross-correlation from higher redshift gas increases relative to lower redshift gas. As the bin number increases, the distribution of lensing source galaxies and structures that produce the lensing effect (the lensing signal) stretch to higher redshifts (see inset in Fig. \ref{figure:plot_1}).
At higher redshifts a greater volume of X-ray emitting hot gas is coincident with the lensing signal and therefore the cross-correlation of the two fields increases. 

Continuing in a similar manner, we decompose the X-ray all-sky map by total halo mass, $M_{200\mathrm{c}}$ (i.e. the total mass encompassed by the radius, $R_{200\mathrm{c}}$, within which the mean mass is 200 times the critical density) and compute the cross-correlation with the total convergence map ($\mathbf{\kappa}(\theta)$; equation \ref{eq:convergence}).
We match the gas particles in the lightcones to the nearest redshift occurrence of the same particle, by particle ID\footnote{A unique identifying number for each particle.}, in the snapshot data. We allow a maximum difference between the redshift of the snapshot and the redshift at which the particle entered the lightcone of $\Delta z = 0.025$, as the snapshots are output with a cadence of $\Delta z =0.05$ for the needed redshift range. 
Using the snapshot data, we identify a gas particle as `belonging' to a halo if the distance ($r$) between the particle and the halo's centre is less than $R_{200\mathrm{c}}$ for the given halo. If a gas particle `belongs' to multiple haloes, we assign the particle to the closest halo in terms of $r/R_{200\mathrm{c}}$. 
We consider only central (field) haloes. 

\begin{figure}
    \includegraphics[width=\columnwidth]{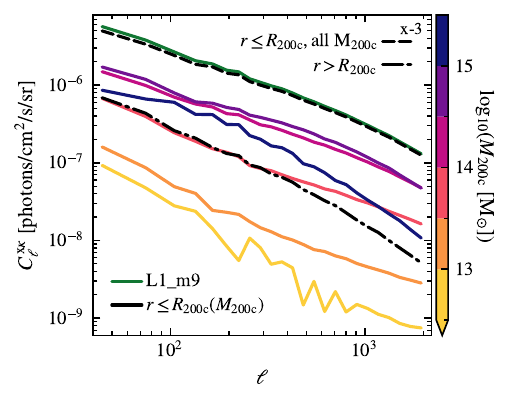}
    \vspace*{-0.5cm}
    \caption{The contribution to the X-ray--lensing cross-correlation of the L1$\_$m9 simulation decomposed by the mass ($M_{200\mathrm{c}}$) of the halo responsible for the X-ray emission. Here, it is assumed that the X-ray emission is only due to hot gas (i.e. uncontaminated by unresolved AGN). The X-ray all-sky map is decomposed into the contributions from gas within $R_{200\mathrm{c}}$ of all haloes ($r \leq R_{200\mathrm{c}},\mathrm{all~} M_{200\mathrm{c}}$, dashed black curve), haloes of a given mass, ($r>R_{200\mathrm{c}}(M_{200\mathrm{c}})$, solid curves) and outside of haloes ($r > R_{200\mathrm{c}}$, dot-dashed black curve). The colour of each solid curve corresponds to the halo mass range shown in the colour bar. Additionally the cross-correlation computed using all gas is overlaid in green. The X-ray emission from gas within haloes of masses $10^{14} \leq M_{200\mathrm{c}} / {\mathrm{M}}_\odot < 10^{15}$ dominates the predicted cross-correlation. 
    }
    \label{figure:halo_mass_decomposition_example}
\end{figure}

Fig. \ref{figure:halo_mass_decomposition_example} compares the contribution to the cross-correlation from the X-ray signal of gas inside $R_{200\mathrm{c}}$ of all haloes ($r \leq R_{200\mathrm{c}},\mathrm{all~} M_{200\mathrm{c}}$, black dashed curve), haloes within a given mass range ($r \leq R_{200\mathrm{c}}(M_{200\mathrm{c}})$, indicated by the colour bar) and gas outside of haloes ($r>R_{200\mathrm{c}}$, dot-dashed curves). 
The cross-correlation computed from all hot gas within the PLC (L1$\_$m9) is overlaid in green.
We only display the third tomographic bin (x-3) as there is little difference between bins (we repeat this figure for all tomographic bins in Appendix \ref{appendix_additional_results}).

From Fig. \ref{figure:halo_mass_decomposition_example} it is clear that the cross-correlation is dominated by gas within $R_{200{\mathrm{c}}}$ of haloes, which accounts for $90-97\%$ of the signal. Specifically, we find that haloes with $M_{200\mathrm{c}}$ $\sim 10^{14} \mathrm{M}_\odot$ dominate for $\ell > 250$, and haloes of mass $M_{200\mathrm{c}} \geq 10^{14} \mathrm{M}_\odot$ dominate for larger angular scales, where $\ell < 250$. X-ray emission from haloes of $M_{200\mathrm{c}} < 10^{13.5} \mathrm{M}_\odot$ accounts for less than $5\%$ of the total cross-correlation. 
As such, we can expect the X-ray--lensing cross-correlation to be most sensitive to baryonic feedback processes that are closely linked to cluster gas fractions. 
The cross-correlation from haloes with $M_{200\mathrm{c}} < 10^{13}~\mathrm{M}_\odot$ is noisier than the cross-correlation for any other mass range shown in Fig. \ref{figure:halo_mass_decomposition_example}. 
At a glance this may appear strange as there are many more haloes with $M_{200\mathrm{c}} < 10^{13}~\mathrm{M}_\odot$ in the lightcone than haloes with $M_{200\mathrm{c}} \ge 10^{13}~\mathrm{M}_\odot$, however because of their low luminosities, only a limited number of these haloes at very low redshifts significantly contribute to the cross-correlation. 

\subsubsection{Sensitivity to cosmology and baryonic feedback}\label{sec:sensitivity_to_cosmology_and_astrophysics}

\begin{figure*}
    \centering
    \includegraphics[width=\textwidth]{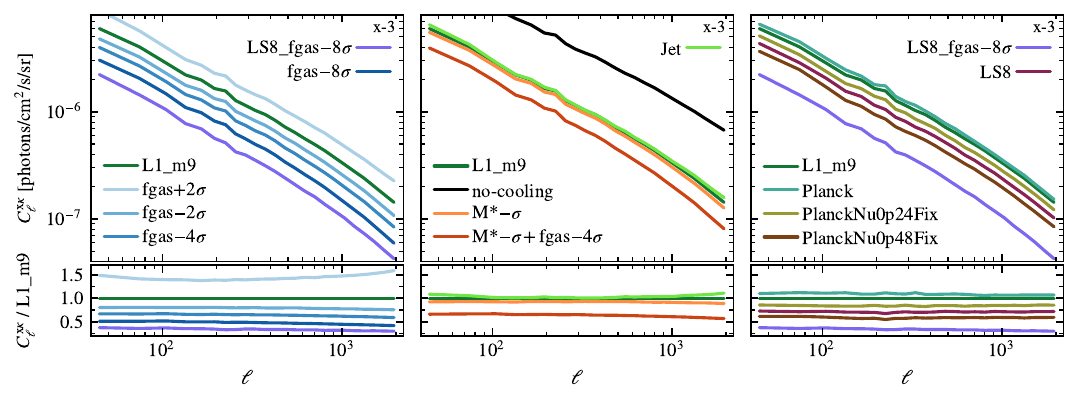}
    \vspace*{-0.5cm}
    \caption{
    The cosmology and baryonic feedback dependence of the predicted X-ray--lensing cross-correlation. 
    In each main panel, the coloured curves correspond to the predicted cross-correlations in the third tomographic bin (x-3) for the different L1$\_$m9  simulations as indicated by the legend (see \S \ref{sec:flamingo_overview} and Table \ref{tab:cosmo_table}). The lower sub-panels compare the variations 
    to the fiducial L1$\_$m9 (dark green) simulation (i.e. $C_{\ell}^{\mathrm{x} \kappa }/\mathrm{L1}\_\mathrm{m9}$). 
    Note we only consider the X-ray emission from hot gas and do not convolve the X-ray emission with the response matrix of any telescope. 
    \textit{Left}: The dependence on the cluster gas fraction ($\mathrm{fgas}$) that the model has been calibrated to. 
    Reducing the cluster gas fraction (mainly by increasing the strength of AGN feedback) suppresses the power of the cross-correlation over all angular scales considered. 
    \textit{Centre}: The predicted cross-correlations for the remaining baryonic feedback variations. 
    Reducing the stellar mass function suppresses the signal on all scales, however this effect is secondary to varying the cluster gas fraction. 
    The choice of AGN feedback model (jet or thermal mode of feedback) has a very minor scale-dependent effect.
    \textit{Right}: The dependence on cosmology and neutrino mass.
    The cross-correlation for the LS8 cosmology is suppressed relative to the fiducial model over all scales shown. Reducing the cluster gas fraction for the LS8 cosmology (LS8$\_\mathrm{fgas}-8\sigma$) further suppresses the signal. 
    Increasing the summed neutrino mass relative to the fiducial value ($0.06$ eV) reduces the power over all scales shown. 
    Comparing between the three panels, it is clear that the effects of varying the cosmology and baryonic feedback are largely degenerate. 
    }
    \label{figure:normalised_diffuse_gas_example}
\end{figure*}

Armed with some intuition about the X-ray--lensing cross-correlation's sensitivity to hot gas within haloes of different masses, we now explore the baryonic feedback and cosmology dependence of this probe. 
We particularly wish to identify the extent to which the cross-correlations can break degeneracies between the effects of varying either the cosmology or baryonic feedback. 

In Fig. \ref{figure:normalised_diffuse_gas_example} we compare the X-ray--lensing cross-correlations for the different L1$\_$m9 simulations, showing the dependence on cosmology and baryonic physics (i.e. subgrid prescriptions). Each of the simulations shown, as indicated by colour, is described further in \S \ref{sec:flamingo_overview}.
Note that here the soft X-ray emission only traces hot gas, i.e. it is uncontaminated by unresolved AGN, and it is not convolved with the response matrix from any telescope. For simplicity only the third tomographic bin (x-3) is shown, but there is little evolution between bins (the cross-correlations in all four bins can be found in Appendix \ref{appendix_additional_results}). 
The left panel of Fig. \ref{figure:normalised_diffuse_gas_example} examines the dependence on the strength of stellar and AGN feedback parametrized through the cluster gas fraction used to calibrate the subgrid prescriptions. The gas fraction is primarily mediated via the strength of AGN feedback. 
We compare the X-ray--lensing cross-correlations for each of the cluster gas fraction variations in FLAMINGO (the gas fraction is varied over the range of $+2\sigma$ to $-8\sigma$ from the observed halo mass - gas fraction relation). 
There is a clear dependence on the AGN feedback strength, where decreasing the cluster gas fraction (increasing the strength of feedback) leads to an almost uniform suppression of the predicted cross-correlation over all angular scales considered. 
The inverse is shown for an increased gas fraction. 
A similar effect is identified by \citet[][]{Shirasaki_2020} (Fig. 5), where, using a halo model formalism, the computed X-ray--lensing cross-correlation decreases almost uniformly when the model allows for a greater amount of energy to be injected into a halo by AGN and supernova feedback. 

The centre panel of Fig. \ref{figure:normalised_diffuse_gas_example} shows the remaining baryonic feedback models. This includes the $-1\sigma$ variations in the stellar mass function (M*$-\sigma$ and M*$-\sigma$\_fgas$-4\sigma$) the kinetic jet model of AGN feedback calibrated on the fiducial cluster gas fraction (Jet) and the adiabatic (no-cooling) simulation. The no-cooling simulation demonstrates that ignoring baryonic feedback and cooling results in a dramatic increase in power over all angular scales. 
Reducing the stellar mass function (SMF) suppresses the predicted cross-correlation on all scales, as seen by the orange curve in Fig. \ref{figure:normalised_diffuse_gas_example}. However, the cross-correlation is less sensitive to variations in the SMF in comparison to the cluster gas fraction. 
This can be understood with our conclusions from Fig. \ref{figure:halo_mass_decomposition_example}. The X-ray component is dominated by hot gas in haloes with mass $M_{200\mathrm{c}} \sim 10^{14} ~\mathrm{M}_\odot$. As such, for the angular scales we consider in this paper, the X-ray--lensing cross-correlation is more sensitive to baryonic feedback processes that can more efficiently alter the cluster gas fraction, i.e. AGN feedback.
The physical processes, such as supernova feedback, which mediate the SMF characteristically impact the distribution of baryonic matter on smaller physical scales than AGN feedback. 
We find that the mode of AGN feedback only has a minor effect with the jet mode of AGN feedback (Jet) having a small increase in power relative to the thermal mode (the fiducial L1$\_$m9 simulation) at both the largest (small $\ell$) and smallest (large $\ell$) angular scales shown. 
As discussed by \citet[][]{Joey_2023}, the two AGN feedback modes redistribute and expel gas from clusters in different ways, which results in the cluster temperature and density profiles for each AGN feedback mode slightly diverging within cluster cores. Such changes in the core temperatures and densities could drive the minor differences we see in the X-ray--lensing cross-correlations. 

We now turn our attention to the dependence on cosmology.
The right panel of Fig. \ref{figure:normalised_diffuse_gas_example} compares the cross-correlations for the fiducial D3A cosmology (L1$\_$m9), the `Planck' cosmology with varying neutrino masses; $0.06$, $0.24$ and $0.48$ eV (Planck, PlanckNu0p24Fix and PlanckNu0p48Fix respectively) and the `lensing' cosmological model for calibrated on the fiducial and $-8\sigma$ cluster gas fractions (LS8 and LS8$\_$fgas$-8\sigma$, respectively).
We see a clear dependence on cosmology and the summed neutrino mass.
The near uniform offset of the cross-correlations for each cosmology shown relative to the prediction for the fiducial D3A cosmology are ranked in order of the $\sigma_8$ cosmological parameter (see Table \ref{tab:cosmo_table}). 

Comparing between the sub panels of Fig. \ref{figure:normalised_diffuse_gas_example}, it becomes apparent that the effects of cosmology and baryonic physics are largely degenerate, at least for the angular scales considered in this study, as both result in offsets from the fiducial model that are of similar magnitude and are nearly independent of angular scale. 

\subsubsection{Sensitivity to box size and resolution}\label{sec:sensitivity_to_boxsize_and_resolution}

\begin{figure}
    \includegraphics[width=\columnwidth]{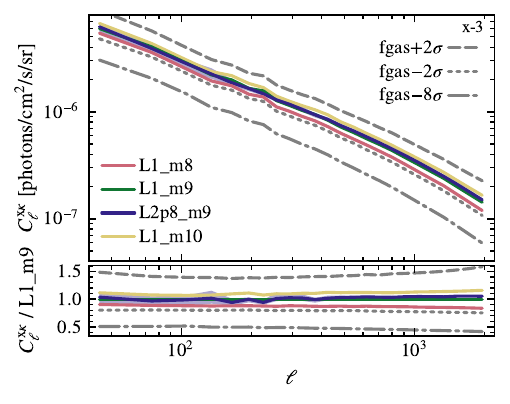}
    \vspace*{-0.5cm}
    \caption{
    The dependence of the predicted X-ray--lensing cross-correlation on the simulation box size and numerical resolution. For clarity we show only the third tomographic bin and assume the soft X-ray emission is uncontaminated by unresolved AGN.  The blue shaded region and solid curve correspond to the scatter and median for each of the 8 individual observers (lightcones) for the L2p8$\_$m9 simulation. The remaining solid curves correspond to a single lightcone for a given resolution model at a fixed box side length of $1~\mathrm{Gpc}$. The grey curves correspond to varying the cluster gas fraction by $+2\sigma$ (dashed), $-2\sigma$ (dotted) and $-8\sigma$ (dot-dashed) from the fiducial L1$\_$m9 model. In the bottom panel each simulation is shown relative to L1$\_$m9. The cross-correlation is converged with box size but not with resolution, as shown by the suppressed (elevated) signal of the high (low) resolution, L1$\_$m8 (L1$\_$m10), model relative to the intermediate resolution (L1$\_$m9 and L2p8$\_$m9) models. The dependence on resolution is likely indirect, reflecting residual differences in cluster gas profiles despite the recalibration. However, the differences between the different resolutions are smaller than the differences between the feedback variations.}
    \label{figure:boxsize_resolution_example}
\end{figure}

In Fig. \ref{figure:boxsize_resolution_example} we examine the dependence on the simulation box size and resolution. 
Here the soft X-ray emission only originates from hot gas, i.e. it is uncontaminated, and it has not been convolved with the response matrix from any telescope. Again, for clarity only the third tomographic bin (x-3) is shown as there is little difference between the four tomographic bins (we compare all four tomographic bins in Appendix \ref{appendix_additional_results}).
The blue shaded region and curve correspond to the median and scatter of the X-ray--lensing cross-correlations for the 8 different lightcones of the larger box size (L2p8$\_$m9) simulation. The remaining solid coloured curves are the predictions for a singular lightcone for a simulation with a given resolution (indicated by the legend), whilst the grey curves (dot, dashed and dot-dashed) indicate different cluster gas fraction variations of the fiducial L1$\_$m9 simulation. 
We show each simulation relative to L1$\_$m9 in the bottom panel. 
Recall that the subgrid prescriptions (e.g. the baryonic feedback models) have been calibrated on the same observed cluster gas fractions for each resolution; low (L1$\_$m10), intermediate (L1$\_$m9 and L2p8$\_$m9) and high (L1$\_$m8). 
Therefore, the different resolutions should be considered as separate models of baryonic physics. 

From Fig. \ref{figure:boxsize_resolution_example} we see that the predicted X-ray--lensing cross-correlations are converged with box size, but not with resolution. 
The features seen in shaded region for $\ell \lesssim 400$ indicate that cosmic variance -- while decreasing towards smaller scales -- is driven by the presence of a few, rare, massive clusters which have a disproportionate impact on the power of the cross-correlation. This is supported by the conclusions from Fig. \ref{figure:halo_mass_decomposition_example}, where a few hundred haloes of mass $M_{200\mathrm{c}} > 10^{15}~\mathrm{M}_\odot$ (that is $\lesssim2\%$ of haloes more massive than $M_{200\mathrm{c}} > 10^{13}~\mathrm{M}_\odot$) account for $15-20\%$ 
of the X-ray signal in the cross-correlation for $\ell \lesssim 400$. 
Relative to the cross-correlation for the intermediate-resolution (L1$\_$m9) simulation, the high-resolution (L1$\_$m8) simulation has less power on all scales and shows a general trend decreasing more for smaller angular scales. This trend is mirrored by the low-resolution (L1$\_$m10) prediction. 

This lack of convergence with resolution is likely not directly caused by the mass resolution itself, 
but is instead due to the imperfect recalibration of the subgrid models and the differences in the thermodynamic profiles for fixed gas fractions \citep[see relevant discussions by][]{FLAMINGO_2023, FLAMINGO_Calibration_2023}.
As shown in \citet[][]{FLAMINGO_Calibration_2023}, at redshift $z=0.1$, the median gas fraction ($f_{\mathrm{gas},M_{500\mathrm{c}}}$) of haloes with mass $M_{500\mathrm{c}}>1.5\times10^{14}~\mathrm{M}_\odot$ for L1$\_$m8 is less than that of L1$\_$m9. Note, that L1$\_$m8 is calibrated to reproduce the cluster gas fractions for haloes with masses between $M_{500\mathrm{c}} = 10^{13.5}$ and $10^{13.75} ~\mathrm{M}_\odot$, therefore the divergence from L1$\_$m9 occurs outside the calibrated halo mass range. Additionally, $f_{\mathrm{gas},M_{500\mathrm{c}}}$ of haloes with $M_{500\mathrm{c}}>5\times10^{14}~\mathrm{M}_\odot$ for the fgas$-2\sigma$ variation of L1$\_$m9 is comparable to the gas fractions of L1$\_$m8. Given that both the gas fraction (Fig. \ref{figure:normalised_diffuse_gas_example}) and haloes within these mass ranges  (see Fig. \ref{figure:halo_mass_decomposition_example}) are significant for the cross-correlation, it is understandable that the X-ray--lensing cross-correlation of the high-resolution simulation is suppressed relative to the fiducial L1$\_$m9 simulation and is similar to the fgas$-2\sigma$ simulation (further discussion can be found in Appendix \ref{appendix_scaling_relations}).
If we compare the differences between the cross-correlations for the different resolutions with those between the different baryonic feedback variations and cosmologies in Fig. \ref{figure:normalised_diffuse_gas_example}, we find that the effect of varying the mass resolution is secondary to both cosmology and baryonic feedback. 

\subsection{Comparing FLAMINGO feedback variations to observations} \label{sec:comparison_with_direct_measurements}

Having examined the sensitivity of the X-ray--lensing cross-correlation to both cosmology and baryonic feedback, we now turn our attention to making comparisons to observational data. Through these comparisons we explore which, if any, of the baryonic feedback variations in FLAMINGO are compatible with the measurements of F24. To compare the predictions from the FLAMINGO suite with these measurements, we now include unresolved AGN and convolve the X-ray emission (from both hot gas and unresolved AGN) with the ROSAT response matrix before computing the ROSAT--DES-Y3 X-ray--lensing cross-correlation.

For the remainder of this section we will only utilise the FLAMINGO simulations with variations of the cluster gas fraction (fgas$-N\sigma$) for the fiducial D3A cosmology and the fgas$-8\sigma$ variation for the lensing cosmology: LS8$\_$fgas$-8\sigma$. We opt to exclude the stellar mass function variations (M$^* - N\sigma$) as the X-ray--lensing cross-correlation is significantly more sensitive to changes in the cluster gas fraction (see Fig. \ref{figure:normalised_diffuse_gas_example}). 
Additionally, we do not include the adiabatic (no-cooling) simulation because it is not realistic and, given the discussion in \S \ref{sec:diffuse_gas_results}, we omit the simulations using the jet-like AGN feedback (Jet) as there is only a minor deviation from the cross-correlation of the fiducial L1$\_$m9 simulation.  

\begin{figure}
    \centering
    \includegraphics[width=\columnwidth]{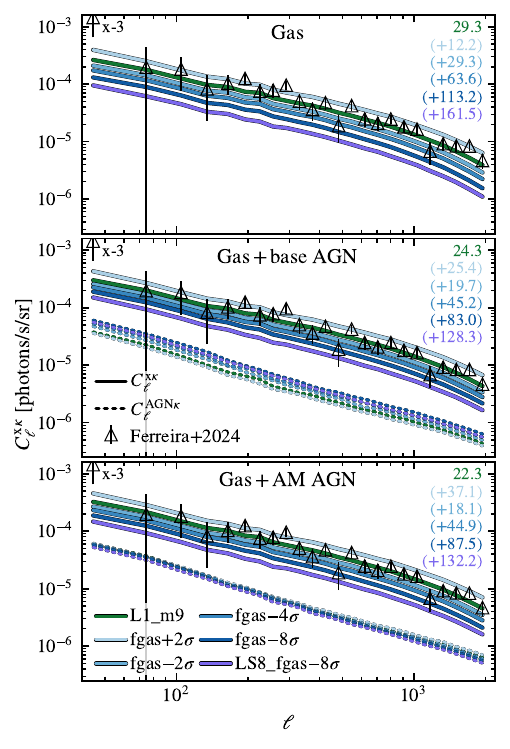}
    \vspace*{-0.5cm}
    \caption{
    The predicted ROSAT--DES-Y3 X-ray–lensing cross-correlation, in the third tomographic bin (x-3), for the gas fraction variations (fgas$\pm N\sigma$) and fiducial L1$\_$m9 simulation (solid curves) compared to the measurements of F24 (black points and error bars). All X-ray emission is convolved with the ROSAT response matrix. \textit{Top}: The predicted cross-correlation when only the X-ray emission from hot gas is included. \textit{Centre}: We include the X-ray emission from both hot gas and unresolved AGN given by the base BHs. \textit{Bottom}: The same as the centre panel, but the AGN contamination is given by the AM BHs. The base AGN and AM AGN are described in \S \ref{sec:BH_luminosities_and_selection}. The dotted curves give the cross-correlation of the X-ray emission from each set of unresolved AGN and DES-Y3 all-sky maps ($C_{\ell}^{\mathrm{AGN} \kappa}$) for the simulation of the corresponding colour. The $\chi^2$ value of the cross-correlation, in the third tomographic bin, for the fiducial L1$\_$m9 simulation (dark green) and the observational data is printed on the right hand side of each panel. The values in parentheses are the $\Delta \chi^2$ (see text) for each cluster gas fraction (indicated by colour). The inclusion of unresolved AGN increases the strength of the cross-correlation.}
    \label{figure:convolved_native_agn_example}
\end{figure}

In Fig. \ref{figure:convolved_native_agn_example} we compare the predicted ROSAT--DES-Y3 X-ray--lensing cross-correlations for each of the cluster gas fraction variations to the measurements of F24 (black points and error bars). Note that all X-ray emission (from both hot gas and unresolved AGN) is convolved with the ROSAT response matrix and for clarity we only show the cross-correlations in the third tomographic bin. 
The solid curves show the cross-correlations for each gas fraction variation (indicated by colour) when we only account for the X-ray emission from hot gas (top panel) and when we include the X-ray emission from unresolved AGN given by the base BHs (centre panel) or AM BHs (bottom panel). Both models for unresolved AGN are described in \S \ref{sec:BH_luminosities_and_selection}. 
The dotted curves give the AGN X-ray--lensing cross-correlation, $C_{\ell}^{\mathrm{AGN}\, \kappa}$, which is the cross-correlation between the X-ray emission from unresolved AGN and the FLAMINGO lensing all-sky maps. 
Finally, for each simulation we compute the $\chi^2$ value between the predicted and observed X-ray--lensing cross-correlations in each tomographic bin. The $\chi^2$ value for the fiducial L1$\_$m9 simulation, computed for the third tomographic bin, is printed on the right hand side of each panel. The values listed below in parentheses indicate the difference between the $\chi^2$ values for each gas fraction variation (indicated by colour) and the fiducial L1$\_$m9 simulation; $\Delta \chi^2 = \chi^2 - \chi_{\mathrm{L1}\_\mathrm{m9}}^2$, in the third tomographic bin. A positive (negative) $\Delta \chi^2$ value indicates worse (better) agreement with the data. 
We repeat Fig. \ref{figure:convolved_native_agn_example} for all four tomographic bins in Appendix \ref{appendix_additional_results}. 

We summarise our comparisons of the cross-correlations predicted by FLAMINGO and observations in Table \ref{tab:x2_native_AGN}, which presents the sum of the $\chi^2$ values in each tomographic bin and the number of standard deviations ($N_{\sigma}$) between each predicted ROSAT-DES-Y3 X-ray--lensing cross-correlation and the measurements of F24.

\begin{table}
\centering
\renewcommand{\arraystretch}{1.3}
    \caption{
        The $\chi^2$ values and number of standard deviations, $N_{\sigma}$, between the predicted ROSAT--DES-Y3 X-ray--lensing cross-correlations for the cluster gas fraction variations of the L1$\_$m9 simulation shown in Fig. \ref{figure:convolved_native_agn_example} and the measurements of F24. The $\chi^2$ and $N_{\sigma}$ values are for the dataset as whole, combining the values computed in each tomographic bin (24 data points per bin) with the covariance matrices given by F24. 
        From left to right, the summary statistics shown are computed for when the cross-correlation is uncontaminated (Gas), contaminated by the unresolved AGN given by the Base AGN or from the AM AGN (AM).
    }\label{tab:x2_native_AGN}
        \begin{tabular}{l p{0.5cm}p{0.5cm} p{0.5cm}p{0.5cm} p{0.5cm}p{0.5cm}}
    \hline
    \multicolumn{1}{l}{Identifier} & \multicolumn{2}{c}{Gas} & \multicolumn{2}{c}{Base AGN} & \multicolumn{2}{c}{AM} \\
    &
    {\(\chi^2\)} & {\(N_\sigma\)} &
    {\(\chi^2\)} & {\(N_\sigma\)} &
    {\(\chi^2\)} & {\(N_\sigma\)} \\
    \hline
    
    L1$\_$m9               & $112.1$	&$1.1$     & $ 98.2$   & $0.2$     &$ 91.4$	&  $0.0$\\
    fgas$+2\sigma$         & $147.7$	&$3.3$     & $168.8$   & $4.4$     &$193.6$	&  $5.5$\\
    fgas$-2\sigma$         & $194.4$	&$5.6$     & $154.9$   & $3.6$     &$143.9$	&  $3.1$\\
    fgas$-4\sigma$         & $291.3$	&$9.5$     & $228.6$   & $7.1$     &$220.8$	&  $6.7$\\
    fgas$-8\sigma$         & $429.9$	&$13.8 $   & $336.0$   & $11.0$    &$340.7$	&  $11.1$\\
    LS8$\_$fgas$-8\sigma$  & $564.1$	&$17.3 $   & $462.6$   & $14.7$    &$465.2$	&  $14.8$\\
    
    \hline
        \end{tabular}
\end{table}

In the top panel of Fig. \ref{figure:convolved_native_agn_example}, we find the same dependencies on the strength of baryonic feedback (parametrised through the cluster gas fraction used to calibrate the subgrid prescriptions) as for the idealised counterparts (as depicted in Fig. \ref{figure:normalised_diffuse_gas_example}).
When we neglect AGN contamination, the fiducial L1$\_$m9 model fits the observed data points more closely than the other cluster gas fraction variations, as demonstrated by the positive $\Delta \chi^2$ values. Moreover, simulations calibrated to lower (higher) cluster gas fractions than the fiducial L1$\_$m9 simulation underestimate (overestimate) the observational data, with the size of the discrepancy increasing as the model deviates further from the fiducial cluster gas fractions. 
As shown in Table \ref{tab:x2_native_AGN}, in this gas-only scenario only the fiducial L1$\_$m9 simulation is compatible (i.e. $N_{\sigma}=1.1$) with the observed dataset, whilst all other simulations considered are offset by at least $3.3 \sigma$.

As shown in the centre and bottom panels of Fig. \ref{figure:convolved_native_agn_example}, the X-ray emission from unresolved AGN increases the amplitude of the predicted cross-correlations on all angular scales (as discussed in \S \ref{sec:results_the_fiducial_model}). 
The increase in power from AGN-contamination further improves the already good agreement between the fiducial L1$\_$m9 simulation and the observational data points, as summarised in Table \ref{tab:x2_native_AGN}. 
Moreover, the inclusion of AGN-contamination reduces the discrepancy between observations and simulations calibrated to yield lower cluster gas fractions (which have an increased strength of baryonic feedback), as indicated by the smaller $\Delta \chi^2$ values of the AGN-contaminated cross-correlations compared to those listed for the gas-only cross-correlations in the top panel of Fig. \ref{figure:convolved_native_agn_example}.  
Yet, as summarised in Table \ref{tab:x2_native_AGN}, this increase in power from AGN-contamination is insufficient to fully reconcile the stronger baryonic feedback variations with observations. We find the fgas$-4\sigma$ ($N_\sigma \geq 6.7$), fgas$-8\sigma$ ($N_\sigma \geq 11$) and LS8$\_$fgas$-8\sigma$  ($N_\sigma \geq 14.7$) variations to be highly incompatible with the measurements of F24. 
While the fgas$-2\sigma$ simulation is inconsistent with the data at the $N_\sigma \ge 3.1$ level, we will show in \S\ref{sec:abundance_matching_results} that the more flexible $\Delta$AM model can bring it into agreement. The inclusion of AGN-contamination from the base and AM AGN results in the predicted fgas$-2\sigma$ cross-correlations being offset from the observed dataset by $3.6\sigma$ and $3.1\sigma$, respectively.
Last, the inclusion of AGN-contamination results in the fgas$+2\sigma$ simulation (calibrated to a higher cluster gas fraction and weaker baryonic feedback strength) further overestimating the observed data points, such that it is offset by $\geq4.4 \sigma$. Hence, even with the inclusion of AGN contamination the fiducial L1$\_$m9 simulation remains the only simulation fully consistent with the measurements of F24.

Fig. \ref{figure:convolved_native_agn_example} demonstrates that the contribution of unresolved AGN to the total cross-correlation increases for simulations calibrated to lower cluster gas fractions (stronger prescriptions of baryonic feedback). This is due to the gas only X-ray--lensing cross-correlation being more sensitive to changes in the cluster gas fraction than the AGN X-ray--lensing cross-correlations (dotted curves). 
Contamination by unresolved AGN (for both the base AGN and AM AGN) therefore marginally reduces the sensitivity of the X-ray--lensing cross-correlation to changes in the strength of baryonic feedback (cluster gas fraction).

Fig. \ref{figure:convolved_native_agn_example} shows that the base AGN X-ray--lensing cross-correlation, $C_{\ell}^{\mathrm{AGN} \kappa}$, increases with feedback strength whilst the AM AGN X-ray--lensing cross-correlation remains far more consistent across all the simulations shown. The greater variation seen in the base AGN X-ray--lensing cross-correlation is expected as the AM BH luminosities are consistently sampled from the same luminosity functions (described in \S \ref{sec:BH_luminosities_and_selection}) and, compared to the base BHs, are less effected by the differences between each simulations subgrid prescriptions that influence the seeding, growth, feedback and accretion rates of BHs. 
However, in general we do not find a large difference between the contributions of the X-ray emission from the base and AM AGN to the cross-correlation. On average, across all tomographic bins, we find that for the fiducial L1$\_$m9, fgas$+2\sigma$, $-2\sigma$, $-4\sigma$, $-8\sigma$ and LS8$\_$fgas$-8\sigma$ simulations the X-ray emission from the base AGN account for $\approx7,\,4,\,10,\,14,\,20$ and $25\%$ of the cross-correlations' signal, whilst the AM AGN contribute $\approx10,\,8,\,13,\,15,\,19$ and $22\%$ of the signal, respectively. We show the contribution to the cross-correlation from AGN-contamination for individual tomographic bins in Appendix \ref{appendix_additional_results}. 
The expected (and potential maximum) increase in the signal of the predicted cross-correlation due to contamination from unresolved AGN is not negligible and should thus be taken into account. 

The final effect we consider in relation to the impact of AGN point sources, both resolved and unresolved by ROSAT, comes from the binary mask itself. As stated in \S \ref{sec:AGN_contaminated_all_sky_map_method}, when we construct contaminated X-ray all-sky maps, we follow the binary masking scheme of F24 and simply mask all pixels containing a resolved point source (i.e.\ any pixel containing an AGN point source with a count rate  $> 0.02~\mathrm{photon ~s}^{-1}$). The binary mask reduces the signal of the cross-correlation, which is primarily driven by the masking of the more X-ray bright resolved point sources, as intended. If we recompute the AGN-contaminated cross-correlation for the fiducial L1$\_$m9 simulation without masking these resolved point sources the amplitude of the cross-correlation increases by between $\approx2 \%$ and $\approx10 \%$ (for contamination by either the base or AM AGN) across all four tomographic bins. In addition to this desired primary effect, there are undesired secondary effects of the binary mask that further affect the amplitude of the cross-correlation. These secondary effects stem from X-ray bright AGN being biased to be located within the central regions of gas-rich and likely X-ray bright haloes, where both the uncontaminated X-ray and the lensing signals peak. The binary mask is therefore correlated with the underlying lensing maps and masking these resolved point sources also covers the X-ray emission from hot gas, which can further reduce the cross-correlation signal \citep[as noted by][]{Posta_2024}. However, we find that these secondary effects are negligible in comparison to the desired primary impact of masking the X-ray emission from the resolved point sources (see Appendix \ref{appendix_binary_mask}).

\subsubsection{Varying the contamination by unresolved AGN}\label{sec:abundance_matching_results}

As indicated by Fig. \ref{figure:convolved_native_agn_example}, contamination from unresolved AGN can have a non-negligible impact on the X-ray--lensing cross-correlation. 
To further examine the effects of AGN contamination and account for the uncertainty associated with our abundance matching approach, we now utilise the $\Delta\mathrm{AM}$ BHs introduced in \S \ref{sec:abundance_matching_motivation_and_method}. 
The $\Delta\mathrm{AM}$ BHs are similar to the AM BHs, except that we perturb the BH Luminosity -- maximum circular velocity relation (the maximum circular velocity is a robust proxy for halo mass) to alter the impact of contamination from unresolved $\Delta\mathrm{AM}$ AGN on the X-ray--lensing cross-correlation. 
Recall that for each population of $\Delta\mathrm{BHs}$, we induce a log-normal scatter onto the maximum circular velocities of all haloes prior to re-assigning the BH luminosities by the rank order of the selected haloes' maximum circular velocities. The level of log-normal scatter is given by a free parameter, $\sigma_r$. 

\begin{figure}
    \centering
    \includegraphics[width=\columnwidth]{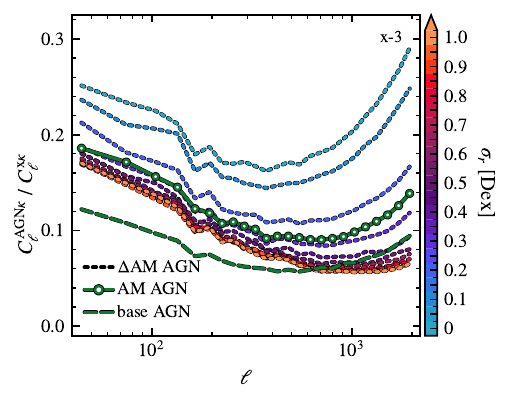}
    \vspace*{-0.5cm}
    \caption{The contribution from unresolved AGN to the predicted ROSAT--DES-Y3 X-ray--lensing cross-correlation for the fiducial L1$\_$m9 simulation. 
    Each curve shows the fraction of the predicted cross-correlation that is attributed to the cross-correlation of the X-ray emission from unresolved AGN and the lensing signal ($C_{\ell}^{{\mathrm{AGN}} \kappa} / C_{\ell}^{{\mathrm{x}} \kappa}$) for different values of  $\sigma_r$, the log-normal scatter in the abundance matching of AGN luminosities and halo circular velocities (indicated by colour). The tomographic bin is indicated in the upper right corner. The contributions from unresolved AGN as given by the base BHs and AM BHs respectively (as shown in Fig. \ref{figure:convolved_native_agn_example}) are overlaid for comparison. A larger value of $\sigma_{r}$ increases the intrinsic scatter of the BH Luminosity–halo mass relation, which reduces the contribution from unresolved AGN to the X-ray--lensing cross-correlation. AGN contamination is maximised when $\sigma_{r}=0$ dex.}
    \label{figure:AM_params_example}
\end{figure}

In Fig. \ref{figure:AM_params_example} we use the fiducial L1$\_$m9 simulation to demonstrate how the contribution from unresolved AGN to the measured cross-correlation varies with $\sigma_{r}$. 
Here, we show the cross-correlation of the X-ray emission from only unresolved $\Delta\mathrm{AM}$ AGN and the lensing signal, $C_{\ell}^{\mathrm{AGN} \kappa}$, as a fraction of the `full' X-ray -- lensing cross-correlation (i.e. the X-ray component includes the X-ray emission from both hot gas and unresolved AGN),  $C_{\ell}^{{\mathrm{x}} \kappa}$. For clarity we only show the third tomographic bin. Each dotted curve corresponds to a different population of $\Delta \mathrm{AM}~\mathrm{BHs}$ parametrised by $\sigma_{r}$, which is indicated by the colour bar. For reference, we overlay the contributions from unresolved AGN given by the base BHs (dashed green curve) and the AM BHs (green curve with circular markers).

Fig. \ref{figure:AM_params_example} affirms that the level of AGN contamination can be increased relative to that from the AM AGN by decreasing the intrinsic scatter of the BH Luminosity -- maximum circular velocity relation. Incremental changes in $\sigma_r$ alter the level of AGN contamination on all angular scales, however, the effect is greater on smaller angular scales. The level of AGN contamination is maximised when we set $\sigma_r=0$ dex, i.e.\ when we minimise the scatter in the BH Luminosity -- maximum circular velocity (and halo mass) relation. We find that the relation between $\sigma_r$ and the level of AGN contamination from $\Delta \mathrm{AM}~\mathrm{AGN}$  saturates for $\sigma_r>0.4$ dex. For larger $\sigma_r$ the ranking (or distribution) of the maximum circular velocities becomes so `shuffled' by the induced log-normal scatter that any underlying (or pre-existing) correlation between the unperturbed (i.e.\ before including log-normal scatter) maximum circular velocities and BH luminosities is erased. 

\begin{figure}
    \centering
    \includegraphics[width=\columnwidth]{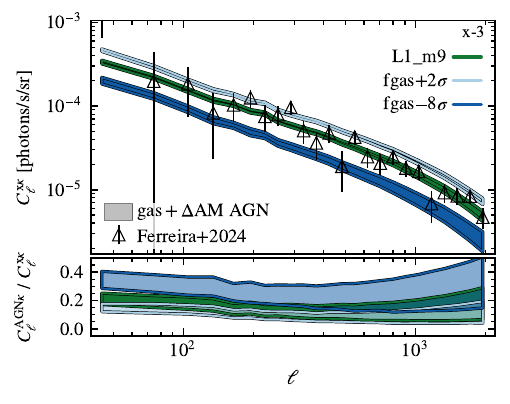}
    \vspace*{-0.5cm}
    \caption{The ROSAT--DES-Y3 X-ray–lensing cross-correlations, in the third tomographic bin (x-3), for the fgas$+2\sigma$ and fgas$-8\sigma$ variations of the L1$\_$m9 simulation compared to the measurements of F24 (black points and error bars). 
    All X-ray emission is convolved with the ROSAT response matrix. The shaded regions correspond to the range of X-ray--lensing cross-correlations contaminated by $\Delta \mathrm{AM}$ AGN with $0 \leq \sigma_r \leq 1$ dex (see description in the text). The sub-panels show the ratio of the X-ray--lensing cross-correlations when only the X-ray emission from unresolved AGN is included ($C_{\ell}^{\mathrm{AGN} \kappa}$) and when both the X-ray emission from hot gas and unresolved AGN are accounted for ($C_{\ell}^{\mathrm{x} \kappa}$). }
    \label{figure:delta_example}
\end{figure}

Fig. \ref{figure:delta_example} compares the range of $\Delta \mathrm{AM}~\mathrm{AGN}$ contaminated cross-correlations, in the third tomographic bin, for the fiducial L1$\_$m9,  fgas$+2\sigma$ and fgas$-8\sigma$ simulations to the measurements of F24. The shaded regions outline the range of $\Delta \mathrm{AM}~\mathrm{AGN}$ contaminated cross-correlations that corresponds to varying $\sigma_r$ between $1.0$ dex (minimum contamination) and $\sigma_r=0.0$ dex (maximum contamination). The bottom panel shows the contribution from unresolved AGN to the `full' contaminated cross-correlation, as shown in Fig. \ref{figure:AM_params_example}.  
As indicated by the increasing spread of the shaded regions in Fig. \ref{figure:delta_example}, the cross-correlations of simulations calibrated to reproduce lower cluster gas fractions are more sensitive to changes in the level of AGN contamination (and therefore $\sigma_r$). This is expected given that the contribution from unresolved AGN to the cross-correlation increases for lower cluster gas fraction models, as re-affirmed in the bottom panel of Fig. \ref{figure:delta_example}.   

\begin{figure}
    \centering
    \includegraphics[width=\columnwidth]{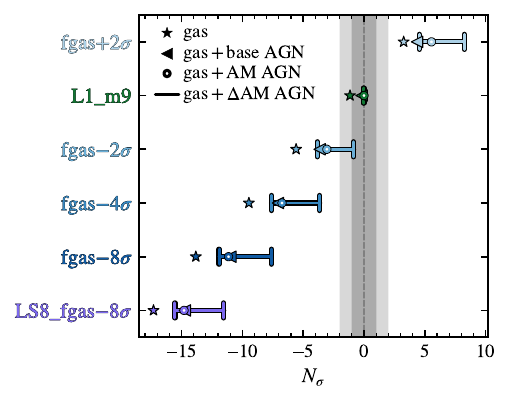}
    \vspace*{-0.5cm}
    \caption{
    The number of standard deviations ($N_{\sigma}$) between the FLAMINGO X-ray--lensing cross-correlations and the measurements of F24 when unresolved AGN are neglected (Gas; star markers) and when the X-ray emission from unresolved AGN is given by: the base AGN (triangle markers), AM AGN (circle markers) or $\Delta{\mathrm{AM~AGN}}$ (error bar) models are included. Values of $N_{\sigma}$ are computed from the sum of the $\chi^2$ values across all tomographic bins.
    The minimum (maximum) $N_{\sigma}$ values, i.e. the left (right) tick on the error bar, for the $\Delta{\mathrm{AM~AGN}}$ contaminated cross-correlations correspond to the AGN contamination being minimised (maximised) within the confines of the  $\Delta{\mathrm{AM~AGN}}$ model (see \S \ref{sec:BH_luminosities_and_selection}). A positive $N_{\sigma}$ value indicates that the predicted cross-correlation overestimates the observations and the inverse is true for negative values. }
    \label{figure:AM_summary_stat}
\end{figure}

Finally, for each of the $\Delta \mathrm{AM}~\mathrm{AGN}$ contaminated cross-correlations we compute the sum of the $\chi^2$ values in each tomographic bin and the number of standard deviations ($N_{\sigma}$) with respect to the measurements of F24 (as in Table \ref{tab:x2_native_AGN}). 
In Fig. \ref{figure:AM_summary_stat} we visualise the range of $N_{\sigma}$ values computed for each of the simulations listed in Table \ref{tab:x2_native_AGN} (additionally, the $\chi^2$ and $N_{\sigma}$ are given in Table \ref{tab:x2_delta_AM} of Appendix \ref{appendix_additional_results}). We overlay the $N_{\sigma}$ values computed for the $\mathrm{AM}~\mathrm{AGN}$ (circular markers), base AGN (triangular markers) and uncontaminated (gas only; star markers) cross-correlations, which are listed in Table \ref{tab:x2_native_AGN}. The grey shaded regions overlaid in Fig. \ref{figure:AM_summary_stat} outline $N_{\sigma}\pm2$ and $N_{\sigma}\pm1$. In Fig. \ref{figure:AM_summary_stat} we assign a negative (positive) sign to $N_{\sigma}$ values to indicate that the FLAMINGO cross-correlation underestimates (overestimates) the observations. The minimum $N_{\sigma}$ values (left most tick) of each error bar correspond to the $\Delta \mathrm{AM}~\mathrm{AGN}$ cross-correlation when AGN contamination is minimised ($\sigma_r=1.0$ dex) within the confines of our abundance matching approach. Conversely, the maximum (right most tick) values of the error bar are given by the $\Delta \mathrm{AM}~\mathrm{AGN}$ cross-correlation when AGN contamination is maximised ($\sigma_r=0$ dex).  Note that changing $\sigma_r$ does not have a uniform effect on the cross-correlation (as demonstrated in Fig. \ref{figure:AM_params_example}) because the sensitivity of the X-ray -- lensing cross-correlation to changes in $\sigma_r$ has some dependence on the strength of baryonic feedback (see relevant discussions of Fig. \ref{figure:delta_example}). 

The large number of standard deviations spanned by the error bars in Fig. \ref{figure:AM_summary_stat} indicate that perturbing the intrinsic scatter in the BH Luminosity -- circular velocity relation can have significant consequences for the resultant X-ray -- lensing cross-correlation. However, even if we consider the full range of $\Delta \mathrm{AM}~\mathrm{AGN}$ contaminated cross-correlations, only the fiducial L1$\_$m9 and fgas$-2\sigma$ simulations are compatible with the measurements of F24. 
The fiducial L1$\_$m9 simulation remains within $2\sigma$ of the observed data for all parametrisations of the $\Delta \mathrm{AM}$ AGN, whilst fgas$-2\sigma$ is only compatible with F24 when the contamination from unresolved $\Delta \mathrm{AM}$ AGN is parametrised by $\sigma_r \lesssim 0.17$ dex. The remaining cluster gas fraction variations (i.e.\ fgas$+2\sigma$,fgas$-4\sigma$, fgas$-8\sigma$ and LS8$\_$fgas$-8\sigma$) are all incompatible with the observations. 
The fgas$-2\sigma$ simulation agrees best with the data when we maximise the contamination from the unresolved $\Delta \mathrm{AM}$ AGN (i.e.\ $\sigma_r=0$ dex), at which point fgas$-2\sigma$ is offset from the observations by $\approx 0.9\sigma$. 
In this maximum contamination scenario (i.e.\ $\sigma_r=0$ dex), the fgas$-4\sigma$, fgas$-8\sigma$ and LS8$\_$fgas$-8\sigma$ simulations deviate from the observations by $3.6, 7.6$ and $11.5 \sigma$, respectively. Larger values of $\sigma_r$ result in greater discrepancies, with the exception of fgas$-2\sigma$, which most matches F24 best when the AGN contamination is minimised. However, even for $\sigma_r=1$ dex fgas$-2\sigma$ is offset by $4.5 \sigma$.

In summary, if we account for the X-ray emission from unresolved AGN predicted by a wide range of abundance matching scenarios, only the fiducial L1$\_$m9 FLAMINGO simulation and, if we push the contamination by AGN to very large values, also the fgas$-2\sigma$ variation, are compatible with the F24 observations.

\subsection{Comparing FLAMINGO cosmology variations to observations}\label{sec:comparison_with_direct_measurements_cosmo}

We now repeat the comparisons from \S \ref{sec:comparison_with_direct_measurements} between FLAMINGO and the observations of F24, but for the variations of the L1$\_$m9 simulation that do not assume the fiducial D3A cosmology and are calibrated to the fiducial cluster gas fractions. This includes the lensing LS8 cosmology simulation, the Planck cosmology simulation, and its two massive-neutrino variations PlanckNu0p24Fix and PlanckNu0p48Fix. 
As noted in \S \ref{sec:flamingo_overview}, the particle lightcones for these variations of the fiducial L1$\_$m9 simulation do not contain any information on the hot gas within the lightcone. As such, we cannot reliably re-compute an all-sky map for which the X-ray emission from hot gas is convolved with the ROSAT response matrix. However, the impact of the ROSAT response matrix on the resultant X-ray--lensing cross-correlation is reasonably consistent for each of the FLAMINGO cluster gas fraction variations considered in this work (see Appendix \ref{appendix_diffuse_gas_lightcones}). Given this consistency, we can estimate the effect of convolving the X-ray--lensing cross-correlation with the ROSAT response matrix by assuming it to have the same relative effect as for the fiducial cosmology. 

For the Planck, PlanckNu0p24Fix, PlanckNu0p48Fix and LS8 simulations we estimate the gas-only, ROSAT convolved, X-ray--lensing cross-correlations as 
\begin{equation}
   C_{\ell}^{{\mathrm{gas}} \, \kappa} \approx R_{\mathrm{L1\_m9}} \, N_{\ell}^{{\mathrm{gas}} \, \kappa}\,, 
   \label{eq:cosmo_gas_approx}
\end{equation}
where $N_\ell$ indicates the cross-correlation for the case when the X-ray emission is not convolved with the ROSAT response matrix (e.g.\ the cross-correlations shown in Fig.~\ref{figure:normalised_diffuse_gas_example}), and $R_{\mathrm{L1\_m9}} = C_{\ell}^{{\mathrm{gas}} \, \kappa} \, /\, N_{\ell}^{{\mathrm{gas}} \, \kappa}$ is the ratio of the convolved and non-convolved gas-only cross-correlations for the fiducial L1$\_$m9 simulation (see Appendix \ref{figure:response_function_appendix}). 
We then add the AGN-only X-ray--lensing cross-correlation ($C_{\ell}^{{\mathrm{AGN}} \, \kappa}$) convolved with the ROSAT response matrix to approximate the AGN-contaminated cross-correlations as 
\begin{equation}
   C_{\ell}^{{\mathrm{x}} \, \kappa} \approx R_{\mathrm{L1\_m9}} \, N_{\ell}^{{\mathrm{gas}} \, \kappa} \, + C_{\ell}^{{\mathrm{AGN}} \, \kappa}.
   \label{eq:cosmo_gas_agn_approx}
\end{equation} 

We estimate the AGN-contaminated X-ray--lensing cross-correlations, as given by equation \ref{eq:cosmo_gas_agn_approx}, for the cosmology variations using the X-ray emission from the base, AM and $\Delta$AM AGN, as done for the cluster gas fraction variations in \S \ref{sec:comparison_with_direct_measurements}.
The resultant AGN-contaminated cross-correlations follow the trends identified for the idealised counterparts shown in Fig.~\ref{figure:normalised_diffuse_gas_example}. The amplitude of the cross-correlation decreases with increasing neutrino mass and for a lower S$8$ parameter. 

\begin{figure}
    \centering
    \includegraphics[width=\columnwidth]{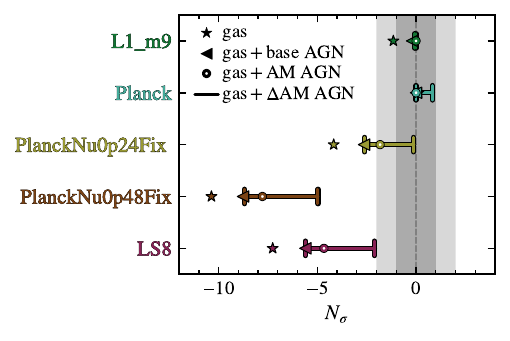}
    \vspace*{-0.5cm}
    \caption{ The same as Fig. \ref{figure:AM_summary_stat} but for the different cosmology and neutrino mass variations of the fiducial L1$\_$m9 simulation.}
    \label{figure:AM_summary_stat_cosmo}
\end{figure}

In Fig.~\ref{figure:AM_summary_stat_cosmo} we show the number of standard deviations, $N_{\sigma}$, between the predicted cross-correlations and the measurements of F24 (analogous to Fig. \ref{figure:AM_summary_stat}). 
If we neglect unresolved AGN, the Planck simulation is found to be in excellent agreement ($N_\sigma \approx 0.1$) with F24, whereas the PlanckNu0p24Fix, PlanckNu0p48Fix and LS8 simulations are offset by $\approx 4.2, 10.4$ and $7.3 \sigma$ respectively. 
When the X-ray emission from unresolved AGN is given by the base FLAMINGO (AM) BHs, we find the Planck, PlanckNu0p24Fix, PlanckNu0p48Fix and LS8 simulations are offset from the observed cross-correlation by $\approx 0.0 \sigma$ ($\approx 0.0 \sigma$),  $2.6 \sigma$ ($1.8 \sigma$), $8.8 \sigma$ ($7.8 \sigma$) and $5.6\sigma$ ($4.7\sigma$), respectively. In these scenarios the PlanckNu0p48Fix and LS8 simulations are inconsistent with observations. 
However, as shown in Fig. \ref{figure:AM_summary_stat_cosmo}, for the $\Delta$AM AGN model it is possible to bring the LS8 simulation to within $\approx 2.1 \sigma$ of the observed cross-correlations by maximising the level of AGN-contamination (i.e.\ $\sigma_r=0$ dex). 
The Planck cosmology variation remains consistent with the observed cross-correlation for the full range of $\Delta$AM AGN-contaminated cross-correlations, whilst PlanckNu0p24Fix is discrepant when the $\Delta$AM AGN are parametrised by $\sigma_r \gtrsim 0.4$ dex. The PlanckNu0p48Fix simulation cannot be reconciled with the observations, which is not surprising as this simulation's cosmology was extreme by design \citep[e.g. see][]{Elbers_2024}. 
We find that the results shown in Fig. \ref{figure:AM_summary_stat_cosmo} do not change for when the assumed impact of convolving the X-ray--lensing cross-correlation with the ROSAT response matrix, (given by $R_{\mathrm{L1\_m9}}$) is varied to match differences in $R_{\mathrm{L1\_m9}}$ when it is computed for each of the FLAMINGO cluster gas fractions (see Appendix \ref{figure:AM_summary_stat_cosmo_appendix}).

In conclusion, Fig. \ref{figure:AM_summary_stat_cosmo} reaffirms that in comparison to the changes in the cluster gas fraction, the predicted X-ray--lensing cross-correlation is insensitive to the differences between the fiducial FLAMINGO D3A and the FLAMINGO Planck cosmologies (see Fig. \ref{figure:normalised_diffuse_gas_example}). 
Moreover, given the uncertainty in the contributions from unresolved AGN, the cross-correlation does not provide strong constraints on cosmology. Hence, whilst we may be able to select between the different levels of baryonic feedback strength in FLAMINGO (\S \ref{sec:comparison_with_direct_measurements}), a deeper understanding of the X-ray emission from unresolved AGN is required to also distinguish between FLAMINGO's different cosmology models. This comes with the caveat that the effects of varying the cosmology and baryonic physics are largely degenerate (see \S \ref{sec:diffuse_gas_results}). Therefore, even if all the relevant AGN could be resolved observationally, the greater sensitivity to changes in baryonic feedback would likely prevent the X-ray--lensing cross-correlation from placing strong constraints on cosmology.

\section{Discussion and outlook}\label{sec:discussion}

In the recent literature there is a discrepancy in the required strength of AGN feedback and cluster gas mass fractions: kSZ measurements and stacks of optically-selected clusters observed by eROSITA both require stronger, more efficient AGN feedback to expel gas from low-mass clusters than indicated by gas fractions inferred from pre-eROSITA X-ray measurements and X-ray scaling relations in low redshift clusters.
In this section we examine if the FLAMINGO X-ray--cosmic-shear cross-correlations presented in \S \ref{sec:comparison_with_direct_measurements} and \ref{sec:comparison_with_direct_measurements_cosmo} provide new insights.

The joint modelling of kSZ measurements by \citet[][]{Schaan_2021} (from the Atacama Cosmology Telescope (ACT) \citep[specifically the maps of][]{Naess_2020}, \textit{Planck} and the Baryon Oscillation Spectroscopic Survey \citep[BOSS; as given by][]{Ahn_2014}) with cosmic shear data from KiDS-1000 \citep[][]{Asgari_2021} and DES \citep[in particular the data summarised by][]{Sevilla-Noarbe_2021} by \citet[][]{Schneider_2022} and \citet[][]{Bigwood_2024} find a preference for stronger baryonic feedback (and lower cluster gas fractions) than prescribed by hydrodynamical simulations with cluster gas fractions akin to those of FLAMINGO's fiducial L1$\_$m9  simulation, which was calibrated on cluster gas fractions inferred from pre-eROSITA X-ray measurements \citep[see][]{Kugel_2024}. 
This conclusion is reaffirmed by \citet[][]{McCarthy_2024} who finds, through comparisons of kSZ profiles measured from samples of haloes matched in mass via galaxy-galaxy lensing measurements \citep[from both KiDS-1000 and DES-Y3 data, see][]{Amon_2023}, that the fgas$-8\sigma$ FLAMINGO simulation is consistent with \citet[][]{Schaan_2021}, whilst the fiducial L1$\_$m9 simulation is discrepant, at a $7\sigma$ level in each halo mass samples.
Additionally, \citet[][]{Siegel_2025} find that the fgas$-8\sigma$ simulation is consistent with their joint analysis of SDSS/DESI+ACT kSZ profiles \citep[given by][]{Schaan_2021, Guachalla_2025, Hadzhiyska_2025}, eROSITA X-ray halo gas masses \citep[][]{Bulbul_2024}, and mean halo masses given by galaxy-galaxy lensing measurements \citep[from regions overlapping with KiDS-1000, DES-Y3 and HSC][]{Giblin_2021,Gatti_2021, Li_2022}. The fiducial L1$\_$m9 simulation is found to be a combined (from each halo mass sample) $>8\sigma$ discrepant from the observed kSZ profiles.
Similar findings are presented by \citet[][]{Bigwood_2025b}, whose systematic comparison of simulations to observed kSZ profiles \citep[given by][]{Schaan_2021, Guachalla_2025, Hadzhiyska_2025} finds that the fiducial L1$\_$m9 FLAMINGO simulation is offset by $3.4\sigma$ from observations, whilst fgas$-8\sigma$ is consistent (at a level of $1.2\sigma$).  
This preference for reduced cluster gas fractions derived from kSZ observations is supported by the gas fractions inferred from the stacking of optically selected eROSITA clusters by \citet[][]{Popesso_2024_gas_frac}. Specifically, \citet[][]{Popesso_2024_gas_frac} find the gas fractions of low-mass clusters ($M_{500\mathrm{c}} \approx 10^{13.5-14.5}\mathrm{M}_{\odot}$) to be lower than those of the fiducial L1$\_$m9 simulation and instead to be more consistent with the fgas$-8\sigma$ simulation \citep[see comparison in figure 5 of][]{Siegel_2025}.  

\citet[][]{Eckert_2025} show that the X-ray luminosity and X-ray temperature scaling relations for low-redshift galaxy groups \citep[from the \textit{XMM-Newton} Group AGN Project, X-GAP; see][]{Eckert_2024} are inconsistent with the stronger AGN feedback required by kSZ measurements and gas fractions from optically-selected clusters. \citet[][]{Eckert_2025} find that their scaling relations are consistent with the fiducial L1$\_$m9 FLAMINGO simulation, whilst the fgas$-8\sigma$ simulation is discrepant at the $5.7\sigma$ level. Moreover, \citet[][]{Eckert_2025} caution that observed cluster masses and gas fractions estimates are subject to large systematic uncertainties, which, combined with selection biases, may be contributing to this discrepancy.
This incompatibility with more extreme AGN feedback is seconded by \citet[][]{Seppi_2026} who directly compared core-excised X-GAP observations with mock (XXM-Newton) observables created by forward modelling X-GAP analogues in FLAMINGO. Their joint analysis of core-excised mean temperatures, velocity dispersions, number of groups, X-ray luminosity and X-ray temperature scaling relations shows that fgas$-2\sigma$ is consistent with the X-GAP observations (at the $0.8\sigma$ level), while the fiducial L1$\_$m9, fgas$-4\sigma$ and fgas$-8\sigma$ simulations disagree with X-GAP at the $3.1\sigma$,  $2.4\sigma$ and $4.4\sigma$ level, respectively. Despite utilising the same low-redshift X-GAP dataset as \citet[][]{Eckert_2025}, \citet[][]{Seppi_2026} find a preference for stronger feedback than the fiducial L1$\_$m9 simulation, which is attributed to the full end-to-end forward modelling of observations and the handling of covariances between observables. 
Nonetheless, both sets of X-GAP scaling relations find common ground in being incompatible with the more extreme feedback scenarios favoured by kSZ measurements and gas fractions inferred from the stacking of optically-selected eROSITA clusters. 


Across all of the X-ray--lensing cross-correlations considered in this paper, the fiducial L1$\_$m9 simulation is consistently in best agreement with observations. Specifically, the un-contaminated (i.e. gas-only) and AGN-contaminated cross-correlations are offset from F24 by $\lesssim 1.1\sigma$ and $\lesssim 0.2\sigma$, respectively. Conversely, we do not find any scenario in which the observed cross-correlation is compatible with the strength of baryonic feedback that is favoured by kSZ measurements (i.e. the fgas$-8\sigma$ simulation). As such, the feedback strength inferred from the FLAMINGO X-ray--lensing cross-correlation is in excellent agreement with that inferred from the scaling relations of \citet[][]{Eckert_2025}.  Moreover, the strength of baryonic feedback favoured by \citet[][]{Seppi_2026} (e.g. fgas$-2\sigma$) can be reconciled with the predicted cross-correlations given sufficient AGN-contamination, e.g.\ as given by the $\Delta\mathrm{AM}$ AGN for $\sigma_r \lesssim0.17\,$dex.

To this point, we have not considered the cosmology and neutrino mass variations of the fiducial L1$\_$m9 simulation discussed in \S \ref{sec:comparison_with_direct_measurements_cosmo}. The FLAMINGO Planck cosmology is of particular interest here because, as shown in Fig \ref{figure:AM_summary_stat_cosmo}, when we deliberately increase the contribution from unresolved AGN (reduce $\sigma_r$ for the $\Delta$AM AGN) the cross-correlation for the Planck simulation further overestimates the observed cross-correlation, yet remains consistent with the measurements of F24 at the $< 2\sigma$ level.
Adopting cosmologies similar to the FLAMINGO Planck cosmology could, to some degree, counteract the decrease in the cross-correlation's amplitude from increasing the strength of baryonic feedback relative to the fiducial L1$\_$m9 simulation (as shown in Fig.~\ref{figure:normalised_diffuse_gas_example}) and improve agreement between stronger feedback models and the observed cross-correlation.
However, as indicated in Fig \ref{figure:normalised_diffuse_gas_example}, the cross-correlation's increase in amplitude due to adopting the Planck cosmology (i.e.\ larger values for the $\Omega_\mathrm{m}$, $\sigma_8$ and $S_8$ parameters; Table \ref{tab:cosmo_table}) is small in comparison to the change from adopting the baryonic feedback strength required by the fgas$-8\sigma$ simulation. As such, a cosmology with a much stronger clustering of matter than the FLAMINGO Planck cosmology would be required to reconcile the feedback strength of the fgas$-8\sigma$ simulation with the observed cross-correlation. Adopting FLAMINGO's Planck cosmology would likely further improve the agreement of the fgas$-2\sigma$ simulation with the observed cross-correlation and further align the predicted cross-correlations with the expectations based on \citet[][]{Seppi_2026}.
The inverse would also be true: adopting cosmologies that decrease the amplitude of the cross-correlation relative to the fiducial D3A cosmology (e.g.\ the LS8 simulation) would result in a preference for weaker baryonic feedback than in the fiducial L1$\_$m9 simulation. This is demonstrated in Fig.~\ref{figure:AM_summary_stat} by LS8$\_$fgas$-8\sigma$ being in worse agreement with the observations than fgas$-8\sigma$. 

Altogether, our results suggest that conclusions drawn from the X-ray--lensing cross-correlations are compatible with those inferred from the X-GAP cluster scaling relations. However, there is tension with the feedback strength required by kSZ measurements, and with the cluster gas fractions inferred from the stacking of optically-selected eROSITA clusters.

\section{Summary and conclusion}\label{sec:summary_and_conclusions}

In this work we used the FLAMINGO suite of cosmological hydrodynamical simulations to explore the efficacy of the X-ray--cosmic-shear cross-correlation as a probe of large-scale structure and baryonic feedback. We computed the cross-correlation between synthetic X-ray and cosmic shear full-sky maps built from FLAMINGO's suite of full-sky particle lightcones and \textsc{HEALPix} maps, as described in \S \ref{sec:methods}. We compared the predictions from FLAMINGO with the measurements of F24. The X-ray all-sky maps were constructed in the soft band (0.5-2.0 keV) both with and without convolving the X-ray emission with the effective detector area of the ROSAT response matrix. 

We explored the origin and robustness of the X-ray--lensing cross-correlation by decomposing the X-ray component of this probe by both halo mass and redshift before examining the cross-correlation's sensitivity to systematic variations in the prescribed baryonic feedback, cosmology, simulation box size, and resolution (\S \ref{sec:diffuse_gas_results}). We found that the X-ray component is most sensitive to the distribution of hot gas in low-redshift haloes with masses $10^{14}\leq M_{200\mathrm{c}}/\mathrm{M}_{\odot}\leq10^{15}$ (\S \ref{sec:decomposition_results}) and to changes in the strength of AGN feedback. However, we found that the effects of variations in cosmology (specifically the $\sigma_{8}$ parameter) and baryonic physics are largely degenerate (\S \ref{sec:sensitivity_to_cosmology_and_astrophysics}). 

We compared the predictions from FLAMINGO with the cross-correlation, detected by F24, between the cosmic shear data from the Dark Energy Survey (year 3 data release) and the X-ray all-sky map from ROSAT (\S \ref{sec:comparison_with_direct_measurements} and \S \ref{sec:comparison_with_direct_measurements_cosmo}). Here, the key finding is that the fiducial FLAMINGO model, calibrated on pre-eROSITA cluster gas fractions, is in excellent agreement with the observed cross-correlation, whilst simulations calibrated to yield higher or significantly lower cluster gas fractions are ruled out.
Additionally, we used an abundance matching approach (as described in \S \ref{sec:abundance_matching_motivation_and_method}) to demonstrate that X-ray contamination from unresolved AGN should not be neglected in these comparisons. We found that increasing the contribution from unresolved AGN to the cross-correlation, which was accomplished by reducing the intrinsic scatter in the BH Luminosity -- halo maximum circular velocity relation, allows for the FLAMINGO simulations with moderately stronger feedback (fgas$-2\sigma$; calibrated to a lower cluster gas fraction) than the fiducial model to be reconciled with the observations (\S \ref{sec:abundance_matching_results}).

In greater detail, our main findings are as follows:
\begin{itemize}

    \item  The X-ray component of the X-ray--lensing cross-correlation is dominated by the X-ray emission from hot gas within galaxy clusters at low redshifts (see \S \ref{sec:decomposition_results}).
    Decomposing a lightcone for the fiducial L1$\_$m9 FLAMINGO simulation by the redshift of the X-ray emitting gas reveals that hot gas at $z\gtrsim0.5$ is inconsequential for the cross-correlation. The contribution from hot gas decreases with redshift, however this redshift dependence is weaker for the cross-spectra in more distant (higher-redshift) tomographic bins (Fig. \ref{figure:redshift_decomposition}). 
    
    \item If we decompose the same X-ray lightcone by total halo mass, $M_{200\mathrm{c}}$, we find that hot gas within $R_{200\mathrm{c}}$ of haloes accounts for $90-97\%$ of the predicted cross-correlation's signal for $30<\ell<2048$ (Fig.~\ref{figure:halo_mass_decomposition_example}). Moreover, haloes with $M_{200{\mathrm{c}}} \sim 10^{14} ~\mathrm{M}_\odot$ dominate at all angular scales, whilst more massive haloes, $M_{200{\mathrm{c}}} \geq 10^{15} ~\mathrm{M}_\odot$, are significant for $ \ell < 250$ but, their relative contribution diminishes towards smaller scales. The X-ray emission from haloes with $M_{200{\mathrm{c}}}< 10^{13.5} ~{\mathrm{M}}_\odot$ accounts for $\lesssim 5\%$ of the signal.
    
    \item The effects of variations in cosmology and baryonic physics on the X-ray--cosmic-shear cross-correlation are nearly degenerate (\S \ref{sec:sensitivity_to_cosmology_and_astrophysics}). Relative to the fiducial L1$\_$m9 FLAMINGO simulation, varying either the cosmology or prescription of baryonic feedback results in changes to the predicted cross-correlation that are nearly independent of angular scale for $30 < \ell < 2048$ (Fig.~\ref{figure:normalised_diffuse_gas_example}). Stronger AGN feedback (i.e.\ models calibrated to produce lower cluster gas fractions) suppresses the cross-correlation on all angular scales considered in this paper. Increasing the strength of stellar feedback has the same effect, but to a much lesser extent. The implementation of AGN feedback has a limited ($\lesssim10\%$) effect on the cross-correlation, where the cross-correlation for the model using jet-like AGN feedback is slightly elevated relative to the fiducial model, which uses purely thermally-driven AGN  feedback, at both the smallest and largest angular scales considered ($30 < \ell < 75$ and $1720 \lesssim \ell<2048$ respectively). Relative to the fiducial L1$\_$m9 FLAMINGO simulation, increasing the summed neutrino mass or adopting the `LS8' lensing cosmology suppresses the cross-correlation almost uniformly across all angular scales. 
    
    \item The predicted X-ray--lensing cross-correlations are converged with simulation box size but not with the simulation resolution (\S \ref{sec:sensitivity_to_boxsize_and_resolution} and Fig.~\ref{figure:boxsize_resolution_example}). This lack of convergence is the product of imperfect recalibration of the subgrid models at each resolution and differences in the thermodynamic profiles for fixed gas fractions. We find that an order of magnitude change in mass resolution has a $\approx15\%$ effect on the measured cross-correlation, whilst cosmic variance is a $\leq  20\%$ effect and decreases with increasing $\ell$ (Fig. \ref{figure:boxsize_resolution_example}). The effects of cosmic variance and mass resolution are both secondary to the effect of varying the cluster gas fractions in the simulations.

    \item Contamination in the form of X-ray emission from AGN unresolved by ROSAT has a non-negligible impact on the predicted X-ray--lensing cross-correlation. The X-ray emission from unresolved AGN `boosts' its amplitude on all angular scales included in this work (\S \ref{sec:results_the_fiducial_model} and \S \ref{sec:comparison_with_direct_measurements}). The X-ray emission from unresolved AGN given by the `base' black hole (BHs) particles (i.e. AGN as they naturally exist in FLAMINGO) and the X-ray emission from unresolved AGN given by a simple abundance matching approach have a similar impact on the cross-correlation (Fig.~\ref{figure:plot_1} and Fig.~\ref{figure:convolved_native_agn_example}). These abundance matched (AM) AGN are constrained to reproduce the observed quasar luminosity functions of \citet[][]{Shen_2020}. Finally, the cross-correlation between the lensing signal and the unresolved base AGN increases with the simulation's feedback strength, whilst for the AM AGN there is a more subtle decrease in signal with increasing feedback (Fig.~\ref{figure:convolved_native_agn_example}).

    \item Under the strong assumption that X-ray contamination from unresolved AGN is negligible, we find that the fiducial L1$\_$m9 FLAMINGO simulation agrees with the measurements by F24 at the 1.1$\sigma$ level (Fig.~\ref{figure:plot_1} and Table~\ref{tab:fiducial_model_Nsigma_values_table}), whilst all other cluster gas fraction variations of the L1$\_$m9 simulation (fgas$\pm 2\sigma$, fgas$-4\sigma$, fgas$-8\sigma$ and LS8$\_$fgas$-8\sigma$) are discrepant by more than $3.3\sigma$ (see the summary statistics in Table~\ref{tab:x2_native_AGN}). 
    
    \item Accounting for the X-ray emission from unresolved AGN reduces (increases) the disagreement between the measured X-ray--lensing cross-correlation and the FLAMINGO simulations with stronger (weaker) baryonic feedback than the fiducial model. 
    However, the fgas$-2\sigma$ simulation, which has moderately stronger baryonic feedback, is still offset from the measurements of F24 by $3.6\sigma$ and $3.1\sigma$ when AGN-contamination is included using the base and AM AGN models, respectively (Fig.~\ref{figure:convolved_native_agn_example} and Table~\ref{tab:x2_native_AGN}). In these scenarios, even stronger baryonic feedback (i.e.\ the fgas$-4\sigma$, fgas$-8\sigma$ and LS8$\_$fgas$-8\sigma$ simulations) are still highly inconsistent with the observations (i.e.\ discrepant by $\geq 6.7, 11.0~\mathrm{and}~ 14.7 \, \sigma$, respectively).
    
    \item The impact of X-ray emission from unresolved AGN on the X-ray--lensing cross-correlation can be increased or decreased by perturbing the intrinsic scatter in the BH Luminosity -- halo maximum circular velocity relation (\S \ref{sec:abundance_matching_results}). Increasing (decreasing) the scatter reduces (boosts) the impact of contamination from unresolved $\Delta \mathrm{AM}$ AGN on the cross-correlation (Fig.~\ref{figure:AM_params_example} and  Fig.~\ref{figure:delta_example}). Irrespective of the scatter (at least within the confines of our $\Delta \mathrm{AM}$ model), weaker baryonic feedback than prescribed by the fiducial L1$\_$m9 simulation (i.e.\ fgas$+2\sigma$) is clearly irreconcilable with the observed data (discrepant at the $\geq 4.5 \sigma$ level; Fig.~\ref{figure:AM_summary_stat}). Within the range of sampled $\Delta \mathrm{AM}$ contaminated cross-correlations, the fgas$-2\sigma$ simulation can be reconciled with observations to the $0.9 \sigma$ level (Fig.~\ref{figure:AM_summary_stat}). However, AGN contamination cannot increase the amplitude of the cross-correlation sufficiently to reconcile the simulations using even stronger baryonic feedback (fgas$-4\sigma$, fgas$-8\sigma$ and LS8$\_$fgas$-8\sigma$) with the observations (Fig.~\ref{figure:AM_summary_stat}).

    \item The X-ray--lensing cross-correlations predicted by the fiducial L1$\_$m9 FLAMINGO simulation are consistent with the observations  of F24 (Fig.~\ref{figure:delta_example} and Fig.~\ref{figure:AM_summary_stat}) and the X-GAP X-ray scaling relations of  \citet[][]{Eckert_2025}. For reasonable amounts of X-ray emission from unresolved AGN, the increased strength of baryonic feedback preferred by the analysis of \citet[][]{Seppi_2026} (i.e.\ fgas$-2\sigma$) can also be reconciled with the observed X-ray--lensing cross-correlations. Within the confines of the $\Delta \mathrm{AM}$ model for unresolved AGN  it is not possible to obtain agreement between the measurements of F24 and FLAMINGO's most extreme baryonic feedback variation (i.e.\ fgas$-8\sigma$), which appears to be required by recent kinetic Sunyaev-Zel’dovich (kSZ) effect measurements \citep[e.g,][]{McCarthy_2024, Bigwood_2024, Bigwood_2025b, Siegel_2025} and stacks of optically-selected eROSITA X-ray clusters \citep[][]{Popesso_2024_gas_frac}.

\end{itemize}

We conclude that the X-ray--cosmic-shear cross-correlation is already a powerful probe of the distribution of hot gas in clusters relative to the underlying matter density field.
The constraining power of the X-ray--cosmic-shear cross-correlation can most easily be improved by utilising the published and upcoming data releases of the eROSITA-DE All-Sky Survey \citep[e.g. eRASS1,][]{Merloni_2024, Bulbul_2024} from the SRG/eROSITA telescope \citep[][]{Sunyaev_2021, Predehl_2021}. Compared to ROSAT, the eROSITA X-ray data offer greater angular resolution, redshift depth and sensitivity, which can further mitigate the impact of AGN contamination \citep[relevant discussions and comparisons by e.g,][]{Merloni_2024, Zheng_2024}. We show a predicted eROSITA-like X-ray--lensing cross-correlation in Appendix~\ref{appendix_eROSITA}.
As we venture further into the era of stage-IV galaxy surveys, the projected leaps in the precision and volume of the cosmic shear data, such as that offered by Euclid \citep[see relevant overviews in e.g.][]{Euclid_overview_2024, Prat_2025} will complement the eROSITA X-ray surveys and maximise the effectiveness of X-ray--cosmic-shear joint analysis.
X-ray measurements remain one of the most valuable sources of information on the properties of hot gas in clusters. Further exploration of X-ray all-sky measurements in joint analyses with adjacent low-redshift probes \citep[such as the thermal Sunyaev-Zel’dovich effect; e.g.][]{Posta_2024} provides an exciting opportunity that may yield important insights into the interplay between baryonic feedback and large-scale structure.

\section*{Acknowledgements}
We are grateful to the anonymous referee for their very helpful suggestions. We thank David Alonso, Tassia Ferreira and Nora Elisa Chisari for the assistance with the observed X-ray--cosmic-shear data and the useful discussions on the masking of point sources within the ROSAT all-sky maps. 
WM thanks Jeger Broxterman, Ian McCarthy and Rob McGibbon for the insightful discussions and assistance with (re-) computing both the ROSAT all-sky maps and cross-correlations used in this paper. 
This work is partly funded by the Dutch Research Council (NWO) programme Athena 184.034.002.
WM acknowledges support by NWO through the Dark Universe Science Collaboration (OCENW.XL21.XL21.025). 
This work used the DiRAC@Durham facility managed by the Institute for Computational Cosmology on behalf of the STFC DiRAC HPC Facility (www.dirac.ac.uk). The equipment was funded by BEIS capital funding via STFC capital grants ST/K00042X/1, ST/P002293/1, ST/R002371/1 and ST/S002502/1, Durham University and STFC operations grant ST/R000832/1. DiRAC is part of the National e-Infrastructure.

\section*{Data Availability}

 The data supporting the plots within this article are available on reasonable request to the corresponding author. The FLAMINGO simulation data, including the ROSAT-like X-ray all-sky maps created for this paper are publicly available\footnote{\url{https://flamingo.strw.leidenuniv.nl/data.html}}. However, the $\Delta$AM maps AGN point source maps constructed for this paper are not included. Instead, these point source maps can be made available on request. 
 The numerical tools utilised in our analysis to construct the individual X-ray maps and decompose the particle lightcones have been made publicly available as part of the \textsc{LightconeIO} package\footnote{\url{https://lightconeio.readthedocs.io}}. 



\bibliographystyle{mnras}
\bibliography{example} 



\appendix

\section{X-ray emission in FLAMINGO} \label{appendix_diffuse_gas_lightcones}

In this appendix we show the impact of the ROSAT response matrix, along with other key choices relating to computing the X-ray emission from both hot gas and point sources (AGN and XRBs), on the ROSAT-like X-ray all-sky maps and X-ray--lensing cross-correlations described in \S \ref{sec:methods}. 

\begin{figure}
        \centering
	\includegraphics[width=\columnwidth]{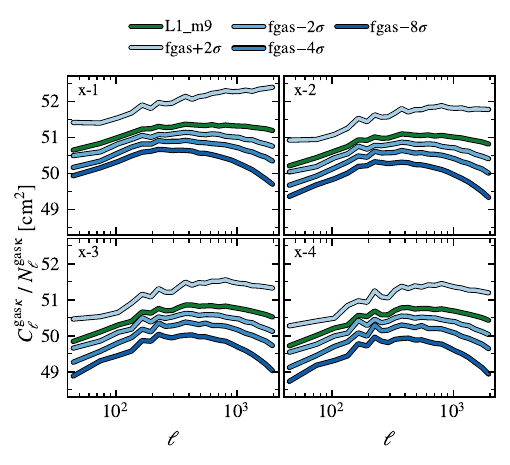}
        \vspace*{-0.5cm}
    \caption{ 
    The ratio of the X-ray--lensing cross-correlation for the L1$\_$m9 cluster gas fraction variations (indicated by colour), in each tomographic bin, when the X-ray emission from hot gas is convolved with the ROSAT response matrix ($C_{\ell}^{{\mathrm{gas}}\, \kappa}$) and when it is not ($N_{\ell}^{{\mathrm{gas}}\, \kappa}$). The tomographic bin is indicated in the upper left corner of each panel.}
    \label{figure:response_function_appendix}
\end{figure}

In Fig. \ref{figure:response_function_appendix} we show the ratio between the X-ray--lensing cross-correlations when the X-ray emission from hot gas is ($C_{\ell}^{{\mathrm{gas}}\, \kappa}$) and is not ($N_{\ell}^{{\mathrm{gas}}\, \kappa}$) convolved with the ROSAT response matrix. We present this ratio in each tomographic bin for the cluster gas fraction variations of the fiducial L1$\_$m9 simulation. 
For a given feedback variation, the ratio between the convolved and `normalised' cross-correlations is fairly consistent across all tomographic bins. Furthermore, the offset of each feedback variation from the fiducial model is consistent across all tomographic bins.

\begin{figure}
    \includegraphics[width=\columnwidth]{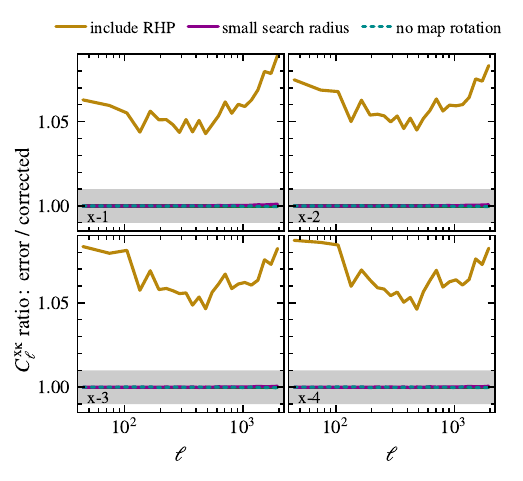}
    \vspace*{-0.5cm}
    \caption{The ratio between the X-ray--lensing cross-correlation of the L1$\_$m9 simulation when only the X-ray emission from hot gas is included and when the X-ray emission from recently heated particles (RHPs) is included (gold curve), the smoothing search radius is not corrected (i.e.\ it is too small: purple curve), and when the all-sky maps are not rotated to avoid replication of structures when integrating the X-ray emission along the line-of-sight (cyan dotted curve). The tomographic bin is indicated in the lower left corner of each panel. The X-ray emission is not convolved with the ROSAT response matrix. }
    \label{figure:smoothing_cl}
\end{figure}

As noted in \S \ref{sec:Xray_diffuse_gas_lightcone_methods}, when constructing new X-ray all-sky maps from a PLC, we make two minor changes to the method used for all other FLAMINGO smoothed maps. First, we make a correction to the scale used to identify if smoothing should be applied to the map and second, we account for the uncertainty in the scale factor when a particle (that has crossed the lightcone) was last directly heated by AGN feedback. 

When constructing smoothed all-sky maps, a particle's position on the sky, distance from the lightcone's origin and SPH smoothing length are used to determine the location and extent to which the particle's value (of the property being mapped) is smoothed onto the sky.
For any given particle, all pixels in the map whose centres lie within a given search radius $\theta_{\mathrm{s}}$ may be updated based on a weighting function as described in Appendix A2 of \citet[][]{FLAMINGO_2023}. The search radius is given by 
\begin{equation}
    \theta_{\mathrm{s}} = \gamma \arctan(h/r)
\end{equation}
where $h$ is the particle's SPH smoothing length, $r$ is the distance from the particle to the observer and $\gamma$ is the radius at which the SPH smoothing kernel falls to zero in units of the smoothing length.
The all-sky smoothed \textsc{HEALPix} maps created on-the-fly erroneously used a search radius computed with $\gamma=1$ when comparing the search radius to the angular radius of a pixel to identify if smoothing should be applied for a given gas particle. 
In this instance $\gamma=1$ is less than the correct value, which results in a smaller search radius. Consequently, a small fraction of particles that should have been smoothed over multiple pixels, instead only contribute to a single pixel. For example, the correct value of $\gamma$ of the fiducial L1$\_$m9 simulation is larger by a factor of 1.9 and therefore particles with $\theta_{\mathrm{s}}=$ 1-1.9 times the maximum pixel radius are not smoothed, but should have been. As the size of the search radius only affects whether or not smoothing is applied for a given particle, it did not affect how, or which, pixels were updated when smoothing was applied. 
In our work we adopt the correct values of $\gamma$, but this correction has a negligible effect on the resultant cross-correlations. 

We exclude X-ray emission from gas particles that have recently been directly heated by AGN feedback (i.e. recently heated particles; RHPs) from our X-ray all-sky maps as they may temporarily have extreme temperatures and densities which lead to them being unrealistically X-ray bright. As described in \S \ref{sec:FLAMINGO_base_Xray_Lightcones}, gas particles are labelled as RHPs if they have been heated by AGN feedback within the last $15\,\text{Myr}$ and have temperatures $10^{-1}\, \Delta T_{\mathrm{AGN}} < T/\,\mathrm{K} < 10^{0.3}\, \Delta T_{\mathrm{AGN}}$, where $\Delta T_{\mathrm{AGN}}$ is a calibrated sub-grid parameter for the AGN feedback temperature increment. 
A \textsc{BFloat16} lossy compression filter\footnote{See description of SWIFT compression filters, \url{https://swift.strw.leidenuniv.nl/docs_will/ParameterFiles/lossy_filters.html}} was applied to the particle lightcones' recorded values of the scale factors at which gas particles were last directly heated by AGN feedback. As such, we cannot reliably identify if $t_{\mathrm{Last\_Feedback}} \leq 15 \,\text{Myr}$, where $t_{\mathrm{Last\_Feedback}}$ is the time elapsed from when a particle crossed the lightcone to when it was last directly heated by AGN feedback.
Instead, due to the limited precision, we find that a time period of $70\,\text{Myr}$ is now required to reliably identify the same gas particles, at redshift $z\leq 0.5$, where, given no loss in precision, $t_{\mathrm{Last\_Feedback}}  \leq 15\,\text{Myr}$.  
When building our X-ray all-sky maps, we label gas particles as RHPs if $t_{\mathrm{Last\_Feedback}} \leq 70\,\text{Myr}$ (instead of $t_{\mathrm{Last\_Feedback}} \leq 15\,\text{Myr}$), whilst having temperatures $10^{-1}\, \Delta T_{\mathrm{AGN}} < T/\,\mathrm{K} < 10^{0.3}\, \Delta T_{\mathrm{AGN}}$ and densities $n_{\mathrm{H}}>10^{-2.25} \, \mathrm{cm}^{-3}$. The minimum density requirement ($n_{\mathrm{H}}>10^{-2.25} \, \mathrm{cm}^{-3}$) is included to counter balance the increased uncertainty in $t_{\mathrm{Last\_Feedback}}$, such that we mitigate incorrectly labelling gas particles as RHPs when they appear to have been recently heated by AGN feedback based on $t_{\mathrm{Last\_Feedback}}$ but do not have temperatures and densities that lead to unrealistically strong X-ray emission. 
This change in the RHP selection criteria has a negligible impact on the results of this paper. Moreover, the impact of changing how a RHP is defined is small compared to the impact of including the X-ray emission from RHPs in the all-sky maps. 

In Fig. \ref{figure:smoothing_cl} we present the ratio between the X-ray--lensing cross-correlation computed for the fiducial L1$\_$m9 simulation, as shown in Fig. \ref{figure:normalised_diffuse_gas_example}, and the cross-correlations when we include the X-ray emission from RHPs (gold curve), do not correct the small search radius (purple solid curve) or do not rotate the all-sky maps whilst integrating along the X-ray emission along the line-of-sight (cyan dotted curve). The ratio is shown for each tomographic bin and the X-ray emission from hot gas was not convolved with the ROSAT response matrix. 
As can be seen, for the range of angular scales we consider in this paper, rotating the X-ray all-sky maps to avoid the replication of structures along the line of sight and correcting the size of the smoothing kernel have almost no impact. However, by comparison, filtering out recently heated gas particles has a more noticeable ($\leq 9.2 \%$) effect on the resultant X-ray--lensing cross-correlation.

\subsection{AGN point sources in FLAMINGO}

\begin{figure}
	\includegraphics[width=0.9\columnwidth]{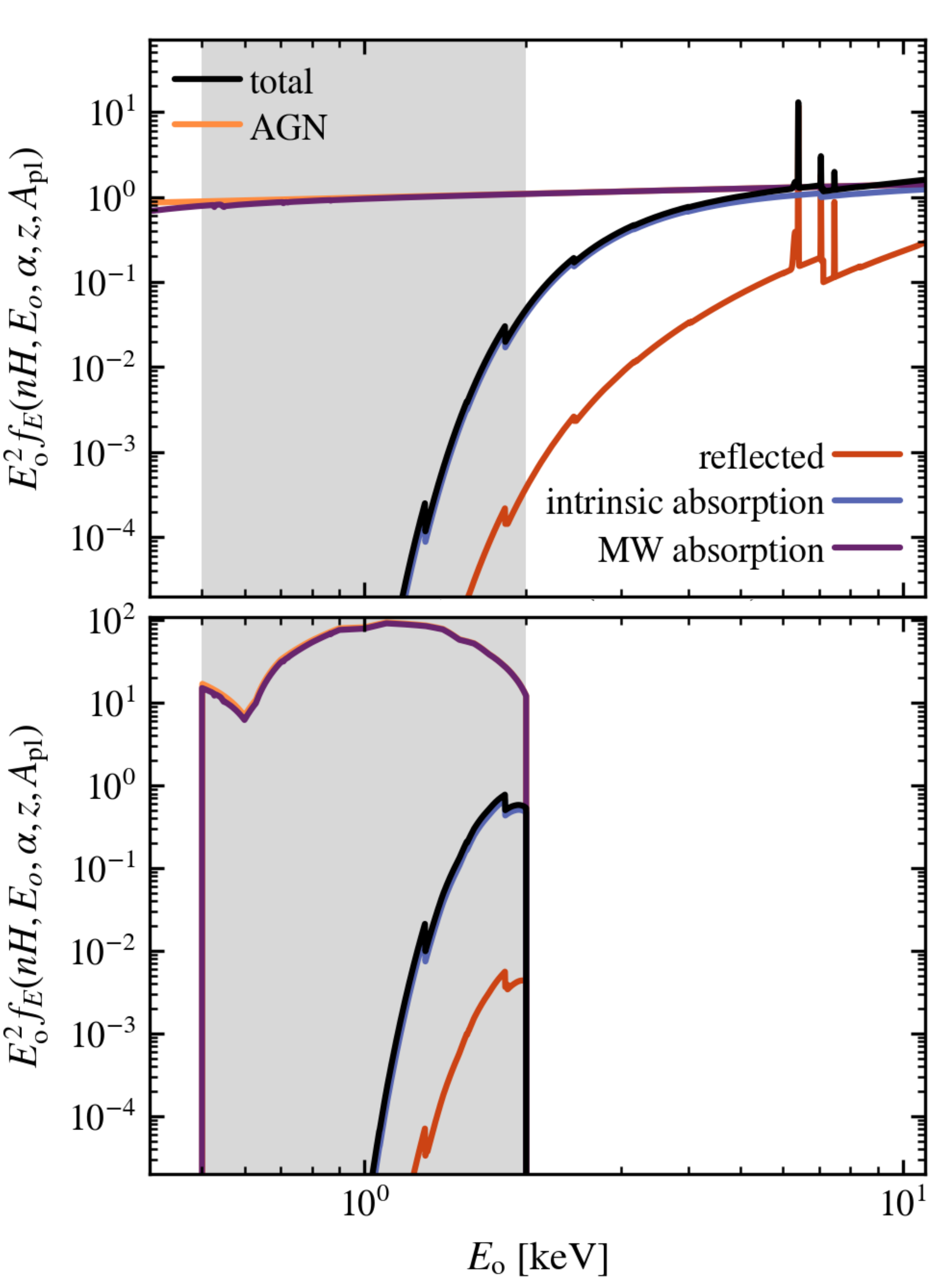}
    \vspace*{-0.25cm}
    \caption{The AGN spectrum template at $z=0.01$ with an intrinsic column density of $10^{23}\,\mathrm{cm}^{-2}$ and normalised such that $A_{\rm{pl}}=1$.
    The shaded region indicates of the range of the ROSAT soft band, 0.5--2 keV.
    \textit{Top}: The template when not convolved with the ROSAT response matrix. \textit{Bottom}: The template when convolved with ROSAT.
    }
    \label{fig:spectra}
\end{figure}

In Fig.~\ref{fig:spectra} we show an example of the AGN spectra template, $f_{\mathrm{E}} \equiv f_{E}(N_\text{H}, E_{o}, \alpha, z, A_{\rm{pl}})$, (described in \S \ref{sec:AGN_lightcone_methods}) as a function of observed photon energy ($E_o$) for an AGN at $z=0.01$ with an intrinsic column density of $10^{23}\,\mathrm{cm}^{-2}$ where we have assumed a power-law index of $\alpha=1.8$ normalised such that $A_{\rm{pl}}=1$. 
The bottom (top) panel shows the template when $f_{\mathrm{E}}$ is (not) convolved with the ROSAT response function and the shaded region indicates the 0.5--2.0 keV ROSAT energy band. 
The total final spectrum, combining the effects of absorption (intrinsic and Galactic) and reflection, is given by the black curve. The prime AGN spectrum (modelled as a power-law) and each of the mentioned effects are indicated by colour. 

\begin{figure*}
    \includegraphics[width=\textwidth]{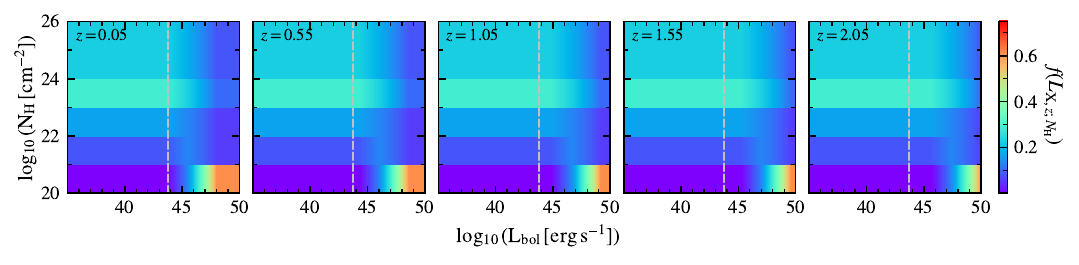}
    \vspace*{-0.5cm}
    \caption{Probability distribution of neutral hydrogen column density ($N_{\mathrm{H}}$), $f(L_{\mathrm{x}, z; N_{\mathrm{H}}})$ (indicated by colour) for a given bolometric luminosity ($L_{\mathrm{bol}}$).}
    \label{figure:nH_function}
\end{figure*}

Fig. \ref{figure:nH_function} shows the probability function of neutral hydrogen column density, $f(L_{\mathrm{x}, z; N_{\mathrm{H}}})$, given by (Section 3, equations 2--6) \citet[][]{Ueda_2014} at redshifts $z=0.05,\,0.55,\,1.05\,1.55$ and $2.05$. The model does not evolve with redshift for $z>2$. The X-ray luminosities in the 2--10 keV band used by \citet[][]{Ueda_2014} have been corrected to bolometric luminosities as described in \S \ref{sec:AGN_lightcone_methods}. 

\begin{figure}
    \includegraphics[width=\columnwidth]{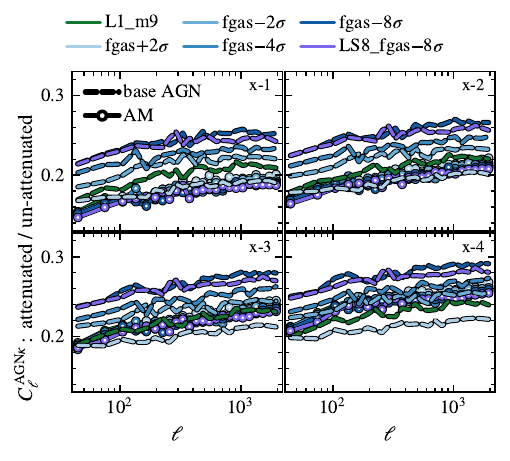}
    \vspace*{-0.5cm}
    \caption{
    The ratio between the AGN only X-ray--lensing cross-correlations, $C_{\ell}^{\mathrm{AGN}\, \kappa}$ when we assume the AGN spectrum accounts for attenuation (i.e.\ $f_{E}(N_\text{H}, E_{o}, \alpha, z, A_{\rm{pl}})$,  as described in the text) and when it is a simple power-law (i.e.\ $f_{\mathrm{pl}} \propto (E_\gamma ~\mathrm{keV}) ^ {\Gamma}$). 
    The cross-correlations only include the X-ray emission from unresolved AGN, using either the base AGN (dashed curves) or  AM AGN (curves with circular markers). The colour indicates the corresponding FLAMINGO simulation. The tomographic bin is indicated in the upper right corner of each panel. }
    \label{figure:attenuation_example}
\end{figure}

Fig. \ref{figure:attenuation_example} shows the ratio between the AGN-only cross-correlations, ${C_{\ell}^{{\rm{AGN}} \kappa}}$, for the case when the AGN spectrum template accounts for attenuation (i.e.\ modelled with $f_{\mathrm{E}}$, see example in Fig.~\ref{fig:spectra}) relative to the case when attenuation is neglected, i.e.\ the spectrum is a simple power-law. Adopting a simple power-law as the AGN spectrum neglects the effects of absorption (both intrinsic and from the Galactic foreground), which leads to greater X-ray emission from all AGN point sources on the sky. This change in X-ray emission is largest for faint AGN with the highest intrinsic column densities. In turn, this increased X-ray emission from faint AGN raises the contribution from unresolved AGN to the X-ray--lensing cross-correlation. 

\begin{figure}
        \centering
	\includegraphics[width=\columnwidth]{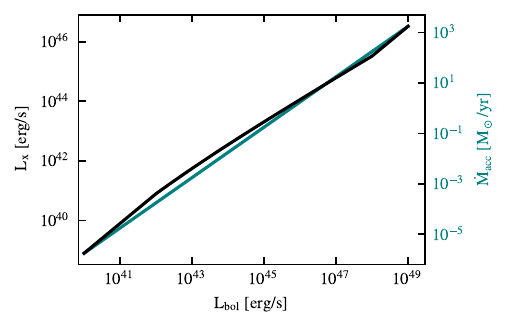}
    \vspace*{-0.5cm}
    \caption{The relation of the X-ray Luminosity($L_{\mathrm{x}}$) and Bolometric Luminosity ($L_{\mathrm{bol}}$) given by the adopted bolometric correction factor (as detailed in the text). Additionally, the relation between the black hole mass accretion rate ($\dot{M}_{\mathrm{accr}}$) and Bolometric Luminosity is shown by the cyan curve and right-most y-axis.}
    \label{fig:Lx_Lbol_Maccr_functional_form}
\end{figure}

As described in \S \ref{sec:AGN_lightcone_methods}, we estimate the photon count rate of AGN from the bolometric luminosities ($L_{\mathrm{bol}}$) of BH particles, which we convert to X-ray luminosities ($L_{\mathrm{x}}$) in the soft band (0.5-–2 keV) using a bolometric correction factor.
We adopted the bolometric correction factor given by equation 5 and the parameters (for the soft X-ray band) in Table 1 of \citet[][]{Shen_2020}.
Additionally, as detailed in \S \ref{sec:AGN_lightcone_methods}, we extended this function to allow for bolometric luminosities both higher and lower than the fit by \citet[][]{Shen_2020}. The resultant relation between X-ray and bolometric luminosity given by the correction function is shown in Fig. \ref{fig:Lx_Lbol_Maccr_functional_form}. 
Furthermore, in Fig. \ref{fig:Lx_Lbol_Maccr_functional_form} we also show the relation between the bolometric luminosity and the mass accretion rate of the BHs ($L_{\mathrm{bol}} = \dot{M}_{\mathrm{accr}} \epsilon_{\mathrm{r}}c^2$).

\begin{figure}
    \includegraphics[width=\columnwidth]{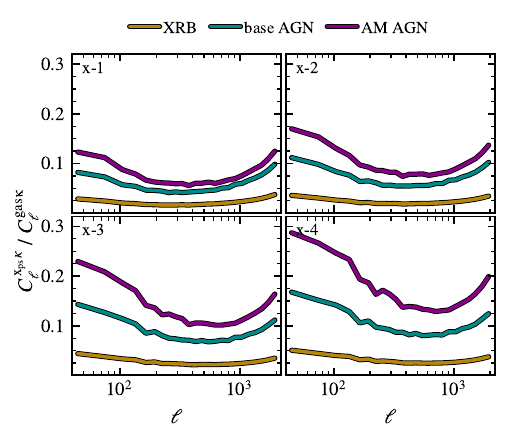}
    \vspace*{-0.5cm}
    \caption{The ratio between the X-ray--lensing cross-correlations of the L1$\_$m9 simulation when the X-ray emission is only from hot gas and when the X-ray emission is only from point sources ($C_{\ell}^{x_{ps}\,\kappa}$): X-ray binaries (XRBs; Gold curve), AM AGN (cyan curves) or base AGN (purple curves).
    The tomographic bin is indicated in the upper left corner of each panel. }
    \label{figure:XRB}
\end{figure}

In Fig. \ref{figure:XRB} we demonstrate, with the fiducial L1$\_$m9 simulation, that X-ray emission from XRBs makes a negligible contribution to the X-ray--lensing cross-correlation in comparison to the X-ray emission from unresolved AGN.
We construct all-sky maps of the X-ray emission from XRBs for redshifts $z\leq 3$ from the all-sky stellar mass \textsc{HEALPix} maps and the star formation rates of gas within $R_{200\mathrm{c}}$ of haloes identified in the halo-lightcone. We estimate the X-ray emission from XRBs as given by equation 5 (or model 5 in table 2) of \citet[][]{Aird_2017} and do not account for any effects relating to attenuation of this X-ray emission. 
The signal of the cross-correlation when the X-ray emission is from only the base AGN (and when only from the AM AGN) is greater than that for when the X-ray emission is purely from XRBs by a factor of $\approx 3-4$ (and $\approx4-5$). The X-ray emission from both the AGN and XRBs are secondary to the X-ray emission from hot gas. 

\section{Quasar Luminosity and Maximum Circular Velocity Functions} \label{appendix_QLF_and_abundance_match}

In this appendix we examine the quasar luminosity function (QLF) in one lightcone of the fiducial L1$\_$m9 simulation and show the convergence of the maximum circular velocity function with resolution, which both relate to the abundance matching approach that we use to model unresolved AGN, as described in \S \ref{sec:BH_luminosities_and_selection}.

\begin{figure*}
        \centering
	\includegraphics[width=\textwidth]{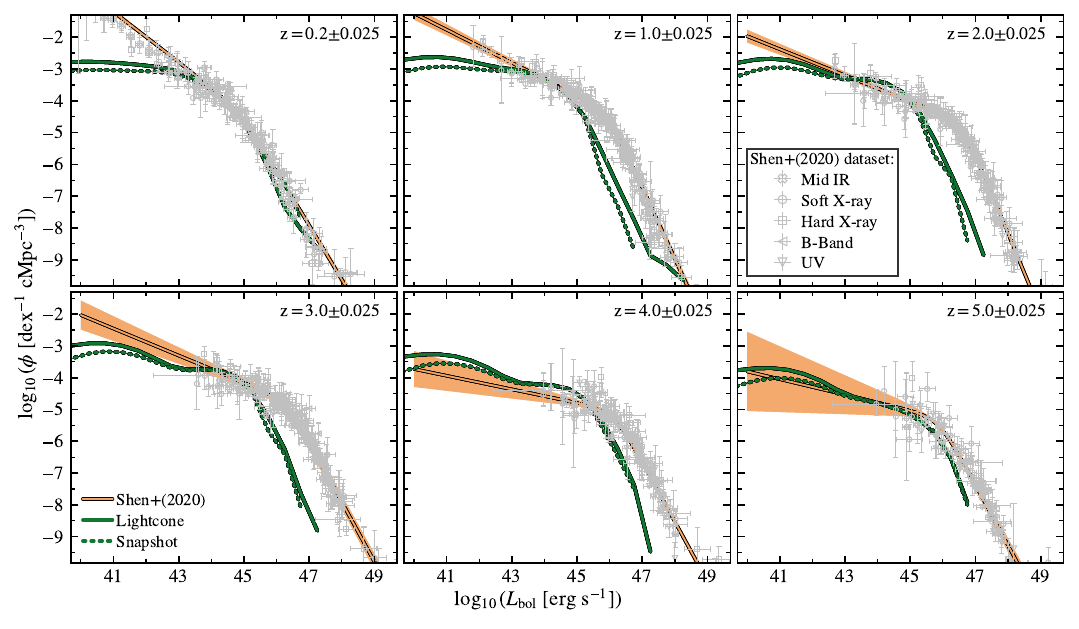}
    \vspace*{-0.5cm}
    \caption{The bolometric quasar luminosity functions for the FLAMINGO L1$\_$m9 simulation computed from the particle lightcone (solid green curve) and the most massive black hole per subhalo from the snapshot data (dotted green curve) compared to the \citet[][]{Shen_2020} fits to observations (solid orange curve) and their compilation of observational datasets (grey markers). The shaded region gives the associated $\pm 1 \sigma$ uncertainty for each fit to the observed QLF. The redshift of each QLF is indicated in the upper right corner of each panel, with the $\pm$ showing the sampled redshift range from the lightcone. }
    \label{fig:Luminosity_function}
\end{figure*}

The QLFs for the suite of FLAMINGO simulations are presented, at multiple redshifts, by \citet[][]{Ding_2025}. Specifically, \citet[][]{Ding_2025} showed that L1$\_$m9 underestimates the abundance of bright quasars for redshifts $z \gtrsim 1$.

In Fig. \ref{fig:Luminosity_function} we confirm that the underabundance of bright and faint quasars shown by \citet[][]{Ding_2025} for the fiducial L1$\_$m9 simulation snapshots (with the simulations full volume) is also present in the lightcones.
Here, we construct QLFs from the base BHs of the L1$\_$m9 PLC using redshift bin widths of $\Delta z = 0.05$ centred on $z = 0.2, 1, 2, 3, 4~\mathrm{and~}5$, within which we use 20 evenly log-spaced bins of the range $10^{40} \leq L_{\mathrm{bol}} \leq 10^{50}~\mathrm{erg}~\mathrm{s}^{-1}$.
We compare the QLFs for the lightcone (solid green curve) with those computed from the snapshots (using the most massive BH per subhalo, as in \citet[][]{Ding_2025}; dotted green curve) and the observations given by \citet[][]{Shen_2020} (solid orange curve, the shaded region indicates $\pm \sigma$). These observed QLFs are the ones we sample from in our abundance matching scheme, described in \S \ref{sec:abundance_matching_motivation_and_method}. Additionally, in Fig. \ref{fig:Luminosity_function}, we overlay the observed datasets compiled by \citet[][]{Shen_2020}.

The lightcone QLFs have a greater abundance of faint quasars than the snapshot QLFs, because the lightcone QLFs include all BH particles in the lightcone whereas the snapshot QLFs include only the most massive BH (MMBH) per subhalo. 
For redshifts $z \leq 1$ the lightcone QLF has an underabundance of bright quasars relative to the snapshots due to cosmic variance. 
Compared to the observations, both sets of FLAMINGO QLFs underestimate the abundance of both bright and faint quasars which, as discussed by \citet[][]{Ding_2025}, is likely due to the limited temporal and spatial resolution of FLAMINGO. 

\begin{figure}
    \includegraphics[width=0.95\columnwidth]{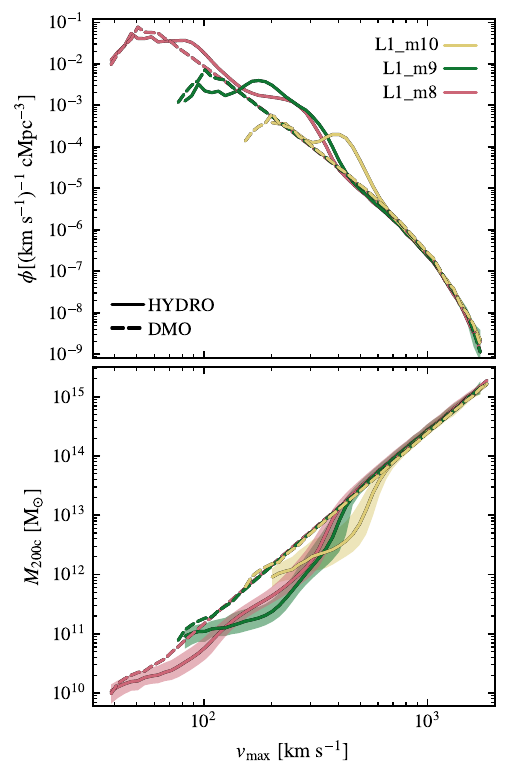}
    \vspace*{-0.5cm}
    \caption{ \textit{Top}: The maximum circular velocity ($v_{\mathrm{max}}$) functions ($\phi$) at redshift $z=0.1$ for both the fiducial hydrodynamic (HYDRO; solid curves) and dark matter only (DMO; dashed curves) $(1~\mathrm{Gpc})^3$ box simulations at each of the available resolutions. These $v_{\mathrm{max}}$ functions include all subhaloes with at least 100 particles.  There is good convergence between the low (L1$\_$m10), intermediate (L1$\_$m9) and high resolutions (L1$\_$m8). \textit{Bottom}: The median (solid and dashed lines) mass of central haloes ($M_{200\mathrm{c}}$) as a function of $v_{\mathrm{max}}$. The shaded regions indicate the $16^{\mathrm{th}}-84^{\mathrm{th}}$ percentiles per $v_{\mathrm{max}}$ bin.}
    \label{figure:Vmax_Function_Mass_relation}
\end{figure}

The top panel of Fig. \ref{figure:Vmax_Function_Mass_relation} presents the maximum circular velocity ($v_{\mathrm{max}}$) function, at redshift $z=0.1$, for the low, intermediate and high resolution 1 Gpc box size simulations (L1$\_$m10, L1$\_$m9 and L1$\_$m8, respectively). For each resolution we show the fiducial dark matter only (DMO; dashed curves) and hydrodynamic (HYDRO; solid curves) variations. The $v_{\mathrm{max}}$ function was computed from all haloes with at least 100 particles in a snapshot at $z=0.1$ using 100 equal width logarithmic bins for $v_{\mathrm{max}}<10^4~\mathrm{km~s}^{-1}$. The shaded regions indicate the $1\sigma$ Poisson error for the corresponding hydrodynamic simulation (indicated by colour).
The bottom panel depicts the total halo mass ($M_{200\mathrm{c}}$) of central haloes as a function of $v_{\mathrm{max}}$ at $z=0.1$ for the same simulations as in the top panel. Each curve indicates the median halo mass,$M_{200\mathrm{c}}$, per velocity bin, whilst shaded regions outline the $16^{\mathrm{th}}$ to $84^{\mathrm{th}}$ percentiles. 
As shown in Fig. \ref{figure:Vmax_Function_Mass_relation}, $v_{\mathrm{max}}$ is well converged with resolution and is a robust indicator of the total dynamical mass within a subhalo, especially for haloes with $M_{200\mathrm{c}}>10^{13}~\mathrm{M}_\odot$.

\section{Supplementary results} \label{appendix_additional_results}

\begin{figure}
        \centering
        \includegraphics[width=\columnwidth]{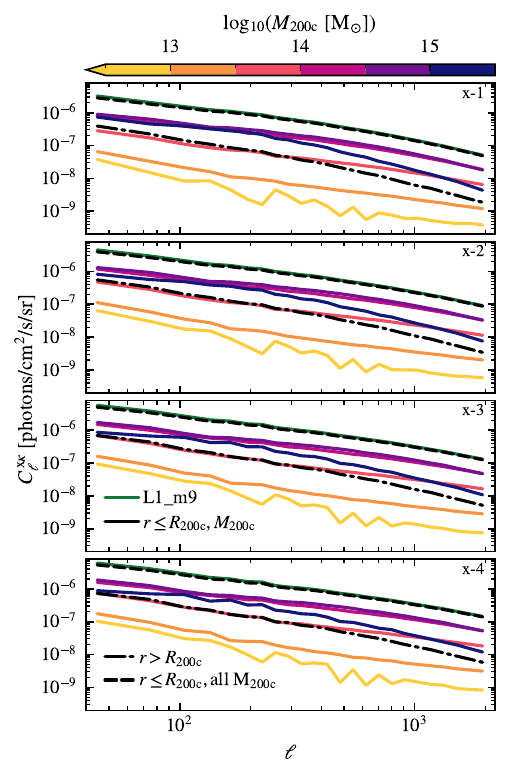}   
        \vspace*{-0.5cm}
    \caption{The same as Fig. \ref{figure:halo_mass_decomposition_example}, but repeated for each tomographic bin, which is indicated in the upper right corner of each panel.}
    \label{figure:halo_mass_decomposition_appendix}
\end{figure}

\begin{figure*}
    \centering
    \includegraphics[width=\textwidth]{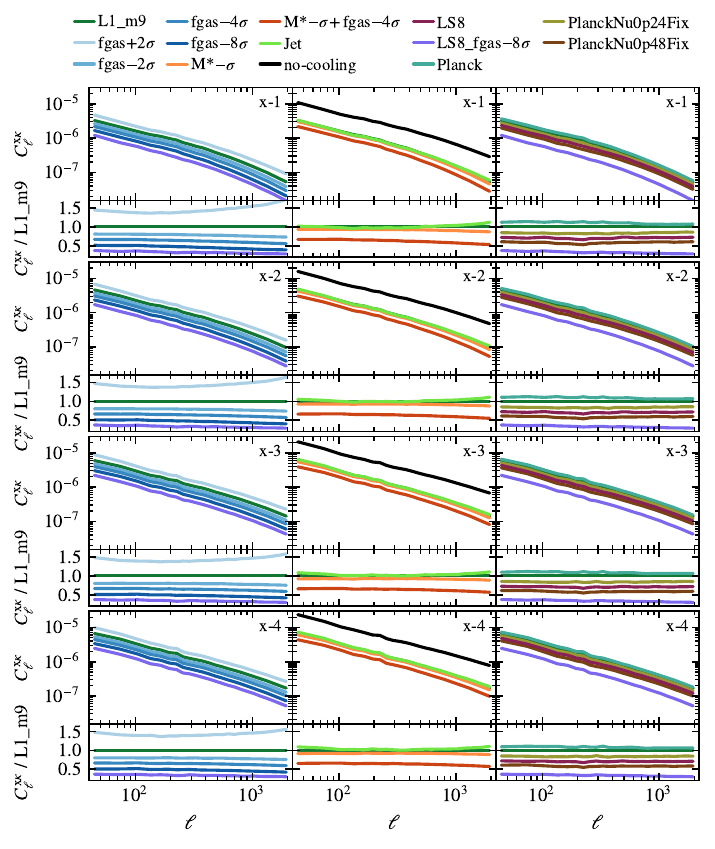}
    \vspace*{-0.5cm}
    \caption{The same as Fig. \ref{figure:normalised_diffuse_gas_example}, but for each tomographic bin, which is printed in the upper right corner of each main panel. The cross-correlations shown in the main panels are therefore in units of ${\mathrm{photons}}/{\mathrm{s}}/\mathrm{cm}^2/{\mathrm{sr}}$}
    \label{figure:normalised_diffuse_gas_appendix}
\end{figure*}

\begin{figure}
        \centering
        \includegraphics[width=\columnwidth]{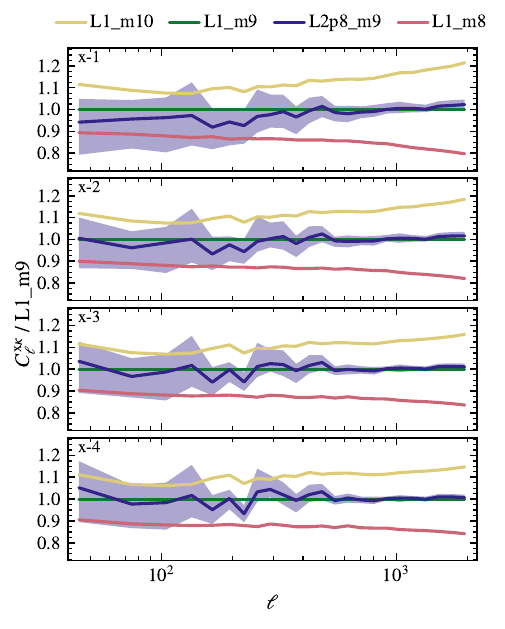}
        \vspace*{-0.5cm}
    \caption{The same as lower panel of Fig. \ref{figure:boxsize_resolution_example}, but repeated for each tomographic bin, which is indicated in the upper left corner of each panel.}
    \label{figure:boxsize_resolution_appendix}
\end{figure}

In this appendix we provide supplementary results and repeat figures presented in \S \ref{sec:results}, but showing all tomographic bins instead of only the third tomographic bin (x-3).
For each of the figures shown, the conclusions drawn from the corresponding figures in the main text (\S \ref{sec:results}) hold when looking at all tomographic bins.

Figs. \ref{figure:halo_mass_decomposition_appendix}, \ref{figure:normalised_diffuse_gas_appendix} and \ref{figure:boxsize_resolution_appendix}, corresponding to Fig. \ref{figure:halo_mass_decomposition_example}, \ref{figure:normalised_diffuse_gas_example} and  \ref{figure:boxsize_resolution_example} in \S \ref{sec:diffuse_gas_results}, respectively, show the idealised cross-spectra (where we only include the X-ray emission from hot gas, do not convolved with the ROSAT response matrix or account for Galactic foreground absorption) in each tomographic bin.

\begin{figure*}
    \centering
    \includegraphics[width=\textwidth]{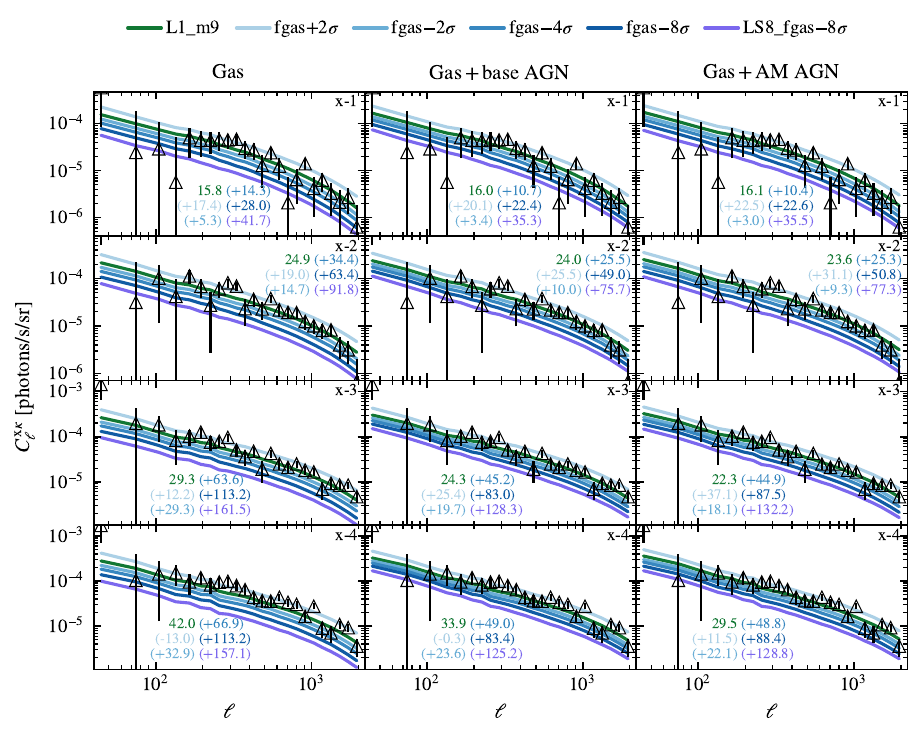}
    \vspace*{-0.5cm}
    \caption{The same as Fig. \ref{figure:convolved_native_agn_example}, but for each tomographic bin, which is indicated in the top right corner of each panel.}
    \label{figure:convolved_abundance_match_appendix}
\end{figure*}

In Fig. \ref{figure:convolved_abundance_match_appendix} we repeat the comparisons between FLAMINGO and the observations shown in Fig. \ref{figure:convolved_native_agn_example}, but in each tomographic bin. Here we show the predicted ROSAT-like X-ray--lensing cross-correlations when we include the X-ray emission from only hot gas (left column) and when the X-ray emission is contaminated by unresolved AGN given by the base BHs (centre column) or AM BHs (right column). For each simulation we compute the $\chi^2$ value between the predicted X-ray--lensing cross-correlation and the measurements of F24 (black points and error bars). In each panel we print the corresponding $\chi^2$ value for the fiducial L1$\_$m9 simulation and in parentheses we show the difference in $\chi^2$ ($\Delta \chi^2$) for each feedback variation and the fiducial L1$\_$m9 simulation.

\begin{figure}
    \centering
    \includegraphics[width=\columnwidth]{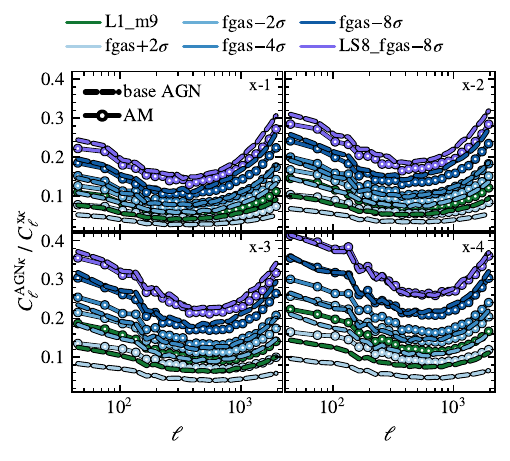}
    \vspace*{-0.5cm}
    \caption{The contribution from unresolved AGN to the cross-correlation when the X-ray emission from AGN is given by the base BHs (dashed curves) or AM BHs (circular markers). The tomographic bin is indicated in the upper right corner of each panel.}
    \label{figure:convolved_abundance_match_appendix_ratio}
\end{figure}

Fig.~\ref{figure:convolved_abundance_match_appendix_ratio} shows the contribution of unresolved AGN to the total predicted cross-correlations, as shown in Fig.~\ref{figure:convolved_abundance_match_appendix}, for both the base FLAMINGO BHs (dashed curves) and abundance matched BHs (circular markers). As discussed in \S \ref{sec:results_the_fiducial_model} and \S \ref{sec:comparison_with_direct_measurements}, we see that the impact of unresolved AGN increases in more distant tomographic bins. 

Table \ref{tab:x2_delta_AM} presents the sum of $\chi^2$ values (top panel) and number of standard deviations ($N_{\sigma}$; bottom panel) of the $\Delta \mathrm{AM}$ AGN contaminated X-ray--lensing cross-correlations (as shown in Figs. \ref{figure:delta_example} and \ref{figure:AM_summary_stat}) with respect to the measurements of F24. Each column lists the summary statistics ($\chi^2$ and $N_{\sigma}$) for each of the baryonic feedback variations shown in Fig. \ref{figure:AM_summary_stat} when the $\Delta \mathrm{AM}$ AGN contaminated cross-correlation is constructed with a given value of $\sigma_r$, the log-normal scatter imposed on the BH Luminosity -- maximum circular velocity relation (see \S \ref{sec:abundance_matching_motivation_and_method}), which is indicated in the top row of the table.

\begin{table*}
\centering
\caption{The $\chi^2$ (top panel) and $N_{\sigma}$ (bottom panel) values (as computed for Table~\ref{tab:x2_native_AGN} discussed in \S\ref{sec:comparison_with_direct_measurements}) for the $\Delta\mathrm{AM}$ AGN contaminated cross-correlations of selected FLAMINGO L1$\_$m9 simulations with respect to the measurements of F24. The populations of $\Delta\mathrm{AM}$ BHs are parametrised by $\sigma_r$, which describes the log-normal scatter imposed on the halo maximum circular velocities during abundance matching as described in \S\ref{sec:abundance_matching_results}. The left-most column lists the FLAMINGO simulation identifier, while each subsequent column gives the corresponding $\chi^2$ (top panel) and $N_{\sigma}$ (bottom panel) values, for the $\sigma_r$ value listed at the top of the column.}
\label{tab:x2_delta_AM}
\begin{tabular}{l rrrrrrrrrrr}
\hline
& \multicolumn{11}{c}{$\sigma_r$ [dex]} \\
Identifier                        & 0.00  & 0.10  & 0.20  & 0.30  & 0.40  & 0.50  & 0.60  & 0.70  & 0.80  & 0.90  & 1.00 \\
\hline
\hline
L1$\_$m9                           	& 92.0 	& 89.4 	& 89.8 	& 92.3 	& 94.1 	& 94.7 	& 95.6 	& 95.9 	& 96.0 	& 96.1 	& 96.3 	\\
fgas$+2\sigma$                     	& 258.2	& 235.2	& 202.3	& 186.7	& 180.3	& 177.2	& 175.3	& 174.0	& 173.4	& 173.1	& 172.5	\\
fgas$-2\sigma$                     	& 107.7	& 115.3	& 134.0	& 146.2	& 152.0	& 154.7	& 156.3	& 157.5	& 157.8	& 158.2	& 158.8	\\
fgas$-4\sigma$                     	& 154.9	& 172.0	& 204.2	& 223.1	& 231.9	& 235.9	& 238.2	& 239.7	& 240.6	& 241.3	& 241.7	\\
fgas$-8\sigma$                     	& 242.2	& 268.0	& 315.7	& 341.0	& 351.7	& 357.9	& 360.2	& 362.5	& 363.4	& 364.4	& 365.2	\\
LS8$\_$fgas$-8\sigma$              	& 353.1	& 384.9	& 440.2	& 467.5	& 478.6	& 485.0	& 488.3	& 489.9	& 491.7	& 492.7	& 493.6	\\
\hline
L1$\_$m9                           	&  0.0 	&  0.0 	&  0.0 	&  0.0 	&  0.0 	&  0.0 	&  0.0 	&  0.0 	&  0.0 	&  0.1 	&  0.1 	\\
fgas$+2\sigma$                     	&  8.2 	&  7.3 	&  5.9 	&  5.2 	&  4.9 	&  4.8 	&  4.7 	&  4.6 	&  4.6 	&  4.6 	&  4.5 	\\
fgas$-2\sigma$                     	&  0.9 	&  1.4 	&  2.5 	&  3.2 	&  3.5 	&  3.6 	&  3.7 	&  3.8 	&  3.8 	&  3.8 	&  3.9 	\\
fgas$-4\sigma$                     	&  3.6 	&  4.5 	&  6.0 	&  6.8 	&  7.2 	&  7.4 	&  7.5 	&  7.5 	&  7.6 	&  7.6 	&  7.6 	\\
fgas$-8\sigma$                     	&  7.6 	&  8.6 	& 10.3 	& 11.1 	& 11.5 	& 11.7 	& 11.8 	& 11.8 	& 11.9 	& 11.9 	& 11.9 	\\
LS8$\_$fgas$-8\sigma$              	& 11.5 	& 12.5 	& 14.1 	& 14.9 	& 15.2 	& 15.3 	& 15.4 	& 15.4 	& 15.5 	& 15.5 	& 15.5 	\\

\hline
\end{tabular}
\end{table*}

\begin{figure}
    \centering
    \includegraphics[width=0.95\columnwidth]{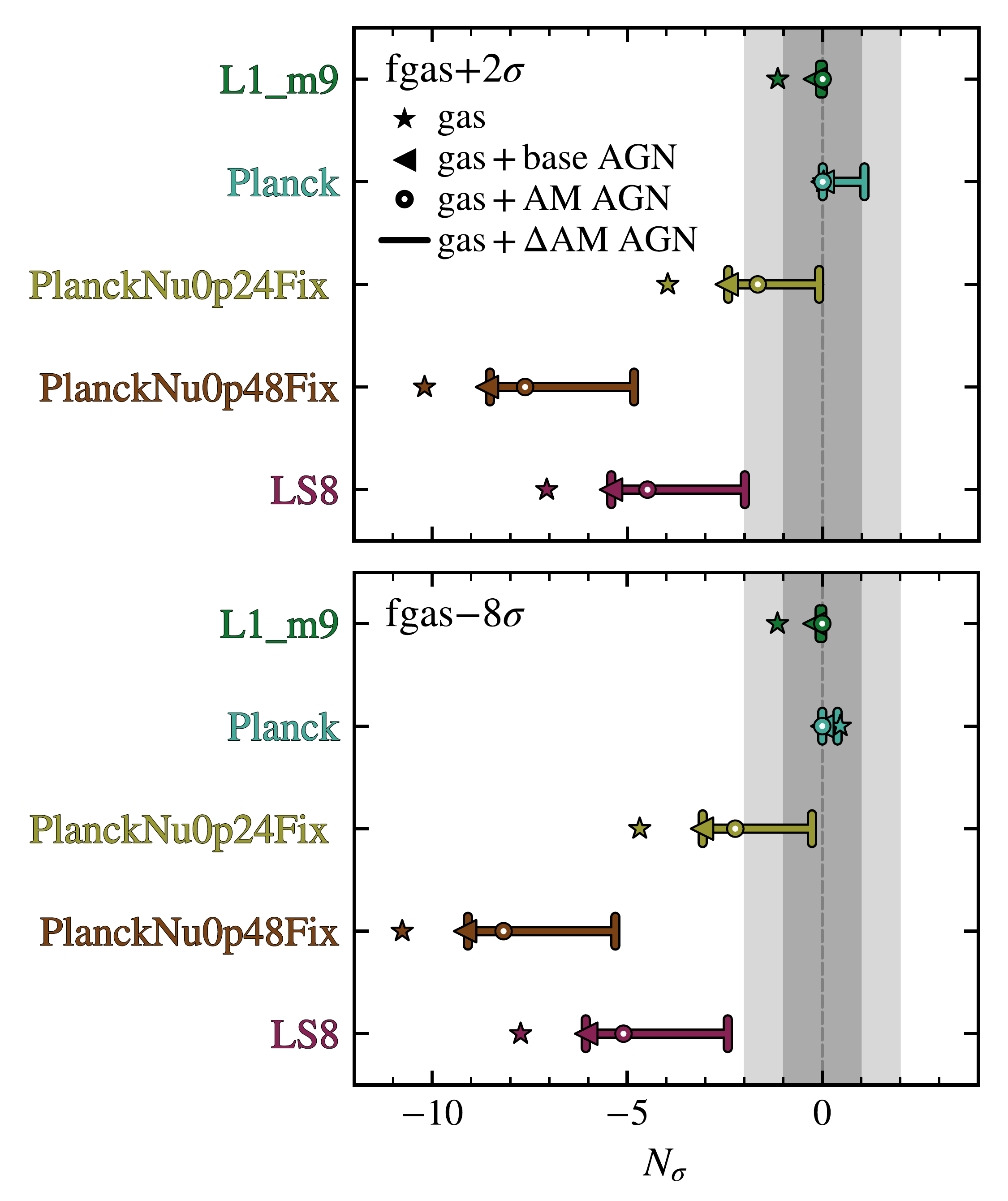}
    \caption{ The same as Fig. \ref{figure:AM_summary_stat_cosmo} but for the case when each cross-correlation is approximated using $R_{\mathrm{L1\_m9}}$ given the fgas$+2\sigma$ (top panel) and fgas$-8\sigma$ (bottom panel) simulations.}
    \label{figure:AM_summary_stat_cosmo_appendix}
\end{figure}

Fig. \ref{figure:AM_summary_stat_cosmo_appendix} repeats Fig. \ref{figure:AM_summary_stat_cosmo}, where we approximate each cross-correlation  as done in \S \ref{sec:comparison_with_direct_measurements_cosmo} (see equations \ref{eq:cosmo_gas_approx} and equations \ref{eq:cosmo_gas_agn_approx}), but now the expected effect of the ROSAT response matrix on the cross-correlation ($R_{\mathrm{L1\_m9}}$) is given by the gas$+2\sigma$ (top panel) and fgas$-8\sigma$ (bottom panel) simulations instead of the fiducial L1$\_$m9 simulation. 
As shown in Fig. \label{figure:response_function_appendix}, convolving the X-ray emission from hot gas with the ROSAT response matrix has a greater impact on the X-ray--lensing cross-correlation for simulations calibrated to yield a higher cluster gas fraction. Hence, Fig. \ref{figure:AM_summary_stat_cosmo_appendix} shows that the conclusions made in \S \ref{sec:comparison_with_direct_measurements_cosmo} are robust to the changes in the expected impact of the ROSAT response matrix, $R_{\mathrm{L1\_m9}}$, that are found between the different cluster gas fraction variations and should therefore be robust for the differences in $R_{\mathrm{L1\_m9}}$ between the different cosmology variations.

\section{Sensitivity to box size and resolution: halo scaling relations} \label{appendix_scaling_relations}

As shown by Fig. \ref{figure:boxsize_resolution_example}, the X-ray--lensing cross-correlation for FLAMINGO is converged with box size but not with resolution.
In this appendix we use the cluster gas fraction -- halo mass and X-ray luminosity -- halo mass relations to demonstrate why the cross-correlation for the high-resolution (L1$\_$m8) FLAMINGO simulation is suppressed relative to the intermediate-resolution (L1$\_$m9) simulation.

\begin{figure}
    \includegraphics[width=\columnwidth]{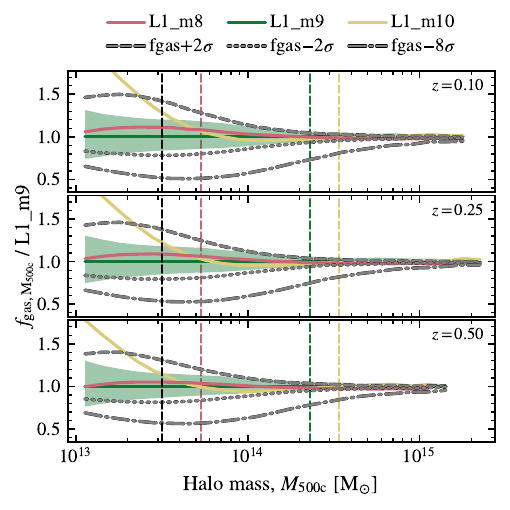}
    \vspace*{-0.5cm}
    \caption{The median cluster gas mass fraction within $R_{500\mathrm{c}}$ as a function of halo mass, $M_{500\mathrm{c}}$, at redshift $z=0.1$ (top panel), $z=0.25$ (middle panel) and $z=0.5$ (bottom panel) for different resolutions and box sizes relative to the result for the L1$\_$m9 simulation. The shaded region outlines the 16th to 84th percentile scatter of the L1$\_$m9 simulation. The $+2\sigma$, $-2\sigma$ and $-8\sigma$ cluster gas fraction variations of the L1$\_$m9 simulation are overlaid in grey (dashed, dotted and dot-dashed, respectively). The vertical dashed lines give the lower (black) and upper (coloured) mass bounds used for the calibration of subgrid models to X-ray observations at redshift $z\approx0.1$ \citep[][]{FLAMINGO_Calibration_2023} of each resolution.}
    \label{figure:boxsize_resolution_fgas_example}
\end{figure}

In Fig. \ref{figure:boxsize_resolution_fgas_example} we examine the convergence of the gas fraction ($f_{\mathrm{gas},M_{500\mathrm{c}}}$)  -- halo mass ($M_{500\mathrm{c}}$) scaling relation with resolution at $z=0.1, 0.25$ and $0.5$ (the X-ray--lensing cross-correlation is dominated by X-ray emission from $z\leq0.5$, see Fig. \ref{figure:redshift_decomposition}). 
The solid curves present the median relations for each resolution, at a fixed box size ($1~\mathrm{Gpc}$), relative to the fiducial L1$\_$m9 simulation for clusters at redshifts $z=0.1$ (top panel), $z=0.25$ (middle panel) and $z=0.5$ (bottom panel).
For comparison we overlay the $+2\sigma$, $-2\sigma$ and $-8\sigma$ cluster gas fraction (fgas$+2\sigma$,fgas$-2\sigma$ and fgas$-8\sigma$ respectively) variations of the L1$\_$m9 simulation.
The high resolution (L1$\_$m8) simulation has a slightly lower gas fraction relative to the fiducial L1$\_$m9 simulation for halo masses $M_{500\mathrm{c}} \gtrsim 1.5\times10^{14}~\mathrm{M}_\odot$, i.e., the halo masses which dominate the X-ray component of the cross-correlation (see Fig. \ref{figure:halo_mass_decomposition_example}).
Given the X-ray--lensing cross-correlation's clear dependence on cluster gas fraction (as shown in \S\ref{sec:sensitivity_to_cosmology_and_astrophysics}) and the juxtaposition of the fgas$-2\sigma$ simulation, it should not be a surprise that the predicted cross-correlation for L1$\_$m8 is suppressed relative to that of L1$\_$m9. The minor divergence of L1$\_$m8 from L1$\_$m9 occurs for masses above the range that L1$\_$m8 was calibrated on ($M_{500\mathrm{c}} = 10^{13.5} - 10^{13.75} ~\mathrm{M}_\odot$). 
These same arguments can be extended to describe the elevation of the cross-correlation for the low-resolution simulation (L1$\_$m10) relative to the fiducial L1$\_$m9 simulation.

\begin{figure}
    \includegraphics[width=\columnwidth]{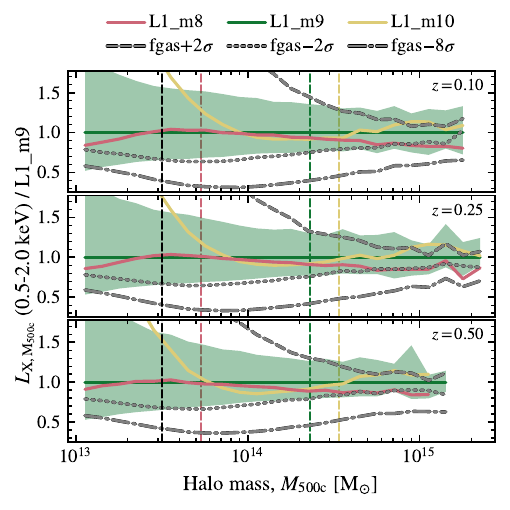}
    \vspace*{-0.5cm}
    \caption{As Fig. \ref{figure:boxsize_resolution_fgas_example}, but for the median soft X-ray luminosity (0.5--2.0 keV). }
    \label{figure:boxsize_resolution_Lx_example}
\end{figure}

In Fig. \ref{figure:boxsize_resolution_Lx_example}, we present the X-ray luminosity ($L_{\mathrm{x},M_{500\mathrm{c}}}$) -- halo mass ($M_{500\mathrm{c}}$) relation, where the X-ray luminosity is in the soft band (0.5 - 2.0 keV). The X-ray luminosity -- halo mass scaling relations are not identical for each model because of the different hot gas profiles that come from the imperfect recalibration of subgrid prescriptions at each resolution.  The layout of Fig. \ref{figure:boxsize_resolution_Lx_example} is the same as that of Fig. \ref{figure:boxsize_resolution_fgas_example}.  
The X-ray luminosities of haloes with $M_{500\mathrm{c}} \gtrsim 5\times10^{14}~\mathrm{M}_\odot$ in the fgas$-2\sigma$ and L1$\_$m8 simulations are very similar for the redshifts shown. Hence, the cross-correlations of the two simulations are similar, because the haloes that dominate the X-ray component of the cross-correlation have similar X-ray luminosities.

\section{Masking Point Sources} \label{appendix_binary_mask}

\begin{figure}
    \includegraphics[width=\columnwidth]{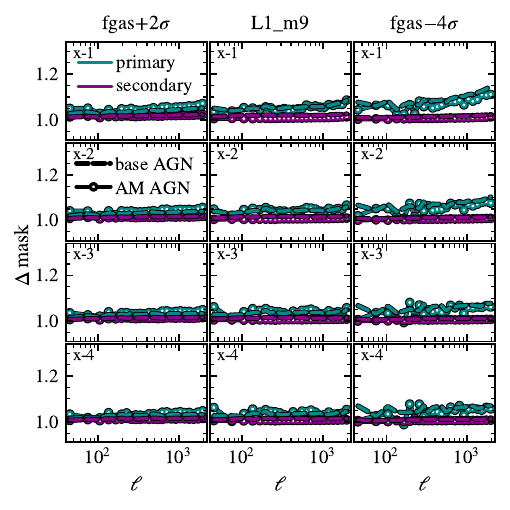}
    \vspace*{-0.5cm}
    \caption{ The relative increase in signal of the X-ray--lensing cross-correlation for the fiducial L1$\_$m9 , fgas$+2\sigma$ and fgas$-8\sigma$ simulations when the effects of the binary mask are neglected, as described in the text.}
    \label{figure:no_mask}
\end{figure}

In Fig. \ref{figure:no_mask} we examine the impact of the binary mask on the X-ray--lensing cross-correlation. Here we show the relative change in signal for the AGN-contaminated cross-correlations for the fiducial (i.e.\ L1$\_$m9), fgas$+2\sigma$ and fgas$-8\sigma$ simulations due to excluding the desired primary (cyan) and undesired secondary (purple) effects of the binary mask as described in \ref{sec:comparison_with_direct_measurements}. Here we approximate the total secondary effects of the binary mask as $\Delta_{\mathrm{mask}} = (C_{\ell}^{{\mathrm{gas}}\,\kappa}+C_{\ell}^{{\mathrm{AGN}}\,\kappa}) \, / \, C_{\ell}^{{\mathrm{x}}\kappa}$, where $C_{\ell}^{{\mathrm{AGN}}\,\kappa}$ is the cross-correlation when the X-ray emission is only given by unresolved AGN. $C_{\ell}^{{\mathrm{AGN}}\,\kappa}$ is computed without a mask, X-ray emission from resolved point sources is zeroed. The change in signal from the mask's primary effect is given as $\Delta_{\mathrm{mask}} = C_{\ell}^{{\mathrm{x}_{\mathrm{all}}}\,\kappa} \, / \, C_{\ell}^{{\mathrm{x}}\kappa}$, where $C_{\ell}^{{\mathrm{x}_{\mathrm{all}}}\,\kappa}$ is for the case when the X-ray emission is only from both resolved and unresolved AGN.

\section{predicted eROSITA cross-correlations} \label{appendix_eROSITA}

As noted in \S \ref{sec:comparison_with_direct_measurements} and \S \ref{sec:comparison_with_direct_measurements_cosmo}, the uncertainty on the impact of AGN contamination reduces the efficacy of the X-ray--lensing cross-correlation as a probe of large-scale structure. 

Adopting higher resolution X-ray data, such as data from the SRG/eROSITA telescope \citep[][]{Sunyaev_2021, Predehl_2021}, can offer an immediate improvement over the ROSAT X-ray measurements used in this paper \citep[see relevant discussions and comparisons of e.g.][]{Merloni_2024, Zheng_2024}. 
The current publicly available half-sky (western hemisphere) X-ray data and catalogues from the eROSITA-DE All-Sky Survey \citep[e.g. the first public data release; eRASS1\footnote{\url{https://erosita.mpe.mpg.de/dr1/}},][]{Merloni_2024, Bulbul_2024} have been used to examine 2-point correlation functions \citep[e.g,][]{Seppi_2024,Comparat_2025}, X-ray angular power spectra \citep[e.g,][]{Lau_2024}, and X-ray cluster scaling relations \citep[e.g,][]{Ramos_2025}.
More importantly, the available eRASS cluster catalogues has a significant overlap with the footprints of the weak-lensing data of the DES-Y3, KiDS \citep[e.g.,KiDS-1000 and KiDS-Legacy samples;][]{Kuijken_2019, Wright_2025} and HSC \citep[e.g.,the year 3 data release][]{Li_2022} catalogues as demonstrated in the eRASS1 mass calibration \citep[e.g,][]{Grandis_2024_WL_eRASS, Kleinebreil_2024}. It is therefore possible to measure an X-ray--cosmic-shear cross-correlation with eROSITA, as all the required data are available.

We compute an eROSITA--DES-Y3 X-ray--cosmic-shear cross-correlation for the fiducial L1$\_$m9 simulation. We first construct eROSITA-like X-ray all-sky maps using the method described in \S \ref{sec:methods}, except that the X-ray emission is convolved with the effective area information given by the publicly available survey-averaged eROSITA\footnote{We select the ARF of TM8 as it excludes TM5 and TM7 which suffer from light leakage \citep[e.g.][]{Predehl_2021}, \url{https://erosita.mpe.mpg.de/dr1/eSASS4DR1/eSASS4DR1_arfrmf/}} auxiliary response files.
Each of the all-sky maps are constructed in the same ROSAT soft X-ray band (0.5--2.0 keV) at an Nside of 16384, which we downsample to an Nside of 8192, matching the eROSITA half-sky maps of \citet[][]{Lau_2024}, prior to masking all AGN with flux (in the soft band, $0.5--2.0$ keV) above $5\times 10^{-14}~\mathrm{erg~s}^{-1}~\mathrm{cm}^{-2}$, matching the flux limits indicated by \citet[][]{Seppi_2022, Merloni_2024, Lau_2024, Toptun_2025}. The resultant X-ray maps are then downsampled again to Nside 1024, matching the resolution of the DES-Y3 lensing maps, before computing the X-ray--lensing cross-correlations as described in \S \ref{sec:cross-corr_method}. 
Finally, we apply a Gaussian beam with a full-width-half-max of $25.7^{\prime\prime}$ to approximate the eROSITA point-spread function \citet[][]{Lau_2024}. 

\begin{figure}
    \includegraphics[width=\columnwidth]{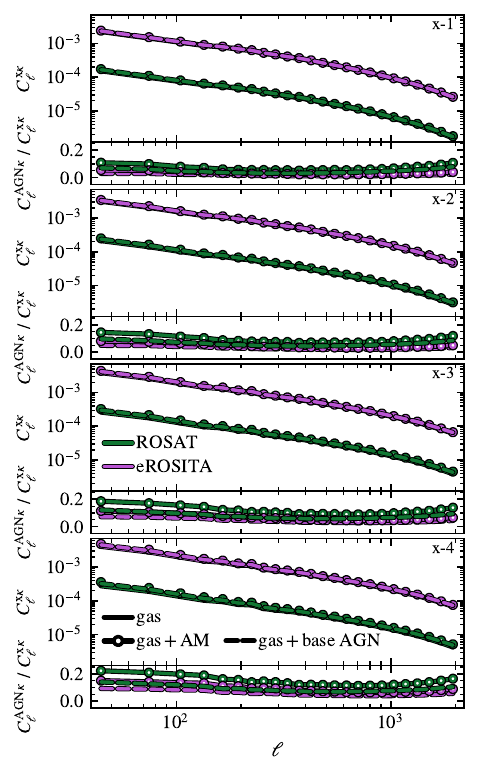}
    \vspace*{-0.5cm}
    \caption{The ROSAT--DES-Y3 (green) and eROSITA--DES-Y3 (purple) X-ray–lensing cross-correlations for the fiducial L1$\_$m9 simulation. The cross-correlations are computed for when the X-ray emission is exclusively from hot gas (solid curve) and for the case when we include the X-ray emission from both hot gas and unresolved AGN, given by either the base AGN (dashed curve) or AM AGN (curve with circular markers). All X-ray emission is in the 0.5--2.0 keV band convolved with either the ROSAT or the eROSITA response matrix, as described in the text. The cross-correlations shown in the main panels are therefore in units of ${\mathrm{photons}}/{\mathrm{s}}/{\mathrm{sr}}$. Each sub-figure corresponds to a different tomographic bin, as indicated in the upper right corner of each main panel. The smaller sub-panels show the ratio of the X-ray--lensing cross-correlations when only the X-ray emission from unresolved AGN is included ($C_{\ell}^{\mathrm{AGN} \kappa}$) and when both the X-ray emission from hot gas and unresolved AGN are accounted for ($C_{\ell}^{\mathrm{x} \kappa}$).}
    \label{figure:eRASS_all_bins}
\end{figure}

The resultant eROSITA-like X-ray--lensing cross-correlations for the fiducial L1$\_$m9 simulation are compared to their ROSAT-like counter-parts in Fig. \ref{figure:eRASS_all_bins}. 
We find that for the gas-only cross-correlations (solid curve), the amplitude of the eROSITA-like cross-correlation is, almost uniformly, greater than that of the ROSAT-like cross-correlation by a factor of $\approx 13.9-14.1$. This increase in signal is mostly due to the difference in the effective area ($A_{\mathrm{eff}}$) of the two telescopes, where for the same (0.5--2.0 keV) soft band $A_{\mathrm{eff}}$ for eROSITA is $\approx15.2$ (ranging between $9.67$ and $64$) times greater than for ROSAT. 
Additionally, as highlighted in the sub-panels for each tomographic bin, the eROSITA-like cross-correlations show a minor reduction in the contribution of unresolved AGN relative to the ROSAT-like cross-correlations. This is caused by the greater sensitivity to point sources and lower flux limit down to which AGN are resolved by eROSITA compared to ROSAT.  
However, any significant improvements due to the greater resolution of eROSITA compared to ROSAT are hamstrung by the requirement that the X-ray all-sky maps are downsampled to match the even lower resolution of the lensing all-sky maps.

\label{lastpage}
\end{document}